\documentclass[10pt,a4paper]{article}
\pdfoutput=1

\usepackage{setspace}
\usepackage[T1]{fontenc}
\usepackage{amsfonts,amsbsy,bm,euscript,mathrsfs}
\usepackage{amssymb,stmaryrd,faktor}
\usepackage[tbtags]{amsmath}
\usepackage[title,titletoc]{appendix}

\usepackage{graphics}
\usepackage{tikz}
\definecolor{hyperref}{RGB}{026,028,087}

\usepackage[bookmarks=true,colorlinks=true,linkcolor=hyperref,citecolor=hyperref,urlcolor=hyperref,bookmarksnumbered]{hyperref}

\newcommand{\MAT}[1]{\begin{pmatrix} #1 \end{pmatrix}}
\newcommand{\ARR}[1]{\begin{matrix} #1 \end{matrix}}
\newcommand{\eq}[1]{\begin{equation}\begin{split} #1 \end{split}\end{equation}}
\newcommand{\al}[1]{\begin{align} #1 \end{align}}
\newcommand{\als}[1]{\begin{subequations}\begin{align} #1 \end{align}\end{subequations}}
\newcommand{\aln}[1]{\begin{align*} #1 \end{align*}}

\newcommand{\ta}[1]{\tag{\theequation} \addtocounter{equation}{1} \label{#1}}
\newcommand{\la}{\label}

\newcommand{\no}{\nonumber}

\newcommand{\ph}{\vphantom}

\newcommand{\X}{\times}
\newcommand{\fr}{\frac}
\newcommand{\fk}{\faktor}
\newcommand{\sfr}[2]{\textrm{\tiny $\fk{#1}{#2}$}}

\newcommand{\sech}{\operatorname{sech}}
\newcommand{\csch}{\operatorname{csch}}
\newcommand{\Tr}{\operatorname{Tr}}

\newcommand{\STr}{\operatorname{STr}}

\newcommand{\sign}{\operatorname{sign}}

\newcommand{\const}{\text{const}}
\newcommand{\diag}{\text{diag}}

\newcommand{\com}[2]{[ #1,\,#2 ]}
\newcommand{\acom}[2]{\{#1 ,\,#2\}}
\newcommand{\del}{\partial}
\newcommand{\dpl}{\partial_+}
\newcommand{\dm}{\partial_-}
\newcommand{\dpdm}{\partial_+\partial_-}
\newcommand{\dpm}{\partial_\pm}
\newcommand{\dmp}{\partial_\mp}

\newcommand{\Z}{\mathbb{Z}}
\newcommand{\R}{\mathbb{R}}
\newcommand{\C}{\mathbb{C}}

\newcommand{\ra}{\rightarrow}
\newcommand{\Ra}{\Rightarrow}

\newcommand{\id}{\mathbf{1}}
\newcommand{\ze}{\mathbf{0}}
\newcommand{\inv}[1]{#1^{-1}}
\newcommand{\nv}[1]{#1^{\vphantom{-1}}}
\newcommand{\td}{\tilde}

\newcommand{\wh}{\widehat}
\newcommand{\wt}{\widetilde}

\newcommand{\tpsuttf}{$\mathbf{\mfpsu\boldsymbol{(}\textbf{2}\boldsymbol{,}\textbf{2}\boldsymbol{|}\textbf{4}\boldsymbol{)}}$}
\newcommand{\Fh}{\wh F}
\newcommand{\adss}[1]{$\textrm{AdS}_{#1} \X S^{#1}$}

\newcommand{\ads}{{\textrm{AdS}}}
\newcommand{\tads}{{\textbf{AdS}}}

\newcommand{\Lag}{\mathcal{L}}

\newcommand{\Act}{\mathcal{S}}

\newcommand{\YL}{\Psi_{\!_L}}
\newcommand{\YR}{\Psi_{\!_R}}
\newcommand{\YLR}{\Psi_{\!_{L,R}}}
\newcommand{\tYL}{\wt\Psi_{\!_L}}
\newcommand{\tYR}{\wt\Psi_{\!_R}}
\newcommand{\tYLR}{\wt\Psi_{\!_{L,R}}}

\newcommand{\pa}{\parallel}
\newcommand{\pe}{\perp}
\newcommand{\PSU}{\text{PSU}}

\newcommand{\SL}{\text{SL}}

\newcommand{\SU}{\text{SU}}
\newcommand{\SO}{\text{SO}}

\renewcommand{\O}{\text{O}}
\newcommand{\U}{\text{U}}
\newcommand{\USp}{\text{USp}}
\newcommand{\eom}{e.o.m.}
\newcommand{\dof}{d.o.f.}

\newcommand{\upm}{{\boldsymbol{\pm}}}
\newcommand{\ump}{{\boldsymbol{\mp}}}
\newcommand{\up}{{\boldsymbol{+}}}
\newcommand{\um}{{\boldsymbol{-}}}

\DeclareFontFamily{U}{mathx}{\hyphenchar\font45}
\DeclareFontShape{U}{mathx}{m}{n}{<5> <6> <7> <8> <9> <10> <10.95> <12> <14.4> <17.28> <20.74> <24.88> mathx10}{}
\DeclareSymbolFont{mathx}{U}{mathx}{m}{n}
\DeclareFontSubstitution{U}{mathx}{m}{n}
\DeclareMathAccent{\widecheck}{0}{mathx}{"71}

\newcommand{\Ac}{\mathcal{A}}

\newcommand{\Dc}{\mathcal{D}}

\newcommand{\Jc}{\mathcal{J}}
\newcommand{\Kc}{\mathcal{K}}

\newcommand{\Nc}{\mathcal{N}}
\newcommand{\Oc}{\mathcal{O}}
\newcommand{\Pc}{\mathcal{P}}
\newcommand{\Qc}{\mathcal{Q}}

\newcommand{\Uc}{\mathcal{U}}
\newcommand{\Vc}{\mathcal{V}}

\newcommand{\Ds}{\mathscr{D}}

\newcommand{\Ts}{\mathscr{T}}
\newcommand{\Xs}{\mathscr{X}}
\newcommand{\Ys}{\mathscr{Y}}

\newcommand{\ISC}{\text{\scriptsize{$I$}}}
\newcommand{\JSC}{\text{\scriptsize{$J$}}}

\newcommand{\mfpsu}{\mathfrak{psu}}

\newcommand{\mfsu}{\mathfrak {su}}
\newcommand{\mfso}{\mathfrak {so}}

\newcommand{\mfu}{\mathfrak {u}}
\newcommand{\mfusp}{\mathfrak {usp}}
\newcommand{\mfa}{\mathfrak a}

\newcommand{\mff}{\mathfrak f}
\newcommand{\mffh}{\wh{\mathfrak f}}

\newcommand{\mfg}{\mathfrak g}
\newcommand{\mfh}{\mathfrak h}

\newcommand{\mfk}{\mathfrak k}
\newcommand{\mfl}{\mathfrak l}
\newcommand{\mfm}{\mathfrak m}
\newcommand{\mfn}{\mathfrak n}

\newcommand{\mfp}{\mathfrak p}

\newcommand{\mfr}{\mathfrak r}

\newcommand{\mft}{\mathfrak t}

\newcommand{\mfA}{\mathfrak A}
\newcommand{\mfB}{\mathfrak B}

\newcommand{\mfI}{\mathfrak I}
\newcommand{\mfJ}{\mathfrak J}

\newcommand{\mfP}{\mathfrak P}

\newcommand{\mfW}{\mathfrak W}
\newcommand{\mfX}{\mathfrak X}
\newcommand{\mfY}{\mathfrak Y}

\renewcommand{\a}{\alpha}
\renewcommand{\b}{\beta}
\renewcommand{\c}{\gamma}
\renewcommand{\d}{\delta}
\newcommand{\e}{\epsilon}
\newcommand{\z}{\zeta}
\newcommand{\h}{\eta}

\renewcommand{\k}{\kappa}
\renewcommand{\l}{\lambda}
\newcommand{\m}{\mu}
\newcommand{\n}{\nu}
\newcommand{\y}{\xi}
\newcommand{\p}{\phi}
\renewcommand{\r}{\rho}
\newcommand{\s}{\sigma}
\renewcommand{\t}{\tau}

\newcommand{\x}{\chi}
\newcommand{\w}{\omega}

\newcommand{\Q}{\Theta}

\renewcommand{\P}{\Phi}
\renewcommand{\S}{\Sigma}
\newcommand{\W}{\Omega}
\newcommand{\ve}{\varepsilon}

\newcommand{\vp}{\varphi}
\newcommand{\vs}{\varsigma}

\newcommand{\coup}{{\rm k}}
\newcommand{\yp}{\xi_\up}
\newcommand{\ym}{\xi_\um}
\newcommand{\ypm}{\xi_\upm}
\newcommand{\ymp}{\xi_\ump}
\newcommand{\Lu}{\Lambda_{_{+}}}
\newcommand{\Ld}{\Lambda_{_{-}}}
\newcommand\mon{\mathbf{0}_{n-2}}
\newcommand\monn{\mathbf{0}_{(n-2)^2}}
\newcommand\ion{\mathbf{E}_{i}}
\newcommand\ionn{\mathbf{E}_{ij}}
\newcommand\mont{\mathbf{0}_{n-3}}
\newcommand\monnt{\mathbf{0}_{(n-3)^2}}
\newcommand\iont{\mathbf{E}_{\ISC}}
\newcommand\mmm{{\rm n}}
\newcommand\xxx{x}
\newcommand\uu{{\mathbf{u}}}
\newcommand\vv{{\mathbf{v}}}
\newcommand\typ{{\tilde{\xi}_+}}
\newcommand\tym{{\tilde{\xi}_-}}
\newcommand\tk{{\tilde{k}}}
\newcommand\tyu{{\tilde {u}}}
\newcommand\tyv{{\tilde {v}}}
\newcommand{\gol}{g_1^{\vphantom{-1}}}
\newcommand{\gtl}{g_2^{\vphantom{-1}}}
\newcommand{\kK}{{\rm K}}

\begin{document}


\phantomsection
\pdfbookmark[1]{Title and Abstract}{pre:Tit}

\thispagestyle{empty}

\begin{center}

\mbox{}

\vspace{-1truecm}
\rightline{Imperial-TP-BH-2012-02}

\vspace{2truecm}
{\Large \bf  Pohlmeyer reduction for superstrings in AdS space}

\vspace{1truecm}

{B. Hoare\,\footnote{\href{mailto:benjamin.hoare08@imperial.ac.uk}{\tt benjamin.hoare08@imperial.ac.uk}}
 and
 A.A. Tseytlin\,\footnote{Also at Lebedev Institute, Moscow. \href{mailto:tseytlin@imperial.ac.uk}{\tt tseytlin@imperial.ac.uk}}}

\vspace{0.5truecm}

{\em Theoretical Physics Group \\
Blackett Laboratory, Imperial College\\
London, SW7 2AZ, U.K.}

\end{center}

\setcounter{footnote}{0}

\vspace{1truecm}

\begin{abstract}
\noindent
The Pohlmeyer reduced equations for strings moving only in the AdS subspace of $\ads_5 \X S^5$ have been used recently in the study of classical Euclidean minimal surfaces for Wilson loops and some semiclassical three-point correlation functions. We find an action that leads to these reduced superstring equations. For example, for a bosonic string in $\ads_n$ such an action contains a Liouville scalar part plus a $K/K$ gauged WZW model for the group $K=\SO(n-2)$ coupled to another term depending on two additional fields transforming as vectors under $K$. Solving for the latter fields gives a non-abelian Toda model coupled to the Liouville theory. For $n=5$ we generalize this bosonic action to include the $S^5$ contribution and fermionic terms. The corresponding reduced model for the $\ads_2 \X S^2$ truncation of the full $\ads_5 \X S^5$ superstring turns out to be equivalent to $\Nc=2$ super Liouville theory. Our construction is based on taking a limit of the previously found reduced theory actions for bosonic strings in $\ads_n \X S^1$ and superstrings in $\ads_5 \X S^5$. This new action may be useful as a starting point for possible quantum generalizations or deformations of the classical Pohlmeyer-reduced theory. We give examples of simple extrema of this reduced superstring action which represent strings moving in the $\ads_5$ part of the space. Expanding near these backgrounds we compute the corresponding fluctuation spectra and show that they match the spectra found in the original superstring theory.
\end{abstract}


\newpage

\phantomsection
\pdfbookmark[1]{Contents}{pre:Con}

\setcounter{secnumdepth}{4}
\setcounter{tocdepth}{4}

\sloppy
\hyphenpenalty=100000

\tableofcontents

\fussy
\hyphenpenalty=200


\newpage

\renewcommand{\theequation}{1.\arabic{equation}}
\setcounter{equation}{0}
\section{Introduction}

The investigation into the Pohlmeyer reduction of superstrings in $\ads_5 \X S^5$ \cite{Grigoriev:2007bu,Mikhailov:2007xr} has advanced significantly over recent years. The reduction, a generalization of the relation between the classical $\O(3)$ sigma model and the sine-Gordon theory \cite{Pohlmeyer:1975nb}, relates the string equations of motion to a classically equivalent set of equations. In the reduction the Virasoro constraints are solved and $\k$-symmetry is fixed, hence the reduced theory manifestly describes only the physical (``transverse'') degrees of freedom of the string.

In general, the Pohlmeyer reduction of strings moving in a symmetric space $\R_{t} \X \fk FG$ is described by a gauged WZW model for a coset $\fk GH$ plus an integrable potential. The Lagrangian for these theories, otherwise known as the symmetric-space sine-Gordon models \cite{D'Auria:1979tb,D'Auria:1980cx, Hollowood:1994vx,Leznov:1982ew,FernandezPousa:1996hi}, was first constructed in \cite{Bakas:1995bm}. One can furthermore find a direct relation between the currents of the string theory and the fields of the reduced theory \cite{Grigoriev:2007bu,Mikhailov:2007xr,Grigoriev:2008jq,Miramontes:2008wt}, such that the construction can be generalized to the superstring. The resulting Pohlmeyer-reduced $\ads_5 \X S^5$ superstring is a fermionic extension of the gauged WZW model for the coset $\fk{\USp(2,2)}{\SU(2)^2} \X \fk{\USp(4)}{\SU(2)^2}$ plus an integrable potential \cite{Grigoriev:2007bu,Mikhailov:2007xr}. The theory was shown to be UV-finite \cite{Roiban:2009vh} and the study of the classical integrable charges suggested the theory was supersymmetric \cite{Schmidtt:2010bi,Schmidtt:2011nr}. In \cite{Goykhman:2011mq,Hollowood:2011fq} it was shown that the action is invariant under some mildly non-local world-sheet supersymmetry transformations.

The reduction procedure of \cite{Grigoriev:2007bu} takes as its starting point a classical solution of the equations of motion for the Green-Schwarz action for superstring theory in $\ads_5 \X S^5$ \cite{Metsaev:1998it}, with both the $\ads_5$ and $S^5$ stress-tensors non-vanishing. Furthermore, the vacuum of the reduced theory corresponds to the BMN point-like string and hence it is natural to compare the S-matrix of BMN light-cone gauge-fixed superstring theory \cite{Beisert:2005tm} (see \cite{Beisert:2010jr,Arutyunov:2009ga} for a review) with that of the reduced theory. While the two S-matrices have the same tensorial structure, the perturbative S-matrix of the reduced theory does not satisfy the Yang-Baxter equation (YBE) \cite{Hoare:2009fs,Hoare:2011fj}. Even so, there is a closely related R-matrix \cite{Beisert:2008tw,Beisert:2010kk}, invariant under the quantum-deformation of $\mfpsu(2|2)^{\oplus 2} \ltimes \R^2$, that satisfies the YBE and it has been conjectured \cite{Hoare:2011fj,Hoare:2011nd} to describe the scattering of solitons \cite{Hollowood:2011fq} in the Pohlmeyer-reduced theory.

This R-matrix appears as a limit of a more general R-matrix, based on the quantum-deformation of the symmetry algebra of the BMN light-cone gauge-fixed superstring, $\mfpsu(2|2)^{\oplus 2} \ltimes \R^3$ \cite{Beisert:2008tw,Beisert:2010kk,Beisert:2011wq,deLeeuw:2011jr}. Solving the crossing equation to compute the phase allows one to construct an interpolating S-matrix \cite{Hoare:2011wr,Hoare:2012fc} depending on two couplings: one is the string tension in the undeformed limit, while the other is the quantum deformation parameter $q$ (related to the coupling of the Pohlmeyer-reduced theory in the limit where the string tension goes to infinity \cite{Hoare:2011fj}). While the realization of unitarity is not yet understood in this interpolating theory (thus far only defined through the S-matrix) it still has a theoretical significance. In a recent work studying the quantum-deformed thermodynamic Bethe ansatz it was shown that when $q$ is a root of unity the spectrum is naturally truncated and the corresponding Y-system has a finite number of integral equations \cite{Arutyunov:2012zt}.

One can also compare the Pohlmeyer-reduced theory with the superstring theory via the computation of one- and two-loop corrections to the partition function for corresponding classical solutions. In contrast to the interpolating picture described above, in this case the one-loop corrections match \cite{Hoare:2009rq,Iwashita:2010tg} if the string tension and the coupling of the reduced theory are proportional to each other. At two loops \cite{Iwashita:2011ha} there is not exact matching, but the result is very suggestive of some non-trivial relation between the two partition functions which remains to be understood.

\

Regardless of possible quantum generalizations, the importance of the Pohlmeyer reduction was recognised in the context of the AdS/CFT correspondence \cite{Tseytlin:2003ii,Mikhailov:2005qv,Mikhailov:2005zd,Mikhailov:2005sy}, and it was used to construct and study various classical string solutions starting with \cite{Hofman:2006xt,Chen:2006gea} (see also \cite{Hollowood:2009tw,Hollowood:2010dt,Hollowood:2011fm}). While these and related papers discussed strings moving in $S^n$, an alternative Pohlmeyer reduction was used also for constructing classical solutions for strings moving entirely in $\ads_n$ (i.e. not probing an additional compact space factor). This alternative reduction procedure was first used for strings in the de Sitter space \cite{Barbashov:1980kz,Barbashov:1982qz,DeVega:1992xc} but can be easily adapted to the $\ads_n$ case. It was used to find string solutions with time-like world-sheets corresponding to solitons of the reduced theory, e.g., of the sinh-Gordon model for $\ads_3$ \cite{Jevicki:2007aa,Jevicki:2008mm,Jevicki:2009uz}. More recently, it has found fruitful applications in the construction of Euclidean minimal surfaces for open strings ending at the AdS boundary \cite{Jevicki:2007aa,Alday:2009ga,Alday:2009yn,Alday:2009dv,Dorn:2009kq,Dorn:2009gq,Jevicki:2009bv,Burrington:2009bh,Dorn:2009hs,Dorn:2010xt,Ishizeki:2011bf,Papathanasiou:2012hx}, which are dual to Wilson loops in planar $\Nc = 4$ super Yang-Mills at strong coupling, and in the study of semiclassical three- and four-point correlation functions for closed-string states \cite{Janik:2011bd,Kazama:2011cp,Kazama:2012is,Caetano:2012ac}.

In general, such descriptions of classical solutions start with equations of motion written using an explicit coordinate parametrization of $\ads_n$. To construct an action when there is an additional compact space sector, it is necessary to use a group-theoretic formulation of the string and reduced theories \cite{Bakas:1995bm,Grigoriev:2007bu}. While for the pure $\ads_n$ reduction some work was done in this direction \cite{Grigoriev:2008jq,Miramontes:2008wt}, no general action was found. Furthermore, it was not clear how such an AdS reduction should be embedded into the general case of the superstring reduction involving not only the $\ads_5 $ space but also the five-sphere and the fermions.

\

This is the question we investigate in this paper. By starting with the Pohlmeyer reduction of strings in $\ads_n \X S^1$, for which the action is known \cite{Grigoriev:2007bu}, and taking the limit in which the $S^1$ part of string stress-tensor (parametrized by $S^1$ angular momentum $\mu$) vanishes, we construct an action for the Pohlmeyer reduction of strings moving only in $\ads_n$. Generalizing this limit to the full Pohlmeyer-reduced $\ads_5 \X S^5$ superstring we include the coupling to the fermionic fields and the contribution of the five-sphere. This is a necessary first step towards a study of the corresponding quantum theory.

In contrast to the standard reduction of \cite{Grigoriev:2007bu} (corresponding to the case of non-zero $S^5$ momentum) which is naturally compared to the superstring in the BMN light-cone gauge with $p^+ \sim \mu$, this reduction may be compared to the superstring in the AdS light-cone gauge \cite{Metsaev:2000yu,Tseytlin:2000na,Giombi:2009gd}.\footnote{Note though that while the AdS light-cone gauge string theory is formulated using a generalized (AdS dependent) diagonal gauge on the 2-d metric, the AdS reduced theory is constructed using the standard conformal gauge.} In the AdS light-cone gauge superstring theory the natural expansion is around a massless geodesic in $\ads_5$. The corresponding $8+8$ physical fluctuations are massless like in flat space. As discussed below, performing the analogous computation in the AdS reduced theory where the counterpart of the massless AdS geodesic is a natural vacuum background we find the same massless spectrum. We also study the spectrum of fluctuations around another simple classical solution of the reduced theory which corresponds to the large-spin limit of the GKP string \cite{Gubser:2002tv,Frolov:2002av,Frolov:2006qe}, and again find full agreement with the superstring theory spectrum.

\

\subsection{Outline of the construction of reduced-theory action for strings in AdS}

As the construction of the action for the Pohlmeyer reduction of strings in $\ads_n$ is somewhat subtle, it is useful to give a brief outline of some key points. As already mentioned, we shall use as our starting point the Pohlmeyer reduction of strings in $\ads_n \X S^1$ \cite{Grigoriev:2007bu}, with $\ads_n$ viewed as the symmetric space $\ads_n = \fk FG = \fk{\SO(n-1,2)}{\SO(n-1,1)}$. As usual, we shall choose the conformal gauge and fix the residual conformal reparametrizations by setting the angle of $S^1$ equal to $\m$ times the world-sheet time. The limit we are interested in is $\m \ra 0$ when the string moves only in $\ads_n$.

In this standard reduction procedure for strings in $\ads_n \X S^1$ the Virasoro constraints are solved by fixing the currents in terms of a generator $T$, which picks out a subgroup $H = \SO(n-1)$ of $G= \SO(n-1,1)$. This gauge-fixing has a trivial $\m \ra 0$ limit: the currents vanish. To find a non-trivial limit we introduce two automorphisms of the algebra $\mff = \mfso(n-1,2)$ depending on $\m$ and an element of the algebra $\mfg = \mfso(n-1,1)$, which picks out a subgroup $K = \SO(n-2)$ of $H$. For a generic non-zero $\m$ the automorphisms are well-defined and one can find an action for the reduced theory in terms of an asymmetrically-gauged WZW model \cite{Miramontes:2008wt}.

In the $\m \ra 0$ limit the action of the automorphisms on $\mfh = \mfso(n-1)$ (with appropriately rescaled generators) gives two copies of the Euclidean algebra, while their action on the generator $T$ (again with an appropriate rescaling) gives two generators $T_\upm$ satisfying $\Tr(T_\upm^2) = 0$. These algebras and generators also appear if one tries to carry out the Pohlmeyer-reduction directly for $\m = 0$ \cite{Grigoriev:2008jq,Miramontes:2008wt}. It is therefore natural that they should show up  in the limiting procedure, through a contraction of the algebra.

Either by (i) partially gauge-fixing, finding an action for non-zero $\m$ and then taking $\m \ra 0$, or by (ii) taking the $\m \ra 0$ limit, then partially gauge-fixing and finding an action\,\footnote{Though these two procedures do not manifestly commute, after an additional partial gauge-fixing in the former method, which is only allowed after $\m$ has been taken to zero, they agree.} one finally arrives at a gauged WZW model for the coset $\fk GK = \fk{\SO(n-1,1)}{\SO(n-2)}$ plus a potential term defined in terms of $T_\upm$. In a particular parametrization of the group-valued field of the gauged WZW model the action can be written as
\eq{\label{introaction}
\Act = \ \fr{\coup}{8\pi} \bigg( & 4 \int d^2 x \; \big[ \dpl \p \dm \p - \fr14 e^{2\p}\big]
                                   + \int d^2x\; e^{2\p} \Tr\big[D_-\ym D_+\yp \big]
\\              & - \Tr \Big[\,\fr{1}{2}\,\int d^2x \ \ \inv k \dpl k \inv k \dm k \
                             - \fr{1}{3}\,\int d^3x \; \ \e^{mnl} \ \inv k \del_m k \ \inv k \del_n k \ \inv k \del_l k
\\              & \qquad \qquad \, + \, \int d^2x \; \ \big( B_+ \dm k \inv k - B_- \inv k \dpl k - \inv k B_+ k B_- +  B_+ B_- \big)\Big]\bigg) \ .}
It is given by the sum of the Liouville action for the field $\p$ and the $\fk KK$ gauged WZW action for $K= \SO(n-2)$ described by the fields $k$ and $B_\pm$. These two parts of the action are coupled through the term depending on the fields $\ypm$ taking values in two $\R^{n-2}$ subspaces of $\mfg = \mfso(n-1,1)$ and transforming as vectors under $K$ (the covariant derivatives $D_\pm$ contain $B_\pm$). The overall coefficient ${\coup}$ is a coupling constant; its choice (e.g., its relation to string tension) depends on an interpretation of the corresponding quantum theory.

Note that while the $\fk KK$ gauged WZW model by itself is topological (describes no dynamical degrees of freedom) \cite{Witten:1991mm,Spiegelglas:1992jg,Alekseev:1995py}, it leads, as we shall see below, to non-trivial dynamics once the coupling to $\ypm$ (and $\p$) is included. The action \eqref{introaction} can be further simplified (using field redefinitions and solving for $\ypm$) to the Liouville part coupled to the WZW action for the group $K$ through a potential term (see \eqref{gwzwc2}, \eqref{gwoc}). The result can be interpreted as a non-abelian Toda model \cite{Leznov:1982ew,Gervais:1992bs} coupled to the Liouville model. Similar models (though without explicit Liouville potential) were discussed in \cite{Bakas:1996np}.

To check that the action \eqref{introaction} indeed describes the Pohlmeyer reduction of strings in $\ads_n$ one may look at the corresponding equations of motion. Eliminating the field $k$ and defining two new fields, $u$ and $v$, equal to $\pm e^{2\p}D_\pm\ypm$, one arrives at the same set of equations as found via the usual Pohlmeyer reduction of strings in $\ads_n$, formulated directly in terms of embedding coordinates (see, e.g., \cite{DeVega:1992xc,Jevicki:2007aa,Dorn:2009kq,Dorn:2009gq}).

In specific cases the equations of motion for the Pohlmeyer reduction of strings in $\ads_n$ can be gauge-fixed to give second-order differential equations for $n-2$ scalar fields derivable from generalized sine/sinh-Gordon (or non-abelian Toda) type actions \cite{Dorn:2009gq,Burrington:2009bh}. Equivalent actions can also be found as specific gauge-fixings of the more fundamental action \eqref{introaction}.

There is increasing evidence that a proper definition of such generalized sine/sinh-Gordon models (preserving, in particular, their integrability at the quantum level) should indeed be in terms of a gauged WZW theory plus a potential. For example, such a definition for the complex sine-Gordon model leads to particular quantum corrections which imply that the Yang-Baxter equation is satisfied beyond the tree level \cite{Dorey:1994mg,Hoare:2010fb}. More recently, it has been shown that the Poisson bracket of this formulation is only mildly non-ultralocal, allowing one to discretize and canonically quantize the theory \cite{Delduc:2012qb,Delduc:2012mk,Delduc:2012vq}.

An important feature of the above construction is that it gives a systematic way to take the $\m \ra 0$ limit. As we shall discuss below, this allows us to generalize it to the full Pohlmeyer-reduced superstring \cite{Grigoriev:2007bu}, including the fermionic fields. One can then check that the spectrum of fluctuations around some simple classical string solutions matches that found directly from the superstring action. Note that in addition to matching of the fluctuation frequencies, one needs to define the quantum reduced theory (i.e. make proper choice of fundamental fields and path integral measure) to have the same number ($8+8$) of degrees of freedom as in the AdS light-cone gauge-fixed superstring.

\

Let us mention that most of the classical applications of this AdS reduction (as in the case of Wilson loops in, e.g., \cite{Alday:2009ga,Alday:2009yn,Ishizeki:2011bf}) deal with space-like world-sheets. However, for technical reasons, in the main part of this paper we will assume that the 2-d string world-sheet is time-like, i.e. has Minkowski signature. Curiously, in this case the resulting reduced theory Lagrangian has some ``ghost''-like terms, i.e. has negative signs in front of the kinetic terms of some fields. While the true reason for this peculiar feature is still to be understood, it turns out that it is absent if the Pohlmeyer reduction is done assuming a space-like (Euclidean signature) string world-sheet: in this case the signs in front of the all kinetic terms in the reduced theory action are positive.

It would be important to study the integrability property of this Euclidean signature version of the full $\ads_5 \times S^5$ Pohlmeyer reduced theory in more detail. In particular, it would be interesting to investigate a possible connection to the ``thermodynamic bubble ansatz'' of \cite{Alday:2009dv} (which was used for the computation of the area of minimal surfaces ending on null polygons). The approach based on the Pohlmeyer reduction may help to find its natural quantum extension.

Another interesting future direction is to study the reduced-theory backgrounds corresponding to the massless geodesic in AdS and the large spin limit of the GKP string. In the massless geodesic case, i.e. expanding around the trivial vacuum of the reduced theory, we find a Lorentz-invariant theory describing $8+8$ massless fields. The S-matrix for these excitations will also be Lorentz invariant, in contrast to the usual BMN light-cone gauge-fixing. For the standard Pohlmeyer reduction of the superstring \cite{Grigoriev:2007bu} there appears to be a rich interpolating structure connecting the Lorentz-invariant massive S-matrix of the reduced theory and the BMN light-cone gauge-fixed superstring S-matrix, and one could investigate if a similar structure appears here.

The background corresponding to the large spin limit of the GKP string can be thought of as an alternative non-trivial ``vacuum'' of the reduced theory. In particular, one may compute quantum corrections to dispersion relations and the S-matrix for the fluctuation modes above this ``vacuum'' and compare to the corresponding string theory results \cite{Basso:2010in,Giombi:2010bj,Basso:2011rc}. This should be technically easier than in the original superstring set-up due to simpler structure of the reduced theory action.

\

\subsection{Structure of the paper}

We start in section \ref{sec:cr} with a discussion of simple examples of the Pohlmeyer reduction in explicit coordinate parametrizations. We explore the $\m \ra 0$ limit of the sinh-Gordon and complex sinh-Gordon models which describe the Pohlmeyer reductions of strings in $\ads_2 \X S^1$ and $\ads_3 \X S^1$. Before taking the limit we use various $\m$-dependent field redefinitions, such that the resulting theories are the Liouville model and the $\SL(2,\R)$ WZW model plus potential. These describe the Pohlmeyer reductions of strings in $\ads_2$ and $\ads_3$ as expected.

In section \ref{sec:gtp} we take the $\m \ra 0$ limit of the group-theoretic formulation of the Pohlmeyer reduction of strings in $\ads_n \X S^1$ to derive the action for the Pohlmeyer reduction of strings in $\ads_n$ \eqref{introaction}. In section \ref{sec:sst} we extend this limiting procedure to the full Pohlmeyer-reduced $\ads_5 \X S^5$ superstring of \cite{Grigoriev:2007bu} and show, in particular, that in the case of the truncation to the $\ads_2 \X S^2$ superstring the corresponding reduced model is equivalent to the $\Nc=2$ super Liouville model.

In section \ref{sec:quafluc} we study the reduced theory description of fluctuations around two backgrounds corresponding to two simple classical string solutions: the massless geodesic in $\ads_5$ and the large-spin limit of the folded spinning string. In both cases we find fluctuation frequencies matching those of string theory.

In section \ref{app:glf} we outline the construction of the reduced theory action for a space-like string world-sheet in which all physical fields have positive kinetic terms. This theory can be found also as a particular analytic continuation of the time-like reduced model.

In section \ref{sum} we summarize the results of the paper and discuss some open questions. In appendix \ref{app:cr} we review in detail the important features of the standard Pohlmeyer reduction of strings in $\ads_n$ formulated in terms of explicit target-space coordinates. In appendix \ref{app:algebras} we discuss some properties of the algebras $\mfso(n-1,2)$ and $\mfpsu(2,2|4)$ and related definitions that are used in the main part of the paper.

\renewcommand{\theequation}{2.\arabic{equation}}
\setcounter{equation}{0}
\section{AdS Pohlmeyer reduction in coordinate parametrization\label{sec:cr}}

In this section we study the relation between the Pohlmeyer reductions of strings in $\ads_n \X S^1$ and $\ads_n$ for $n=2,3$ formulated in terms of explicit target-space coordinates (which we shall refer to as the coordinate parametrization or formulation). We set the angular momentum on the $S^1$ to be proportional to $\m$ and explore how the $\m \ra 0$ limit of the $\ads_n \X S^1$ reduction is related to the $\ads_n$ one. While for $n=2$ this limit appears to go through, albeit with some subtleties, for $n=3$ the construction breaks down when working in this coordinate parametrization, indicating that we need to use a different approach -- starting with the reduced theory in $\ads_3 \X S^1$ formulated in terms of a gauged WZW theory.

The Pohlmeyer reduction for classical strings moving in AdS space formulated in terms of an explicit coordinate parametrization has been extensively studied and applied before \cite{Barbashov:1980kz,Barbashov:1982qz,DeVega:1992xc,Jevicki:2007aa,Jevicki:2008mm,Jevicki:2009uz,Alday:2009ga,Alday:2009yn,Alday:2009dv,Dorn:2009kq,Dorn:2009gq,Jevicki:2009bv,Burrington:2009bh,Dorn:2009hs,Dorn:2010xt,Papathanasiou:2012hx}. As we will repeatedly use the resulting set of equations of motion we have outlined their derivation in appendix \ref{app:cr}, highlighting the features that will be important to us here.

\subsection[Reduction of strings in \texorpdfstring{$\ads_2$}{AdS2} from \texorpdfstring{$\ads_2 \X S^1$}{AdS2 x S1}]{Reduction of strings in $\mathbf{\ads_2}$ from $\mathbf{\ads_2 \X S^1}$\label{sec:crpr2}}

The Pohlmeyer reduction of a classical string moving in $\ads_2 \X S^1$ is well known to be described by the sinh-Gordon equation,
\eq{\label{sinh-Gordon}
\dpdm \p + \fr{\m^2}{2} \sinh 2\p = 0 \ .}
Here $\m$ is proportional to the angular momentum of the string solution on $S^1$, so the $\m \ra 0$ limit corresponds to the case when string moves only in $\ads_2$. However, this limit can na\"ively be taken in many ways. The correct way to take the limit to get the Pohlmeyer reduction of strings in $\ads_2$ is to first shift the field $\p$ by $\log \m$ \cite{Grigoriev:2008jq}
\eq{\label{phi-shift}
\p \ra \p - \log \m \ .}
Expanding out $\sinh$ in terms of exponentials then gives the following equation of motion
\eq{
\dpdm \p + \fr14 e^{2\p} - \fr{\m^4}{2} e^{-2\p} = 0 \ ,}
which in the $\m \ra 0$ limit gives the Liouville equation,
\eq{\label{Liou1eom}
\dpdm \p +\fr14 e^{2\p} = 0 \ .}
The Liouville equation is well known to describe the Pohlmeyer reduction of strings moving in $\ads_2$, see appendix \ref{app:cr}.

A similar limit can also be taken in the sinh-Gordon Lagrangian
\eq{\label{lagsg}
\Lag_{sG} = \dpl \p \dm \p - \fr{\m^2}{2}\cosh2\p \ .}
Indeed, using the shift \eqref{phi-shift} and taking $\m \ra 0$ gives
\eq{\label{Liou1lag}
\Lag_{L} = \dpl \p \dm \p - \fr14 e^{2\p} \ .}

\subsection[Reduction of strings in \texorpdfstring{$\ads_3$}{AdS3} from \texorpdfstring{$\ads_3 \X S^1$}{AdS3 x S1}]{Reduction of strings in $\mathbf{\ads_3}$ from $\mathbf{\ads_3 \X S^1}$\label{sec:crpr3}}

The Pohlmeyer reduction of strings moving in $\ads_3 \X S^1$ is described by the complex sinh-Gordon equations of motion
\eq{\label{complex-sinh-Gordon}
   \dpdm & \p - \tanh\p \sech^2\p\dpl\x\dm\x + \fr{\m^2}{2} \sinh 2\p = 0 \ ,
\\       & \dpl(\tanh^2\p\ \dm\x) + \dm(\tanh^2\p\ \dpl \x) = 0 \ ,}
following from
\eq{\label{lagcsg}
\Lag_{CsG} = \dpl\p \dm \p + \tanh^2\p\ \dpl\x\dm \x - \fr{\m^2}{2}\cosh2\p \ ,}
As before, to take the $\m \ra 0$ limit we need to be careful. In fact we will discuss two approaches. The first, a more na\"ive approach, appears not to be satisfactory, while the second, based on the gauged WZW model, turns out to be consistent.

\

In the first approach we start from the equations of motion \eqref{complex-sinh-Gordon}. As expected, the field $\p$ should again be shifted as in \eqref{phi-shift}, but to find a non-trivial result (i.e. more than just the Liouville equation and a free scalar) the field $\x$ should also be rescaled by $\fk1{4\m}$,
\eq{\label{rescale1}
\x \ra \fr1{4\m} \x \ .}
After performing this shift and rescaling, and then taking the $\m\ra 0$ limit the complex sinh-Gordon equations of motion \eqref{complex-sinh-Gordon} become\,\footnote{Note that to find a non-trivial finite limit of the second equation in \eqref{complex-sinh-Gordon} it should be rescaled by a factor of $\m$.}
\eq{\label{cl1}
\dpdm \p - \fr14 \dpl\x\dm\x \ e^{-2\p} + \fr14 e^{2\p} = 0 \ , \ \ \ \qquad \dpl\dm \x = 0 \ .}
The second equation can be easily solved as follows:\footnote{We shall use the following notation for 2-d coordinates: $x=(\tau,\sigma)$, \ $\xxx^\pm = \fr12(\tau \pm \sigma)$.}
\eq{\label{sol1}
\dpl \x = U(\xxx^+) \ , \ \ \ \qquad \dm \x = V(\xxx^-) \ ,}
so that equations \eqref{cl1} simplify to
\eq{\label{complex-Liouville}
\dpdm \p - \fr14 U V \, e^{-2\p} + \fr14 e^{2\p} = 0 \ , \ \ \ \ \ \ \ \ \ \ \ \ \ \ \ \ \dm U = \dpl V = 0 \ .}
As in the $\ads_2$ case these equations agree with the usual Pohlmeyer reduction of strings moving in $\ads_3$, see appendix \ref{app:cr}.

A problem of this approach becomes apparent if we attempt to take the limit in the Lagrangian \eqref{lagcsg}. Using the shift \eqref{phi-shift} and rescaling \eqref{rescale1}, and taking $\m \ra 0$ gives a finite piece but also a divergent ($\mu^{-2}$) piece
\eq{\label{lagcl1}
\Lag_{CL} =\fr1{16\m^2}\dpl\x\dm\x + \dpl\p\dm\p - \fr14 e^{-2\p} \dpl\x\dm\x - \fr{1}{4} e^{2\p} \ .}

The second approach starts with the gauged WZW formulation of the complex sinh-Gordon model. For now we will use the axially gauged WZW theory for the coset $\fk{\SL(2,\R)}{\U(1)}$ plus an integrable potential. Fixing a gauge\,\footnote{The particular gauge-fixing that we use is ($\s_a$ are the usual Pauli matrices)
\aln{
g = \e^{i \x \s_3} \e^{2\p \s_1} e^{i \x \s_3} \ .}}
on the $\SL(2,\R)$-valued field and expanding in terms of components we find the following Lagrangian
\eq{\label{lagcsggwzw}
\Lag_{CsG_g} = \dpl\p \dm \p & +\sinh^2\p\  \dpl\x\dm \x
\\ & \ \ - a_- \, \sinh^2\p \, \dpl\x - a_+ \sinh^2\p \, \dm \x + a_+a_- \cosh^2\p - \fr{\m^2}{2}\cosh2\p \ .}
As expected, integrating out $a_\pm$ we find \eqref{lagcsg}.

To get a finite and non-trivial result in the $\m \ra 0$ limit let us perform the shift \eqref{phi-shift} and rescale $\x$ and $a_\pm$ by $\m$\,\footnote{Note that this is a different rescaling compared to \eqref{rescale1}. Using \eqref{rescale1} in this second approach gives again a divergent Lagrangian. If we use \eqref{rescale2} in the first approach then we find the Liouville equation of motion and a free scalar. Finally, if we use the rescaling \eqref{rescale2} in the Lagrangian \eqref{lagcsg} we get just the Liouville Lagrangian.}
\eq{\label{rescale2}
(\x,a_\pm) \ra \m \, (\x,a_\pm) \ .}
The resulting Lagrangian is then given by
\eq{\label{lagcsg1gwzw}
\Lag_{CL_g} = \dpl\p \dm \p +\fr14 e^{2\p} \ (\dpl\x-a_+)(\dm \x - a_-) - \fr14 e^{2\p} \ .}
This suggests that $a_\pm$ should be thought of as derivatives of two scalars $a_\pm = \dpm \td \y_\upm$.\footnote{If we na\"ively integrate out the gauge field $a_\pm$ we just find the Liouville Lagrangian -- this should be expected as this is what one gets if the shift \eqref{phi-shift} and rescaling \eqref{rescale2} are implemented in the Lagrangian \eqref{lagcsg}.} Defining the linear combinations, $\y_\upm = \x - \td \y_\upm$, we find the Lagrangian
\eq{\label{lagcL1gwzw}
\Lag_{CL_g} = \dpl\p \dm \p + \fr14 e^{2\p} \dpl \y_\up \dm \y_\um - \fr14 e^{2\p} \ .}

The variational equations following from this Lagrangian are
\eq{\label{cl1a}
\dpdm \p - \fr14 \dpl \y_+ \dm \y_- \ e^{2\p} + \fr14 e^{2\p} = 0 \ , \ \ \ \ \qquad \dpm(e^{2\p}\dmp \y_\mp) = 0 \ .}
If we then set
\eq{
U = e^{2\p} \dpl \y_\up \ , \ \ \ \ \qquad V = e^{2\p} \dm \y_\um \ ,}
and substitute into \eqref{cl1a} we again find the set of equations of motion \eqref{complex-Liouville}, which agree with the usual Pohlmeyer reduction of strings moving in $\ads_3$, see appendix \ref{app:cr}.

This second approach describes a smooth (non-singular) $\m \ra 0$ limit, leading to the equations of motion that are expected from the well-known direct Pohlmeyer reduction of strings moving in $\ads_3$ (see appendix \ref{app:cr}). It is this approach that we will follow and generalize in the following sections.

\renewcommand{\theequation}{3.\arabic{equation}}
\setcounter{equation}{0}
\section{Group-theoretic approach to Pohlmeyer reduction\label{sec:gtp}}

The most natural approach to understanding the Pohlmeyer reduction is based on the geometric description of symmetric space sigma models and integrable extensions of gauged WZW theories. For example, it allows the extension of the bosonic string reduction to the full $\ads_5 \X S^5$ superstring case \cite{Grigoriev:2007bu}. Motivated by the discussion in the previous section, here we also find that this is the appropriate setting to take the limit when the $S^1$ momentum $\m$ goes to zero.

In this section we will explore taking this limit directly in the action of the reduced theory for strings moving in $\ads_n \X S^1$. We will start with the case $n=2$, in which, as in the coordinate formulation of section \ref{sec:crpr2}, the limit goes straightforwardly. Then we will investigate the cases with $n\geq 3$. These cases are more subtle as the reduced theory is now described by a gauged WZW theory.

\subsection[A simple example: \texorpdfstring{$\m \ra 0$}{m -> 0} limit of the reduction for strings in \texorpdfstring{$\ads_2 \X S^1$}{AdS2 x S1}]{A simple example: $\mathbf{\boldsymbol{\m} \ra 0}$ limit of the reduction for strings in $\mathbf{\ads_2 \X S^1}$\label{sec:ads2gt}}

We start by considering the simplest possible example of strings moving in $\ads_2 \X S^1$ and $\ads_2$. The space $\ads_2$ can be written as the symmetric coset space
\eq{
\fr FG = \fr{\SO(1,2)}{\SO(1,1)} \ .}
The Pohlmeyer reduced theory for strings moving in $\ads_2 \X S^1$ is then given by a WZW model for $G = \SO(1,1)$ (as $G$ is abelian here the WZ term vanishes) plus an integrable potential,
\eq{\label{lag1}
\Lag = \Tr\Big[\fr12 \, \inv g \dpl g \inv g \dm g + \m^2 \inv g T g T \Big] \ .} 
Here $T$ is an element of the algebra $\mff = \mfso(1,2)$ satisfying $\Tr(T^2)=-2$.

Parametrizing
\eq{\label{gparamphi}
g = \exp (-2\,\p R) \ ,}
where $R$ is the generator of the subalgebra $\mfg = \mfso(1,1)$, with normalization fixed by $\Tr(R^2)=2$, we find that the Lagrangian \eqref{lag1} reduces to the sinh-Gordon Lagrangian \eqref{lagsg}. Recalling that taking the limit $\m\ra0$ in the sinh-Gordon Lagrangian gave the expected result after a redefinition of $\p$, our aim is to find the corresponding redefinition of the group-valued field $g$ and observe that with the same parametrization \eqref{gparamphi} we find the Liouville Lagrangian \eqref{Liou1lag}.

It turns out that the appropriate transformation is\,\footnote{This transformation is further motivated in the latter parts of this section, in particular in the discussion of the $\m \ra 0$ limit in the reduction procedure (see section \ref{sec:adsngt}).}
\eq{\label{gtrans}
g \ra \m^{R} g \m^{R} \ ,\qquad \qquad \m^{a R} \equiv \exp(a R \log \m) \ ,}
where, after the redefinition, we assume $g$ is finite in the $\m \ra 0$ limit. To take the limit in the Lagrangian we require that
\eq{\label{limitsoft}
\lim_{\m \ra 0} \, \m \, \m^{\pm R} T \m^{\mp R} = \fr12 (T \pm S) \equiv T_\ump \ ,}
where $S$ is orthogonal to both $T$ and $R$ and again normalized so that $\Tr(S^2) = 2$. For an explicit $\mfso(1,2)$ matrix representation for $T$,$S$ and $R$ see appendix \ref{app:alg}.

Then after performing the transformation \eqref{gtrans} and taking the limit $\m \ra 0$ we arrive at
\eq{\label{lag2}
\Lag = \Tr\Big[\fr12 \, \inv g \dpl g \inv g \dm g + \inv g T_\up g T_\um\Big]\ ,}
and parametrizing $g$ as in equation \eqref{gparamphi} we indeed find the Liouville Lagrangian \eqref{Liou1lag} as expected.

\subsection[The general case: \texorpdfstring{$\m \ra 0$}{m -> 0} limit of the reduction for strings in \texorpdfstring{$\ads_n \X S^1$}{AdSn x S1}]{The general case: $\mathbf{\boldsymbol{\m} \ra 0}$ limit of the reduction for strings in $\mathbf{\ads_n \X S^1}$\label{sec:adsngt}}

We will now explore the $\m\ra0$ limit of the Pohlmeyer reduction of strings moving in $\ads_n \X S^1$ for general $n$. As hinted at in the discussion of the $\m \ra 0$ limit of the complex sinh-Gordon equations and Lagrangian in section \ref{sec:crpr3} the situation here becomes more complicated due to the presence of gauge fields in the reduced theory.

In the Pohlmeyer reduction procedure for strings in $\ads_n \X S^1$ one starts by finding a single equation for an $\SO(n-1,1)$ group-valued field with an $\SO(n-1) \X \SO(n-1)$ gauge symmetry. To find a Lagrangian formulation this needs to be partially gauge-fixed to leave an $\SO(n-1)$ gauge symmetry. What we will see in this section is that if one takes the $\m \ra 0$ limit before or after this partial gauge fixing we end up with different set of equations of motion and therefore action, i.e. the $\m \ra 0$ limit does not commute with the partial gauge-fixing.

This discrepancy can be resolved in the same manner as for the smooth $\m \ra 0$ limit taken in the complex sinh-Gordon Lagrangian discussed at the end of section \ref{sec:crpr3}. In particular, considering the action found by first partially gauge-fixing and then taking the limit, we may replace some components of the gauge field by derivatives of new scalar fields that can then be absorbed into already existing fields. It is this resulting action that we claim describes the Pohlmeyer reduction of strings moving in $\ads_n$.

Further evidence for this is provided by observing that with a particular parametrization of the fields the variational equations following from this action can be manipulated to give precisely the set of equations that arise from the coordinate-based Pohlmeyer reduction of strings in $\ads_n$ as described in appendix \ref{app:cr}.

\

For practical purposes we find it useful to work with an explicit matrix representation of the algebra $\mfso(n-1,2)$ and a particular choice of subalgebras in terms of a given basis. For convenience, we summarize the algebraic structure we will use throughout this section:
\begin{itemize}
\item We start with the algebra $\mff = \mfso(n-1,2)$, with generators $\{T,S,N_i,R,M_i,H_i,K_{ij}\}$ as defined in appendix \ref{app:alg}. Note that the generators are normalized as follows
\eq{
&\Tr(N_i N_j) = 2\d_{ij} \ , \qquad \Tr(R^2) = 2 \ , \qquad \Tr(K_{ij}K_{kl}) = -2\d_{ik}\d_{jl} \ ,
\\  &\Tr(T^2) = -2 \ , \ \ \ \ \Tr(S^2) = 2 \ , \ \ \qquad \Tr(M_iM_j) = -\Tr(H_i H_j) = 2 \d_{ij} \ .}
\item The subalgebra $\mfg = \mfso(n-1,1)$ has generators $\{R,M_i,H_i,K_{ij}\}$.
\item The subalgebra $\mfh = \mfso(n-1)$ has generators $\{H_i,K_{ij}\}$.
\item The subalgebra $\mfr = \mfso(1,1)$ has generator $R$.
\item The subalgebra $\mfk = \mfso(n-2)$ has generators $\{K_{ij}\}$; the corresponding group is $K= \SO(n-2)$.
\item The space spanned by $\{H_i\}$ is denoted $\mfl$.
\item The rescaled generators $\{\tilde T,\tilde S,\tilde H_i,\tilde M_i\}$ are defined as $\m\{T,S,H_i,M_i\}$.
\item Two automorphisms of the algebra $\mff$ are defined by $\e_\upm(\mfJ) = \m^{\mp R} \mfJ \m^{\pm R}$.
\item The generators $T_\upm$ are defined as $\lim_{\m\ra 0} \e_\upm(\tilde T) = \fr12 (T \mp S)$.
\item The generators $\S_{\upm\,i}$ are defined as $\lim_{\m \ra 0} \e_\upm(\tilde H_i) = \fr12 (H_i \pm M_i)$.
\item The spaces spanned by $\{\S_{\upm\,i}\}$ are denoted $\mfl_\upm$.
\item The algebras $\mfh_\upm$ are defined to be $\mfk \oplus \mfl_\upm$.\\
It is possible to check that the exponentiation of these algebras gives the Euclidean group $E_{n-2}$.
\end{itemize}

The rescaled generators, $\tilde T,\tilde S,\tilde H_i$ and $\tilde M_i$, will play an important r\^ole as the $\m \ra 0$ limits of these generators and their images under the automorphisms $\e_\upm$ are finite. However, in this limit the set of generators no longer forms a basis for $\mff$. More precisely,
\eq{\label{38anas}
\lim_{\m\ra 0} \e_\upm(\tilde T) = \fr12(T\mp S) \equiv T_\upm \ ,\qquad \qquad &
\lim_{\m\ra 0} \e_\upm(\tilde S) = \fr12(S\mp T) \equiv \mp T_\upm \ , \\
\lim_{\m\ra 0} \e_\upm(\tilde H_i) = \fr12(H_i \pm M_i) \equiv \S_{\upm\,i} \ ,\qquad \qquad &
\lim_{\m\ra 0} \e_\upm(\tilde M_i) = \fr12(M_i \pm H_i) \equiv \pm \S_{\upm\,i} \ ,}
while the remaining generators are invariant under the automorphisms \eqref{automorph}.

\subsubsection[Two ways of taking the \texorpdfstring{$\m \ra 0$}{m -> 0} limit of the reduction procedure]{Two ways of taking the $\mathbf{\boldsymbol{\m} \ra 0}$ limit of the reduction procedure\label{sec:red}}

In this section we consider taking the $\m \ra 0$ limit in the standard Pohlmeyer reduction procedure. We start with strings moving in $\ads_n \X S^1$, choose the conformal gauge ($\sqrt{-h} h^{ab} = \h^{ab}$) and fix the residual conformal diffeomorphisms by setting the angle of $S^1$ to be $\theta = \mu \tau$.

The equations of motion and the conformal gauge (Virasoro) constraints written in terms of the left-invariant Maurer-Cartan one-form $\Jc$
\eq{
& \Jc = f^{-1} d f \ , \ \ \ \ \ \ \ f \in \SO(n-1,2) \ ,
\\ & \Jc = \Ac + \Pc \, \in \mff = \mfso(n-1,2)= \mfg \oplus \mfp \ , \ \ \ \ \ \Ac \in \mfg = \mfso(n-1,1) \ , \ \ \ \ \Pc \in \mfp \ ,}
are
\eq{
\Dc_+ \Pc_- + \Dc_- \Pc_+ = 0 \ , \qquad \Tr(\Pc_\pm \Pc_\pm) = - 2 \m^2 \ .}
Supplementing with the flatness condition for $\Jc$ the full set of equations governing this system are
\eq{\label{eq:gh}
& \Dc_+ \Pc_- = \Dc_- \Pc_+ = 0 \ , \qquad \Tr(\Pc_\pm \Pc_\pm) = - 2 \m^2 \ ,
\\
& \qquad \qquad d\Ac + \Ac \wedge \Ac + \Pc \wedge \Pc = 0 \ .}

To implement the $\m \ra 0$ limit in the reduction procedure we introduce the two automorphisms of the algebra $\mff = \mfso(n-1,2)$
\eq{\label{automorph}
\e_\upm(\mfJ) = \m^{\mp R} \mfJ \m^{\pm R} \ ,}
that were defined in the algebraic setup above.

As was mentioned below \eqref{38anas}, there are already apparent issues with taking $\m \ra 0$ at the level of the algebra, hence for now we continue with finite $\m$. We start the reduction procedure by parametrizing $\Pc_\pm$ in terms of two group $G$-valued fields $\gol, \gtl$ such that the Virasoro constraints are solved
\eq{
\Pc_- = g_1^{-1} \e_\up(\tilde T) \gol \ ,
\qquad
\Pc_+ = g_2^{-1} \e_{\um}(\tilde T) \gtl \ .}
To solve the equations of motion $\Dc_\pm \Pc_\mp = 0$ we write $\Ac_\pm \in \mfg$ in terms of the new fields $A_\pm$
\eq{
\Ac_+ = g_1^{-1} \e_\up(A_+) \gol + g_1^{-1} \dpl \gol \ , \qquad
\Ac_- = g_2^{-1} \e_\um(A_-) \gtl + g_2^{-1} \dm \gtl \ , \qquad A_\pm \in \mfh \ .}
Defining
\eq{g = \gol g_2^{-1} \ ,}
we note that under the $G$ gauge transformations of the string theory the fields $g,A_\pm$ are invariant and the reduced theory equation of motion (found by substituting for $\Pc_\pm$ and $\Ac_\pm$ in the second line of \eqref{eq:gh}) can be written in terms of these variables as
\eq{\label{eq:1}
\dm(g^{-1} \dpl g + g^{-1} \e_\up(A_+) g) & - \dpl \e_\um(A_-)
\\ & + \com{\e_\um(A_-)}{g^{-1} \dpl g +g^{-1} \e_\up(A_+) g}
     + \com{g^{-1} \e_\up(\tilde T) g}{\e_\um(\tilde T)} = 0 \ ,}
or, equivalently, as
\eq{\label{eq:2}
\dpl(g \e_\um(A_-) g^{-1} - \dm g g^{-1}) & - \dm\e_\up(A_+)
\\ & + \com{\e_\up(A_+)}{g \e_\um(A_-) g^{-1} - \dm g g^{-1}}
     + \com{g \e_\um(\tilde T) g^{-1}}{\e_\up(\tilde T)} = 0 \ .}
Equations \eqref{eq:1} and \eqref{eq:2} have an $H^\m_\up \X H^\m_\um$ gauge symmetry where $H^\m_\upm \equiv \exp(\e_\upm(\mfh))$
\eq{\label{ggttrr}
g \ra \inv h_\up g h_\um \ , \qquad \e_\upm(A_\pm) \ra \inv h_\upm \e_\upm(A_\pm) h_\upm + \inv h_\upm \dpl h_\upm \ , \qquad h_\upm(x^\pm) \in H^\m_\upm \ .}

\

Next, we need to (i)  take the $\m \ra 0$ limit,  and  (ii) partially fix the gauge symmetry \eqref{ggttrr}.
 As previously discussed we end up with two different results depending on the order in which we carry out these steps. Let us now consider the two procedures based on these two different orderings.

The first procedure is to partially gauge-fix and then take the limit $\m \ra 0$. For finite $\m$ we can project an element of the algebra onto $\e_\upm(\mfh)$.\footnote{\label{footproj}For an arbitrary element $\h \in \mff$ we can write
\aln{
\h & = \h^T \tilde T + \h^S \tilde S + \h^R R + \h^N_i N_i + \h^M_i \tilde M_i + \h^H_i \tilde H_i + \h^K_{ij} K_{ij} \ ,
\\
\h & = \h^T_\upm \e_\upm(\tilde T) + \h^S_\upm \e_\upm(\tilde S) + \h^R R + \h^N_i N_i + \h^M_{\upm\,i} \e_\upm(\tilde M_i) + \h^H_{\upm\,i} \e_\upm(\tilde H_i) + \h^K_{ij} K_{ij} \ ,}
where
\aln{
\h^T_\upm = \fr1{2\m}(\h^T \pm \h^S) + \fr\m2 (\h^T \mp \h^S) \ , \qquad &
\h^S_\upm = \fr1{2\m}(\h^S \pm \h^T) + \fr\m2 (\h^S \mp \h^T) \ ,
\\
\h^M_{\upm\,i} = \fr1{2\m}(\h^M_i \mp \h^H_i) + \fr\m2(\h^M_i \pm \h^H_i)  \ , \qquad &
\h^H_{\upm\,i} = \fr1{2\m}(\h^H_i \mp \h^M_i) + \fr\m2(\h^H_i \pm \h^M_i)  \ .}
Therefore, the orthogonal projections of $\h$ onto $\mfh$ and $\e_\upm(\mfh)$ are
\aln{
\mfP_{\mfh}(\h) = \h^H_i \tilde H_i + \h^K_{ij} K_{ij} \ ,
\ \ \ \ \ \ \
\mfP_{\e_\upm(\mfh)}(\h) = \h^H_{\upm\,i} \e_\upm(\tilde H_i) + \h^K_{ij} K_{ij} \ .}}
Projecting \eqref{eq:1} and \eqref{eq:2} onto $\e_\um(\mfh)$ and $\e_\up(\mfh)$ respectively we find
\eq{\label{eq:1p}
\dm\big(\mfP_{\e_\um(\mfh)}(g^{-1} \dpl g + g^{-1} \e_\up(A_+) g)\big) & - \dpl \e_\um(A_-)
\\ & + \com{\e_\um(A_-)}{\mfP_{\e_\um(\mfh)}(g^{-1} \dpl g + g^{-1} \e_\up(A_+) g)} = 0 \ ,}
and
\eq{\label{eq:2p}
\dpl\big(\mfP_{\e_\up(\mfh)}(g \e_\um(A_-) g^{-1} - \dm g g^{-1})\big) & - \dm\e_\up(A_+)
\\ & + \com{\e_\up(A_+)}{\mfP_{\e_\up(\mfh)}(g \e_\um(A_-) g^{-1} - \dm g g^{-1})} = 0 \ .}
Therefore, we can use the gauge symmetry \eqref{ggttrr} to fix
\eq{
\e_\up(A_+) = \e_\um(A_-) = \mfP_{\e_\um(\mfh)}(\inv g \dpl g) = \mfP_{\e_\up(\mfh)}(\dm g \inv g) = 0 \ .}
The final set of on-shell gauge-fixed equations of motion are given by
\eq{\label{tomatch1}
\dm(\inv g \dpl g) + \com{\inv g \e_\up(\tilde T) g}{\e_\um(\tilde T)} = 0 \ , \ \ \ \qquad
\mfP_{\e_\um(\mfh)}(\inv g \dpl g) = \mfP_{\e_\up(\mfh)}(\dm g \inv g) = 0 \ .}

It is well known that these equations follow, with a particular on-shell gauge-fixing, from an asymmetrically gauged WZW model plus an integrable potential
\eq{\label{gwzwactasym}
\Act = \ \fr{\coup}{8\pi} \Tr \Big[\fr{1}{2}&\,\int d^2x \ \ \inv{g}\dpl g\ \inv{g}\dm g\ - \fr{1}{3}\,\int d^3x \; \ \e^{mnl} \ \inv{g} \del_m g\ \inv{g}\del_n g\ \inv{g}\del_l g
\\ + & \,\int d^2x \; \ \big( \e_\up(A_+)\dm g\inv{g} - \e_\um(A_-)\inv{g}\dpl g - \inv{g} \e_\up(A_+) g \e_\um(A_-) + A_+ A_- \big)
\\ + &\,\int d^2x \ \, \inv{g} \e_\um(\tilde T) g \e_\up(\tilde T) \Big] \ .}
Let us assume that in the $\m \ra 0$ limit $g$ remains an element of the full group $G$. To analyze the terms involving the gauge field we write
\eq{
A_\pm = B_\pm + \tilde C_\pm , \qquad \text{where} \quad B_\pm \in \mfk \quad \text{and} \quad \tilde C_\pm = C_{\pm\,i}\tilde H_i \in \mfl \ ,}
and take $B_\pm$ and $C_{\pm\,i}$ to remain finite in the $\m \ra 0$ limit. Therefore\,\footnote{Note that here we have used the fact that $\tilde H_i$ scales like $\m$ and hence vanishes in the $\m \ra 0$ limit, while $\e_{\upm}(\tilde H_i) \ra \S_{\pm\,i}$ by construction.}
\eq{A_\pm \xrightarrow[\m\ra0]{} B_\pm \ , \qquad \e_\upm(A_\pm) \xrightarrow[\m\ra0]{} B_\pm + C_\pm \ , \qquad \text{where} \quad B_\pm \in \mfk \quad \text{and} \quad C_\pm = C_{\pm\,i}\S_{\upm\,i} \in \mfl_\upm \ .}

Putting all this together, the $\m \ra 0$ of the action \eqref{gwzwactasym} is
\aln{\Act = \ \fr{\coup}{8\pi} \Tr \Big[ \fr{1}{2} \,\int d^2x & \ \ \inv g \dpl g\ \inv g \dm g\ - \fr{1}{3}\,\int d^3x \; \ \e^{mnl} \ \inv g \del_m g\ \inv g \del_n g\ \inv g \del_l g
\\ \ \ \ \ \ \ + \,\int d^2x &\ \ \big( (B_+ + C_+) \dm g \inv g - (B_- + C_-)\inv g\dpl g - \inv g (B_+ + C_+) g (B_- + C_-) + B_+ B_- \big)
\\ \ \ \ \ + \,\int d^2x &\ \ \inv g T_\up g T_\um \Big] \ .\ta{gwzwactmz}}

\

Let us now consider the second procedure: to first take the $\m \ra 0$ limit and then partially gauge-fix. In the $\m \ra 0$ limit we have seen that the projections used in the previous section become degenerate, see, for example, footnote \ref{footproj}. Therefore, as we now want to work directly in the $\m \ra 0$ limit of \eqref{eq:1} and \eqref{eq:2}, we use the following decomposition of the algebra
\eq{\label{nad}
\mfg = \mfk \oplus \mfl_\up \oplus \mfl_\um \oplus \mfr \ .}
Note that unlike the decompositions used before this is not an orthogonal decomposition.

We define the following algebra elements
\eq{\label{eq:apmdef}
   &  a_+ = g^{-1}(B_++C_+)g + g^{-1}\dpl g \ ,
\\ &  a_- =  g (B_-+C_-) g^{-1} - \dm g g^{-1} \ ,
\qquad \text{where} \quad B_\pm \in \mfk \quad \text{and} \quad C_\pm \in \mfl_\upm \ ,}
along with the following projections of $a_\pm$
\eq{\label{eq:apmproj}
& b_\pm = a_\pm\big|_\mfk \ , \quad c_{\up \pm} = a_\pm\big|_{\mfl_\up} \ , \quad
  c_{\um \pm} = a_\pm\big|_{\mfl_\um} \ , \quad d_\pm = a_\pm\big|_\mfr \ ,}
which relate to the decomposition \eqref{nad}.

Using these definitions the decomposition of \eqref{eq:1} under $\mfg = \mfk \, \oplus \, \mfl_\up \oplus \, \mfl_\um \oplus \, \mfr$ is\,\footnote{Here $\inv g T_\up g$ can be decomposed under $\mft_\up \oplus \mft_\um \oplus \mfn$, where $\mft_\upm$ and $\mfn$ are spanned by $T_\upm$ and $N_i$ respectively. It is this decomposition that the projections in the main text refer to.}
\eq{\label{eq:decomp1}
\mfk : \qquad & \dm b_+ -\dpl B_- + \com{B_-}{b_+} + \com{C_-}{c_{\up+}}\big|_\mfk = 0
\\
\mfl_\up : \qquad & \dm c_{\up+} + \com{B_-}{c_{\up+}} = 0
\\
\mfl_\um : \qquad & \dm c_{\um+} - \dpl C_- + \com{B_-}{c_{\um+}} +\com{C_-}{b_+} + \com{(g^{-1}T_\up g)\big|_\mfn}{T_\um} = 0
\\
\mfr : \qquad & \dm d_+ + \com{C_-}{c_{\up+}}\big|_\mfr + \com{(g^{-1}T_\up g)\big|_{\mft_\up}}{T_\um} = 0 \ ,}
while for \eqref{eq:2} the decomposition is
\eq{\label{eq:decomp2}
\mfk : \qquad & \dpl b_- -\dpl B_+ + \com{B_+}{b_-} + \com{C_+}{c_{\um-}}\big|_\mfk = 0
\\
\mfl_\up : \qquad & \dpl c_{\um-} + \com{B_+}{c_{\um-}} = 0
\\
\mfl_\um : \qquad & \dpl c_{\up-} - \dm C_+ + \com{B_+}{c_{\up-}} +\com{C_+}{b_-} + \com{(gT_\um g^{-1})\big|_\mfn}{T_\up} = 0
\\
\mfr : \qquad & \dpl d_- + \com{C_+}{c_{\um-}}\big|_\mfr + \com{(gT_\um g^{-1})\big|_{\mft_\um}}{T_\up} = 0 \ .}

On-shell the gauge-symmetry \eqref{ggttrr} can be used to fix $C_\pm = 0$ and then $B_\pm = b_\pm = 0$ leaving the following set of equations
\eq{\label{tomatch2}
\dm(\inv g \dpl g) + \com{\inv g T_\up g}{T_\um} = 0 \ , \qquad
(\inv g \dpl g)\big|_\mfk = & \, (\dm g \inv g)\big|_\mfk = 0 \ .}

These equations of motion then follow, with a particular on-shell gauge-fixing, from the following action
\eq{\label{gwzwactmzalt}
\Act = \ \fr{\coup}{8\pi} \Tr \Big[ \fr{1}{2} & \,\int d^2x \ \ \inv g \dpl g\ \inv g \dm g\ - \fr{1}{3}\,\int d^3x \; \ \e^{mnl} \ \inv g \del_m g\ \inv g \del_n g\ \inv g \del_l g
\\ + & \,\int d^2x \; \ \big( B_+ \dm g \inv g - B_-\inv g\dpl g - \inv g B_+ g B_- + B_+ B_- \big)
\\ + & \,\int d^2x \ \; \inv g T_\up g T_\um \Big] \ .}

\subsubsection{Summary\label{sec:sar}}

It is apparent that there is some conflict between these two $\mu \to 0$ limiting procedures, i.e. the two steps involved do not commute. To see this clearly let us return to the first procedure based on first gauge-fixing and then taking the limit. Considering the $\m \ra 0$ limit of the action \eqref{gwzwactasym}, given by \eqref{gwzwactmz}, and varying with respect to $g$ we find the $\m \ra 0$ limit of the equation of motion \eqref{eq:1}, or equivalently \eqref{eq:2}, as expected. These can be decomposed under \eqref{nad} to give \eqref{eq:decomp1} and \eqref{eq:decomp2} respectively, again as one should expect. However, varying with respect to $A_\pm$ we find the constraint equations
\eq{\label{eq:const}
b_+ = B_+ \ , \qquad c_{\up+} = 0 \ , \qquad b_- = B_- \ , \qquad c_{\um-} = 0\ .}
These constraint equations only preserve a part of the $H_\up^0 \X H_\um^0$ gauge symmetry (defined as the $\m \ra 0$ limit of \eqref{ggttrr}) -- under which \eqref{eq:1} and \eqref{eq:2} are invariant. In particular, representing the gauge transformation parameters $h_\upm$ as
\eq{\label{eq:gtparam}
h_\upm = e^{\Xi_{\upm}} \Kc_\upm \ , \qquad \Xi_{\upm} \in \mfl_\upm \ , \quad \Kc_\pm \in K \ ,}
the transformations preserving the constraint equations \eqref{eq:const} are given by restricting
\eq{
\Kc_\up = \Kc_\um = \Kc \ .}
Using these gauge transformations on-shell we can fix $C_\pm = B_\pm = 0$. The gauge-fixed equations of motion are then given by
\eq{\label{tomatch3}
& \dm (g^{-1}\dpl g) +\com{g^{-1}T_\up g}{T_\um} = 0 \ , \qquad
(g^{-1}\dpl g)\big|_{\mfh_\up} = \ (\dm g g^{-1})\big|_{\mfh_\um} = 0 \ .}
As a consistency check, one can show that the $\m \ra 0$ limit of the set of equations \eqref{tomatch1} (with certain rescalings to ensure a finite but non-zero limit) are equivalent to the set of equations \eqref{tomatch3}.

\

It is now fully clear that we find two different sets of equations of motion depending on whether one first takes the $\m \ra 0$ limit or partially gauge-fixes in the reduction procedure. The sets of equations are given by \eqref{tomatch2} in the case of taking the $\m \ra 0$ limit first and by \eqref{tomatch3} in the case of partially gauge-fixing first.

This ambiguity is related to an issue that we have already discussed in section \ref{sec:crpr3}. In the second approach (based on the axially gauged WZW action) described in that section we found that to get the desired result we needed to represent the $\pm$-components of the gauge field as derivatives of two independent scalars. This motivates us to do the same in the action \eqref{gwzwactmz} for $C_\pm$. This is equivalent to gauge-fixing $C_\pm = 0$ using the gauge symmetry described by the parameters $\Xi_\upm$ in \eqref{eq:gtparam}, before using the variational principle to find the equations of motion.

Once this is done the two actions \eqref{gwzwactmz} and \eqref{gwzwactmzalt} become equivalent and we may claim that they should describe the Pohlmeyer reduction of strings moving in $\ads_n$. Indeed, in the next section we will show that using a particular parametrization of $g$ the action \eqref{gwzwactmzalt} leads to the same equations of motion as found in the coordinate formulation of the reduction described in appendix \ref{app:cr}.

\subsection{Properties of the Pohlmeyer-reduced AdS action\label{sec:acpoft}}

In this section we will investigate some properties of the Pohlmeyer-reduced action for strings in AdS space. We will start with the action \eqref{gwzwactmzalt} which is equivalent, as discussed in section \ref{sec:sar}, to \eqref{gwzwactmz} with the gauge-fixing condition $C_\pm = 0$.

\subsubsection{Matching equations of motion of coordinate-based reduction\label{sec:eom}}

To connect with the coordinate reduction described in appendix \ref{app:cr} we introduce a convenient parametrization of the group-valued field $g$
\eq{\label{gparam}
& g = k^{-1} e^{\yp} e^{-2\p\, R} e^{\ym} \ , \qquad \quad \ypm \in \mfl_\upm \ , \qquad k \in K \ .}
Using this parametrization, along with the field redefinition,
\eq{
B_+\ \ra\ k^{-1} B_+ k + \inv k \dpl k \ ,}
the action \eqref{gwzwactmzalt} becomes
\eq{\label{gwzwc}
\Act = \ \fr{\coup}{8\pi}\bigg( & 4 \int d^2 x \; \big[ \dpl \p \dm \p - \fr14 e^{2\p}\big]
     + \int d^2x\; e^{2\p} \Tr\big[D_-\ym D_+\yp \big]
\\ & - \Tr \Big[\,\fr{1}{2}\,\int d^2x \ \ \inv k \dpl k \inv k \dm k\ - \fr{1}{3}\,\int d^3x \; \ \e^{mnl} \ \inv k \del_m k\ \inv k \del_n k\ \inv k \del_l k
\\ & \qquad \qquad \, + \,\int d^2x \; \ \big( B_+ \dm k \inv k - B_- \inv k \dpl k - \inv k B_+ k B_- + B_+ B_- \big)\Big]\bigg)\ .}
where $D_\pm \ypm = \dpm \ypm + \com{B_\pm}{\ypm}$ (or $D_\pm \xi_{\upm\,i} = \dpm \xi_{\upm\,i} - B_{\pm\,ij} \xi_{\upm\,j}$ in explicit component notation, see below). This action contains the Liouville action \eqref{Liou1lag} for the scalar field $\p$ and a gauged WZW action for the coset $\fk KK$ with $K = \SO(n-2)$. These two parts are coupled through the second term involving $\ypm$ taking values in two $\R^{n-2}$ subspaces of $\mfg = \mfso(n-1,1)$ and transforming as vectors under the adjoint action of $K$.

The action \eqref{gwzwc} has a standard $K = \SO(n-2)$ gauge symmetry under which the fields transform as
\eq{\label{gsymtr}
\p \ra \p \ , \quad \ k \ra \inv \Kc k\Kc \ , \quad \ \ypm \ra \inv \Kc \ypm \Kc \ , \quad \ B_\pm \ra \inv \Kc B_\pm\Kc + \inv \Kc \dpm \Kc \ , \quad \ \Kc = \Kc(x) \ .\vphantom{\fr12}}
It is also invariant under 2-d Lorentz symmetry, but being the result of taking the $\m \ra 0$ limit\,\footnote{Recall that in the standard Pohlmeyer reduction of strings in $\ads_n \times S^1$ we set the angle of $S^1$ equal to $\mu \tau$ to fix the residual conformal diffeomorphisms remaining after choosing the conformal gauge. This symmetry is obviously restored in the $\mu \to 0$ limit.} the action \eqref{gwzwc} has also much larger local conformal reparametrization symmetry acting as
\eq{\label{lcr}
& \qquad (\dpl, B_+) \ra \Lu (\dpl, B_+) \ , \qquad (\dm, B_-) \ra \Ld (\dm, B_-) \ ,
\\ & \qquad \, e^{2\p} \ra \Lu \Ld e^{2\p} \ , \qquad \yp \ra {\Ld^{-1}} \yp \ , \qquad \ym \ra {\Lu^{-1}} \ym \ ,
\\ & \xxx^\pm \ra\, f_{_{\pm}} (\xxx^\pm) \ , \qquad \Lu = f'_{_{+}} = \Lu(\xxx^+) \ ,\qquad \Ld =f'_{_{-}} = \Ld(\xxx^-) \ .}
As we will see in section \ref{sec:quafluc} when we study the fluctuations of the action \eqref{gwzwc} this symmetry effectively corresponds to an extra massless mode. This mode should be gauge-fixed if we consider string theory defined directly in $\ads_n$ or it corresponds to a physical mode (associated to the $S^1$) if we consider string theory in $\ads_n\times S^1$ and obtain the reduced theory by the $\mu \to 0$ limit.

\

The equations of motion found by varying the action \eqref{gwzwc} are
\als{
\label{eq1} \p \ : & \qquad \dpl \dm \p + \fr14 e^{2\p} - \fr14 e^{2\p} \Tr[D_- \ym D_+ \yp] = 0 \vphantom{\fr14} \\
\label{eq2} \ypm \ : & \qquad D_\pm (e^{2\p} D_\mp \ymp) = 0 \vphantom{\fr14} \\
\label{eq3} B_+ \ : & \qquad B_- - k B_- \inv k + \dm k \inv k = e^{2\p}\com{\yp}{D_-\ym}\Big|_{\mfk} \vphantom{\fr14} \\
\label{eq4} B_- \ : & \qquad B_+ - \inv k B_+ k - \inv k \dpl k = e^{2\p}\com{\ym}{D_+\yp}\Big|_{\mfk} \vphantom{\fr14} \\
\label{eq5} k \ : & \qquad D_-(\inv k B_+ k + \inv k \dpl k) - \dpl B_- = \inv k\big( \dm B_+ - D_+(k B_- \inv k - \dm k \inv k) \big)k = 0 \ .\vphantom{\fr14}}
Eliminating $k$ by taking $D_+$ of \eqref{eq3} (then using \eqref{eq2} and the second equation in \eqref{eq5}) gives\,\footnote{The same equation can also be found from taking $D_-$ of \eqref{eq4} and using \eqref{eq2} and the first equation in \eqref{eq5}, demonstrating the consistency of the set of equations \eqref{eq3}--\eqref{eq5}.}
\eq{\label{eq6}
F_{+-} \equiv \dpl B_- - \dm B_+ + \com{B_+}{B_-} = e^{2\p}\com{D_+ \yp}{D_- \ym}\Big|_{\mfk} \ ,}
Note that for a special solution $\ymp =0$ the field $B_\pm$ becomes pure gauge and $k$ can be set to a constant, reflecting the topological nature of the $\fk KK$ gauged WZW subsystem of \eqref{gwzwc}.

Defining
\eq{\label{uvypm}
u = e^{2\p}D_+\yp \ , \qquad \qquad v = - e^{2\p}D_-\ym \ ,}
equations \eqref{eq1}, \eqref{eq2} and \eqref{eq6} become
\als{
\label{ep1} & \dpl \dm \p + \fr14 e^{2\p} + \fr14 e^{-2\p} \Tr (u\,v) = 0 \ ,\vphantom{\fr14} \\
\label{ep2} & D_+ v = D_-u = 0 \ , \qquad F_{-+} = e^{-2\p}\com{u}{v}\Big|_{\mfk} \ .\vphantom{\fr14}}
Expanding these equations in components
\eq{\label{foot}
u = u_i \S_{\up\,i} \ , \ \ \qquad v = v_i \S_{\um\,i} \ , \ \ \qquad B_\pm = - \sum_{j>i} B_{\pm\,ij} K_{ij} \ ,}
one finds precisely the set of equations \eqref{17}, \eqref{row1} and \eqref{row2} that arise in the coordinate formulation of the Pohlmeyer reduction of strings moving in AdS space (see, e.g., \cite{Jevicki:2007aa,Dorn:2009kq,Dorn:2009gq} and appendix \ref{app:cr}).

Considering strings moving in $\ads_n$ we expect to find $n-2$ physical (transverse) degrees of freedom. To isolate these we should gauge-fix the residual local conformal reparametrizations \eqref{lcr} under which the fields $u$ and $v$ transform as
\eq{\label{lcruv}
u \ra \Lu^2 \, u \ , \qquad \qquad v \ra \Ld^2 \, v \ .}
The equations of motion for $u$ and $v$ \eqref{ep2} imply that
\eq{\label{lo}
\dpl \Tr(v\tilde v) = \dm \Tr(u\tilde u) = 0 \ , \ \ \ \ \ \text{where} \ \ \ \ \ \tilde u = u_i \S_{\um\,i} \ , \ \ \tilde v = v_i \S_{\up\,i} \ ,}
that is
\eq{\label{holo}
\dpl(v_i v_i) = \dm (u_i u_i) = 0 \ .}
Therefore, using the freedom \eqref{lcruv} we can fix $v_i v_i$ and $u_i u_i$ to be constant, in particular (since the transformations \eqref{lcruv} do not change signs)
\eq{\label{olo}
v_i v_i \ra \sign(v_i v_i) \ , \qquad u_i u_i \ra \sign(u_i u_i) \ .}

\subsubsection{Simplifying the action\label{fsota}}

The action \eqref{gwzwc} has a number of unusual features. In particular, there are a number of fields with ``wrong'' signs for their kinetic terms and it is not clear a priori how one should count degrees of freedom. One of the standard approaches (described in appendix \ref{app:cr}) for dealing with the set of equations \eqref{ep1} and \eqref{ep2}, \eqref{uvypm} is to solve the first-order equations $D_+ v = D_- u = 0$ for the gauge field $B_\pm$. We start by defining the gauge field in terms of two group-valued fields
\eq{\label{bpmk12}
B_+ = - \dpl \nv k_1 k_1^{-1} \ , \qquad B_- = - \dm \nv k_2 k_2^{-1} \ ,\qquad \nv k_{1,2} \in K \ .}
Using the Polyakov-Wiegmann identity to rewrite its gauged WZW part, the action \eqref{gwzwc} takes the compact form
\eq{\label{gwzwccomppre}
\Act = \ \fr{\coup}{8\pi}\bigg( & 4 \int d^2 x \; \big[ \dpl \p \dm \p - \fr14 e^{2\p}\big]
     + \int d^2x\; e^{2\p} \Tr\big[D_-\ym D_+\yp \big]
     + I(k_1^{-1} k \nv k_2) - I (k_1^{-1} \nv k_2)\bigg) \ ,}
where we have introduced the notation
\eq{\la{wwzz}
I(g) \equiv - \Tr\Big[\,\fr12 \,\int d^2x \ \ \inv g \dpl g \inv g \dm g \ - \fr13\, \int d^3x \; \ \e^{mnl} \inv g \del_m g \inv g \del_n g \inv g \del_l g \Big] \ ,}
for a WZW action. Since $D_\pm$ depend only on $\nv k_1,\, \nv k_2$ it is clear that we can decouple $k$ in from \eqref{gwzwccomppre} by doing the field redefinition
\eq{
k = \nv k_1 k' \inv k_2 \ .}
Ignoring the decoupled part $I(k')$ then leaves us with an action for just $\p$, $\ypm$ and  $\nv k_1, \nv k_2$
\eq{\label{gwzwccomp}
\Act = \ \fr{\coup}{8\pi}\bigg(4 \int d^2 x \; \big[ \dpl \p \dm \p - \fr14 e^{2\p}\big]
     + \int d^2x\; e^{2\p} \Tr\big[D_-\ym D_+\yp \big] - I (k_1^{-1} \nv k_2)\bigg) \ ,}
where, from \eqref{bpmk12}, we have $D_+\yp = \del_+\yp - [\dpl \nv k_1 k_1^{-1},\yp ], \ \ D_-\ym = \del_-\ym - [\dm \nv k_2 k_2^{-1},\ym]$.

This action still has a $K$ gauge invariance \eqref{gsymtr}, i.e. $ \ypm \ra \inv \Kc \ypm \Kc,\ \ \nv k_{1,2} \ra \inv \Kc \nv k_{1,2}$, which allows one to fix, e.g., $\nv k_2=\id$ (or equivalently $B_- = 0$). Alternatively, we may introduce new gauge-invariant variables
\eq{
\typ \equiv k_1^{-1} \yp \nv k_1 \ ,\qquad \ \ \ \tym \equiv k_2^{-1}\yp \nv k_2 \ , \qquad \ \ \ \ \ \tk \equiv k_1^{-1} \nv k_2 \ ,}
in terms of which \eqref{gwzwccomp} becomes
\eq{\label{gwzwco}
\Act = \ \fr{\coup}{8\pi}\bigg( 4 \int d^2 x \; \big[ \dpl \p \dm \p - \fr14 e^{2\p}\big]
     + \int d^2x\; e^{2\p} \Tr\big[\tk \ \del_- \tym\ \tk^{-1} \ \del_+\typ \big] - I (\tk)\bigg) \ .}
Since the action \eqref{gwzwccomp}/\eqref{gwzwco} has a simple linear dependence on $\ym$ and $\yp$ / $\tym$ and $\typ$ one can solve for these fields explicitly and get a ``reduced'' effective action for $\p$ and $\tk$ only:
\eq{\label{gwzwcoa}
\Act = \ \fr{\coup}{8\pi}\Big( 4 \int d^2 x \; \big[ \dpl \p \dm \p - \fr14 e^{2\p}\big]
     + \int d^2x\; e^{-2\p} \Tr\big[ \tk^{-1} \, \tyv \, \tk \, \tyu \big] - I (\tk)\Big) \ ,}
where $\tyu=\tyu(x^+)$ and $\tyv=\tyv(x^-)$ are fixed integration matrix functions. This action has the form of a non-abelian Toda model coupled to a Liouville scalar.\footnote{A similar action (but without the $e^{2\p}$ term) was discussed in \cite{Bakas:1996np} where it was noted that it can be itself interpreted as a {\it conformal} non-abelian Toda model.} An equivalent action \eqref{gwcoc} can also be found  by choosing the $B_- = 0$ ($k_2 = \id$) gauge in the set of reduced equations found in the coordinate parametrization (see appendix \ref{app:cr}).

The action \eqref{gwzwcoa} describes $1+\fr12(n-2)(n-3)$ degrees of freedom, that is $\fr12(n-3)(n-4)$ more than the $n-2$ we expect from string theory. To eliminate these extra fields (by solving their equations of motion) it is useful to parametrize the fields $\tyu$ and $\tyv$ in terms of the new variables $ \nv{ \hat k_1},\,\nv{ \hat k_2}, \, U, \, V$ as\,\footnote{Using the residual conformal reparametrization freedom as in \eqref{lcruv}--\eqref{olo} $U$ and $V$ can be fixed to be equal to constants (e.g., $\sign U$ and $\sign V$ respectively).}
\eq{\label{eqeqeq}
\tyv =  V\, \hat k_1 \,  \S_{\um\,1} \, \hat k_1^{-1}  \ , \qquad & \tyu =  U \,\hat k_2 \, \S_{\up\,1} \, \hat k_2^{-1}
\\ V = V(x^-) \ , \quad \nv{\hat k_1} = \nv{\hat k_1}(x^-) \in K \ , \qquad &
   U = U(x^+) \ , \quad \nv{\hat k_2} = \nv{\hat k_2}(x^+) \in K \ .}
The WZW part of the action \eqref{gwzwcoa} is invariant under the field redefinition\,\footnote{Note that by the Polyakov-Wiegmann identity $I(\nv g_1 \nv g_2) = I(\nv g_1) + I(\nv g_2) + \Tr \int d^2x \ \inv g_1 \dpl \nv g_1 \dm \nv g_2 \inv g_2 $ and that $I(g(x^+))=I(g(x^-))=0$.}
\eq{
\tilde k\ \ra\ \ \nv{\hat k_1}(x^-)\, \tilde k\, \hat k_2^{-1}(x^+) \ ,}
and hence $\nv{\hat k_{1,2}}$ can be absorbed into the field $k$, so that \eqref{gwzwcoa} is equal to
\eq{\label{gwzwc2}
\Act = \ \fr{\coup}{8\pi}\Big( 4 \int d^2 x \; \big[ \dpl \p \dm \p - \fr14 e^{2\p}\big]
     + \int d^2x\; e^{-2\p} U\,V \Tr\big[ \tk^{-1} \, \S_{\um\,1} \, \tk\, \S_{\up\,1} \big] - I (\tk)\Big) \ .}
An equivalent action appears in \eqref{gwoc}.

It is useful to define the subgroup of $K = \SO(n-2)$ that commutes with $\S_{\upm\,1}$ as $\widecheck K = \SO(n-3)$ and the corresponding algebra as $\check \mfk = \mfso(n-3)$. Furthermore, we introduce the notation $\check \mfk^c$ for the orthogonal complement of $\check \mfk$ in $\mfk$. We can then parametrize the group-valued field $\tilde k$ as
\eq{\label{possibly}
\tilde k = \hat k \check k \ , \quad \hat k = e^{2\Xs} \ , \qquad \check k \in \widecheck K \ , \quad \Xs \in \check \mfk^c \ .}
Projecting the equation of motion for $k$ from the action \eqref{gwzwc2} onto $\bar \mfk$ we find
\eq{\label{checkintermsofhat}
\dm(\check k^{-1} \dpl \check k) = - \dm(\check k^{-1} \hat k^{-1} \dpl \hat k \check k)\Big|_{\bar \mfk} \ ,}
which can implicitly be used to solve for $\check k$ in terms of $\hat k$. It is important to note that solving for $\check k$ in this way will give a non-local function of $\hat k$.\footnote{Although we will not need it here, this is a procedure that can be performed perturbatively if the group elements $\check k$ and $\hat k$ are parametrized as exponentials of elements of the algebra. For example, using \eqref{possibly} along with $\check k = e^{2 \Ys}$, where $\Ys \in \check \mfk$ we find $\dm\dpl \Ys = \fr12\dm\com{\Xs}{\dpl \Xs} + \Oc(\Xs^4)$. As the leading order piece of $\Ys$ is $\Oc(\Xs^2)$ it will not contribute to the quadratic fluctuation action discussed in section \ref{sec:quafluc}.} In this way we find an effective action for just the $n-2$ degrees of freedom, $\p$ and $\Xs$ -- it is given by \eqref{gwzwc2} with \eqref{possibly} and $\check k$ defined as a function of $\hat k$ through \eqref{checkintermsofhat}. It is important to note that $\check k$ is thus a non-local function of $\hat k$ and therefore the action \eqref{gwzwc2} viewed as a functional of $\hat k$ is non-local.

\

Assuming that $U$ and $V$ have been fixed using the residual conformal diffeomorphisms, this action describes $n-2$ scalar fields with second-order differential equations. This is the expected number of degrees of freedom for string theory in $\ads_n$. However, it appears that the action contains ghost-like degrees of freedom -- the $n-3$ scalar fields represented by $\Xs$ in \eqref{possibly} have the opposite sign for their kinetic terms as compared to the Liouville field $\p$. One approach to alleviating this could be to apply the imaginary rotation\,\footnote{An alternative is to rotate $\phi$ but this would make the action complex.}
\eq{\label{wr}
\Xs \ra i \Xs \ , \qquad U \ra i U \ , \qquad V \ra i V \ .}
This amounts to changing the group of the WZW action in \eqref{gwzwc2} from $\SO(n-2)$ to $\SO(n-3,1)$. This procedure can be easily generalized to the action \eqref{gwzwc} as long as one suitably modifies the generators $\S_{\upm\,1}$, and in both cases one finds a real action.

These ghost-like signs in the action are a persistent issue in the Pohlmeyer reduction of strings in $\ads_n$. They appear also when studying the equations of motion coming from the coordinate-based reduction, outlined in appendix \ref{app:cr}. For $\ads_n$ with $n \geq 4$ (when $K = \SO(n-2)$ becomes non-trivial) having solved for the gauge field, one is led to an action with opposite-sign kinetic terms -- see, for example, \eqref{4thlag}, \eqref{5thlag} and \eqref{gwzwc2ads5}.

\

In the above discussion we have assumed a time-like string world-sheet. It turns out that these ghost-like degrees of freedom are a consequence of this, and if we started instead with a space-like world-sheet all the degrees of freedom would have the correct signs for their kinetic terms. Indeed, the imaginary rotation in \eqref{wr} is part of the transformation required to change from the Pohlmeyer reduction for a time-like world-sheet to that for a space-like world-sheet. While the ultimate resolution of this issue for the Minkowski-signature world-sheet reduction is unclear to us at the moment, we include a more detailed discussion of the above rotation and the space-like string world-sheet reduction case in section \ref{app:glf}. It is worth emphasizing again that it is the 2-d Euclidean-signature reduction that was discussed recently in connection with construction of classical minimal surfaces ending at the boundary of AdS.

\

As an illustration of the above discussion regarding the action \eqref{gwzwc2} let us consider explicitly the cases corresponding to strings in $\ads_4$ and $\ads_5$. For $\ads_4$ the group $K$ is just $\SO(2)$ and we can parametrize $\hat k$ in terms of a scalar $\b$ as
\eq{
\hat k = e^{2\b K_{12}} \ .}
Expanding the action \eqref{gwzwc2} we get precisely the Lagrangian \eqref{4thlag} that is found from the analogous procedure at the level of the equations of motion in appendix \ref{app:sc}.

In the case of $\ads_5$ we may parametrize $\hat k$ in terms of two scalar fields $\b,\c$ as
\eq{
\hat k = e^{-\c K_{23}} e^{2\b K_{12}} e^{\c K_{23}} \ ,}
which is of the form \eqref{possibly}. Unlike the $\ads_4$ case here $\widecheck K$ is no longer trivial. Therefore, we need to solve for $\check k$ in terms of $\b$ and $\c$ using \eqref{checkintermsofhat} and then expand the action \eqref{gwzwc2}. The result is the following non-local action
\eq{\label{gwzwc2ads5}
\Act = \ \fr{\coup}{2\pi} \int d^2 x \; \Big[ \dpl\p\dm\p - \dpl\b\dm\b + \fr14\dpl\c\dm\c
     - \fr14 (\sec2&\b \,\dpl \c) \, \fr{\dm}{\dpl}\, (\sec2\b\,\dpl \c)
\\ & - \fr14 (e^{2\p} + U V \cos2\b \, e^{-2\p})\Big] \ .}
The corresponding variational equations are
\eq{\label{eqeqeqnl}
& \dpl\dm \p +  \fr14(e^{2\p} - U V \cos2\b\, e^{-2\p}) = 0\ ,
\\ & \dpl\dm\b - \fr12 \sec 2\b \tan 2\b \, \dpl \c \,\fr{\dm}{\dpl}(\sec2\b \, \dpl \c) + \fr14 e^{-2\p} U V \sin2\b = 0 \ ,
\\ & \big(1+2 \tan2\b\, \dpl \b \fr1\dpl\big)\big[\dpl (\cos 2\b \,\dm \c) - \dm(\sec2\b \,\dpl\c)\big] = 0 \ .}
These equations are equivalent to the set of equations \eqref{eqeqeqeq} found in appendix \ref{app:sc} via an analogous procedure at the level of the equations of motion.

Generalizing to $\ads_n$, the variational equations following from the action \eqref{gwzwc2}, with an appropriate parametrization of $\hat k$ should be equivalent to those derived in appendix \ref{app:sc} (see \eqref{eqeq1} and \eqref{eqeq2}).

\renewcommand{\theequation}{4.\arabic{equation}}
\setcounter{equation}{0}
\section{The \texorpdfstring{$\mathbf{\boldsymbol{\m} \ra 0}$}{m -> 0} limit of Pohlmeyer-reduced \texorpdfstring{$\mathbf{\ads_5 \X S^5}$}{AdS5 x S5} superstring\label{sec:sst}}

In this section we generalize the $\m \ra 0$ limit of the Pohlmeyer reduction of string theory in $\ads_n \X S^1$ to the full $\ads_5 \X S^5$ superstring. The world-sheet action for classical bosonic strings moving in $\ads_5 \X S^5$ can be written in terms of a symmetric space sigma model with the coset space represented as
\eq{
\ads_5 \X S^5 \cong \fr{\SO(4,2)}{\SO(4,1)} \X \fr{\SO(6)}{\SO(5)} \ .}
This coset can be also represented, using algebra isomorphisms, as
\eq{\label{eq:ucoset}
\ads_5 \X S^5 \cong \fr{\SU(2,2)}{\USp(2,2)} \X \fr{\SU(4)}{\USp(4)} \ .}
It turns out that to include fermions this latter representation is more appropriate. Indeed, the numerator groups in \eqref{eq:ucoset} are then enlarged to the supergroup $\PSU(2,2|4)$ so that the two-dimensional world-sheet sigma model is based on the supercoset
\eq{\label{eq:sucoset}
\fr{\PSU(2,2|4)}{\USp(2,2) \X \USp(4)} \ .}
We will work with a particular $8 \X 8$ matrix representation of the superalgebra $\mffh = \mfpsu(2,2|4)$, which is described in detail in appendix \ref{app:psu224}. Also outlined in this appendix is the $\Z_4$ grading of $\mfpsu(2,2|4)$ that plays an important r\^ole in both the superstring theory and its Pohlmeyer reduction.

Let us consider the field $f \in \PSU(2,2|4)$, and define the left-invariant Maurer-Cartan one-form $\Jc = \inv f d f \in \mffh$. Under the $\Z_4$ grading discussed in appendix \ref{app:psu224} this Maurer-Cartan one-form can be decomposed as
\eq{
\Jc = \Ac + \Qc_1 + \Pc + \Qc_3 \ , \qquad & \Ac = \Jc_0 \in \mfg = \mffh_0 \ , \quad \Qc_1 = \Jc_1 \in \mffh_1 \ ,
\\  & \Pc = \Jc_2 \in \mfp = \mffh_2\ , \quad \Qc_3 = \Jc_3 \in \mffh_3 \ .}
Fixing the conformal gauge we are meant to impose the Virasoro constraints
\eq{\label{eq:p1vir}
\STr(\Pc_\pm \Pc_\pm) = 0 \ ,}
while the variational equations coming from the superstring world-sheet action are
\al{\label{eq:p1fulleom1}
& \Dc_+ \Pc_- + \Dc_- \Pc_+ + \com{\Qc_{1 \, -}}{\Qc_{1 \, +}} + \com{\Qc_{3 \, +}}{\Qc_{3 \, -}} = 0 \ ,
\\ \label{eq:p1fulleom2}
& \qquad \quad \com{\Pc_+}{\Qc_{1 \, -}} = 0 \ , \quad \com{\Pc_-}{\Qc_{3 \, +}} = 0 \ .}
Interpreted as first-order equations for the components of $\Jc$ they should be supplemented by the flatness equation for the Maurer-Cartan one-form $\Jc$. In the superstring context, this is an equation taking values in $\mffh$ and therefore it can be decomposed under the $\Z_4$ grading as follows
\als{\label{eq:p1decompmc1} \vphantom{\mffh_1}
& \mfp \ : && \Dc_- \Pc_+ - \Dc_+ \Pc_- + \com{\Qc_{1 \, -}}{\Qc_{1 \, +}} + \com{\Qc_{3 \, -}}{\Qc_{3 \, +}} = 0 \ ,
\\  \label{eq:p1decompmc2} \vphantom{\mffh_1}
& \mffh_1 \ : && \Dc_- \Qc_{1 \, +} - \Dc_+ Q_{1 \, -} + \com{\Pc_-}{\Qc_{3 \, +}} - \com{\Pc_+}{\Qc_{3 \, -}} = 0 \ ,
\\  \label{eq:p1decompmc3} \vphantom{\mffh_1}
& \mffh_3 \ : && \Dc_- \Qc_{3 \, +} - \Dc_+ Q_{3 \, -} + \com{\Pc_-}{\Qc_{1 \, +}} - \com{\Pc_+}{\Qc_{1 \, -}} = 0 \ ,
\\  \label{eq:p1decompmc4} \vphantom{\mffh_1}
& \mfg \ : && d \Ac + \Ac \wedge \Ac + \Pc \wedge \Pc + \Qc_1 \wedge \Qc_3 + \Qc_3 \wedge \Qc_1 = 0 \ .}
Combining the equation for $\Pc$ \eqref{eq:p1fulleom1} and the flatness equation projected on $\mffh_2$ \eqref{eq:p1decompmc1} gives the following first-order equations for $\Pc_+$ and $\Pc_-$
\eq{\label{eq:p1eomv2}
\Dc_+ \Pc_- + \com{\Qc_{3 \, +}}{\Qc_{3 \, -}} = 0 \ , \qquad
\Dc_- \Pc_+ + \com{\Qc_{1 \, -}}{\Qc_{1 \, +}} = 0 \ .}

These equations, the Virasoro constraints \eqref{eq:p1vir}, the variational equations \eqref{eq:p1fulleom1} and the flatness equation \eqref{eq:p1decompmc1} are invariant under the $G$-gauge symmetry
\aln{
f \ra fg \quad \Ra \quad \Jc \ra \inv g \Jc g + \inv g d g \ \quad \Ra \quad  \vphantom{\mffh_1}
& \mfg \ \, : \qquad \Ac \ra \inv g \Ac g + \inv g d g \ ,
\\\vphantom{\mffh_1}
& \mffh_{1,3} \ : \quad \, \Qc_{1,3} \ra \inv g \Qc_{1,3} g \ ,
\\\vphantom{\mffh_1}
& \mfp \ \, : \qquad \Pc \ra \inv g \Pc g \ .\ta{eq:jft1}}

\subsection[Generalized reduction procedure for \texorpdfstring{$\m \neq 0$}{m != 0}]{Generalized reduction procedure for $\mathbf{\boldsymbol{\m}\boldsymbol{\neq}0}$}

The Pohlmeyer reduction of the above system of equations is carried out in full detail in \cite{Grigoriev:2007bu}. Here we would like to generalize that procedure to admit the limit $\m \ra 0$. The reduction involves solving the first-order equations \eqref{eq:p1eomv2}, the fermionic equations~\eqref{eq:p1fulleom2} and the Virasoro constraints \eqref{eq:p1vir} by introducing new variables parametrizing the physical degrees of freedom. The equations of motion of the Pohlmeyer-reduced theory are then given by the flatness equation projected onto $\mffh_0$, $\mffh_1$ and $\mffh_3$, i.e. equations \eqref{eq:p1decompmc4}, \eqref{eq:p1decompmc2} and \eqref{eq:p1decompmc3} respectively.

As in the bosonic construction we need to identify some specific features of the algebra, in particular certain generators. The maximal abelian subalgebra, $\mfa$, of $\mfp$ is two-dimensional and we choose the following basis
\eq{\label{eq:achoice}
T_A=\fr{i}{2} \, \diag(1,1,-1,-1,0,0,0,0) \ ,
\ \ \ \ \ \ \ \ \ \
T_S=\fr{i}{2} \, \diag(0,0,0,0,1,1,-1,-1) \ .}
We also introduce
\eq{
T = T_A + T_S = \fr{i}{2} \,\diag(1,1,-1,-1,1,1,-1,-1)\ ,}
which induces a further $\Z_2$ decomposition of the algebra $\mfpsu(2,2|4)$
\eq{
\mffh = \mffh^\pa \oplus \mffh^\pe \ ,}
described in detail in appendix \ref{app:psu224}. $T$ defines the subalgebra $\mfh$ of $\mfg = \mffh_0$ that commutes with it, namely
\eq{
\mfh \equiv \mffh_0^\pe = \mfsu(2) \oplus \mfsu(2) \oplus \mfsu(2) \oplus \mfsu(2) \ , \ \ \ \ \ \ \ [\mfh, T]=0 \ .}

By analogy with the bosonic construction, in order to have a non-trivial $\m \ra 0$ limit we need to identify the generator $R = {\rm T}_0^\pa(\fr12,0,0,0,0,0,0,0)$, as given in appendix \ref{app:psu224}, and define the algebra automorphisms
\eq{
\e_\upm(\mffh) = \m^{\mp R} \, \mffh\,  \m^{\pm R} \ .}
The key property of $R$, as in the bosonic construction of section \ref{sec:adsngt}, is that it is an element of the bosonic algebra $\mfusp(2,2)$ (i.e. it is a generator of the gauge group associated with the AdS symmetric space) that does not commute with $T$. The final comment of an algebraic nature is to identify the subalgebra of $\mfh = \mfsu(2)^{\oplus\, 4}$ that commutes with $R$. We will denote this $\mfk$ and in this case it is $\mfsu(2)^{\oplus\, 3}$.

\

To generalize the procedure of \cite{Grigoriev:2007bu} we start by parametrizing $\Pc_\pm$ in terms of two $\USp(2,2) \X \USp(4)$ group-valued fields,
\eq{
\Pc_- = p_{1-} \inv g_1 \e_\up(T_A) \gol + p_{2-} \inv g_1 \e_\up(T_S) \gol \ , \qquad
\Pc_+ = p_{1+} \inv g_2 \e_\um(T_A) \gtl + p_{2+} \inv g_2 \e_\um(T_S) \gtl \ ,
\\ \nv g_{1,2} \in \USp(2,2)\X \USp(4) \ ,}
where $p_{1,2\pm}$ are functions of the world-sheet coordinates. Substituting these expressions into the Virasoro constraints \eqref{eq:p1vir} implies that $p_{1+}=p_{2+}=p_+$ and $p_{1-}=p_{2-}=p_-$. Thus
\eq{\label{eq:p1cur}
\Pc_- = p_- \inv g_1 \e_{\up}(T) \gol \ , \qquad
\Pc_+ = p_+ \inv g_2\e_{\um} (T) \gtl \ .}

The world-sheet action for the \adss5 superstring possesses a local fermionic $\k$-symmetry \cite{Metsaev:1998it} which allows one to eliminate $16$ of the $32$ fermionic \dof~in the superalgebra $\mfpsu(2,2|4)$. A discussion of the $\k$-symmetry in the context of the Pohlmeyer reduction can be found in \cite{Grigoriev:2007bu}. Here we note that the Pohlmeyer-reduced theory describes just the physical \dof~and therefore the $\k$-symmetry gauge should be fixed. Following \cite{Grigoriev:2007bu} we do this by choosing the on-shell gauge
\eq{\label{eq:zzero}
\Qc_{1\,-} = \Qc_{3\,+} = 0 \ ,}
so that the fermionic equations of motion are satisfied. This actually leaves some residual part of the $\k$-symmetry that we will use later on. Equations \eqref{eq:p1eomv2} then simplify to
\eq{\label{eq:p1boskafix}
\Dc_+ \Pc_- = 0 \ , \qquad \Dc_- \Pc_+ = 0 \ ,}
which are precisely the same equations as in \eqref{eq:gh} that were solved in the bosonic string reduction discussed in section \ref{sec:adsngt}. These imply that $p_\pm$ are functions of $x^\pm$ and therefore the residual symmetry in the conformal gauge, represented by the conformal reparametrizations, can then be used to set
\eq{
p_+ = p_- = \m = {\rm const.}}
Following the bosonic construction, equations \eqref{eq:p1boskafix} are solved by taking
\eq{\label{eq:bosfields2}
\Ac_+ = \inv g_1 \dpl \gol + \inv g_1 \e_\up(A_+) \gol \ , \qquad
\Ac_- = \inv g_2 \dm \gtl + \inv g_2 \e_\um(A_-) \gtl \ ,
\\ \qquad \qquad \qquad \quad \, A_\pm \in \mffh_0^\pe \equiv \mfh = \mfsu(2)^{\oplus 4} \ .}
Defining
\eq{\label{eq:definitions}
g = \gol \inv g_2 \ , \qquad \td \Qc_1 = \gtl \Qc_{1\,+} \inv g_2 \ , \qquad \td \Qc_3 = \gol\Qc_{3\,-} \inv g_1 \ ,}
and substituting \eqref{eq:p1cur} (with $p_\pm = \m$), \eqref{eq:zzero} and \eqref{eq:bosfields2} into \eqref{eq:p1decompmc2} and \eqref{eq:p1decompmc3} we find
\eq{
&\dm \td \Qc_1 + \com{\e_\um(A_-)}{\td \Qc_1} - \m\com{\e_\um(T)}{\inv g \td \Qc_3 g} = 0 \ , \\
&\dpl\td \Qc_3 + \com{\e_\up(A_+)}{\td \Qc_3} - \m\com{\e_\up(T)}{g \td \Qc_1 \inv g} = 0 \ .}
This in turn implies
\eq{
\dm \e_\up(\td \Qc_1) + \com{A_-}{\e_\up(\td \Qc_1)} - \m\com{T}{\e_\up(\inv g \Qc_3 g)} = 0 \ \quad \Ra \quad \dm \e_\up(\td \Qc_1)^\pe + \com{A_-}{\e_\up(\td \Qc_1)^\pe} = 0 \ , \\
\dpl\e_\um(\td \Qc_3) + \com{A_+}{\e_\um(\td \Qc_3)} - \m\com{T}{\e_\um(g \Qc_1 \inv g)} = 0 \ \quad \Ra \quad \dpl\e_\up(\td \Qc_3)^\pe + \com{A_+}{\e_\um(\td \Qc_3)^\pe} = 0 \ .}
The residual part of the $\k$-symmetry can be thus fixed by setting
\eq{
\e_\up(\td \Qc_1)^\pe = \e_\um(\td \Qc_3)^\pe = 0 \ .}
Finally, we can define the fermionic fields of the Pohlmeyer-reduced theory
\eq{\label{ylyrdef}
\YR = \e_\up(\td \Qc_1)^\pa \in \mffh_1^\pa \ , \qquad \YL = \e_\um(\td \Qc_3)^\pa \in \mffh_3^\pa \ .}

\

Thus far the \eom~\eqref{eq:p1fulleom1} and \eqref{eq:p1fulleom2} and the Virasoro constraints \eqref{eq:p1vir} of the \adss5 superstring have been solved through writing the currents in terms of a new set of fields $\{g,A_\pm,\YR,\YL\}$ describing the physical \dof~of the system. Substituting this change of variables into the remaining components of the flatness equation, i.e. equations \eqref{eq:p1decompmc4}, \eqref{eq:p1decompmc2} and \eqref{eq:p1decompmc3}, gives the following set of \eom~for the reduced theory
\al{\label{eq:p1redeom} \vphantom{\mffh}
& \dm(\inv g \dpl g + \inv g \e_\up(A_+) g) - \dpl \e_\um(A_-) + \com{\e_\um(A_-)}{\inv g \dpl g + \inv g \e_\up(A_+) g}
\\\vphantom{\mffh}
& \no \qquad \qquad \qquad \qquad \qquad \qquad \qquad \qquad \quad + \m^2 \com{\inv g \e_\up(T) g}{\e_\um(T)} + \m \com{\inv g \e_\up(\YL) g}{\e_\um(\YR)} = 0 \ ,
\\\vphantom{\mffh}
&\label{eq:p1redeomf1} \dm \YR + \com{A_-}{\YR} - \m \com{T}{\e_\up(\inv g \e_\up(\YL) g)} = 0 \ ,
\\
&\label{eq:p1redeomf2} \dpl\YL + \com{A_+}{\YL} - \m \com{T}{\e_\um(g \e_\um(\YR) \inv g)} = 0 \ .}
Again, as in the bosonic case, these equations have an $H_\up^\m \X H_\um^\m$ gauge symmetry given by the transformations \eqref{ggttrr} for the bosonic fields and
\eq{\label{eq:p1hxhgauge}
\YR \ra \e_\up(h_\um)^{-1} \YR \e_\up(h_\um)^{\vphantom{-1}} \ , \qquad \YL \ra \e_\um(h_\up)^{-1} \YL \e_\um(h_\up)^{\vphantom{-1}} \ ,}
for the fermionic fields.

\

To write down a Lagrangian whose variational equations give the \eom~\eqref{eq:p1redeom} the $H_\up^\m \X H_\um^\m$ gauge symmetry needs to be partially fixed. Furthermore, as in our earlier discussion of the bosonic case, one gets apparently different results depending if this is done before or after taking the $\m \ra 0$ limit. The construction turns out to be an obvious generalization of the one in section \ref{sec:red}. In particular, we find that the two different results are related by an off-shell gauge-fixing. Here for computational simplicity we will consider the case when we gauge-fix before taking the $\m \ra 0$ limit. We will then need to gauge-fix to zero the analogs of the $C_\pm$ fields in the bosonic construction in order to find the correct $\m \ra 0$ limit of the Pohlmeyer-reduced $\ads_5 \X S^5$ superstring theory. This also allows us to establish the equivalence with the action that would be found by taking the $\m \ra 0$ limit first and then gauge-fixing.

Starting with finite $\m$ and projecting \eqref{eq:p1redeom} onto $\e_\um(\mfh)$ and using the fermionic equations of motion and the $H_\um^\m$ gauge symmetry we observe that on-shell one can fix
\eq{
A_- = \mfP_{\e_\um(\mfh)}\big(\inv g \dpl g - \fr12\com{\com{T}{\YR}}{\YR}\big) = 0 \ .}
Similarly, it is possible to show that at the same time using the $H_\up^\m$ gauge symmetry one can also fix (on the equations of motion)
\eq{
A_+ = \mfP_{\e_\up(\mfh)}\big(\dm g \inv g + \fr12\com{\com{T}{\YL}}{\YL}\big) = 0 \ .}

The resulting set of equations can then be found (after a particular on-shell gauge-fixing) from the action
\eq{\label{eq:rtaction}
\Act = \fr{\coup}{4\pi} \STr\Big[\fr{1}{2} & \int d^2x \; \inv g \dpl g \inv g \dm g - \fr{1}{3} \int d^3x \; \e^{mnl} \inv g \del_m g \inv g \del_n g \inv g \del_l g
\\   + & \int d^2x \; \big(\e_\up(A_+) \dm g \inv g - \e_\um(A_-) \inv g \dpl g - \inv g \e_\up(A_+)\, g\, \e_\um(A_-) + A_+A_-\big)
\\   + & \int d^2x \; \big(\YL T \dpl \YL + \YL T \com{A_+}{\YL} +\YR T \dm \YR + \YR T \com{A_-}{\YR}\big)
\\   + & \int d^2 x\;\big(\m^2 \; \inv g\, \e_\up(T)\, g\, \e_\um(T) + \m \; \inv g\, \e_\up(\YL)\, g\, \e_\um(\YR) \big) \Big] \ .}
In summary, the fields in this action all live in various subalgebras of $\mfpsu(2,2|4)$ or subgroups of $\PSU(2,2|4)$. The supertrace is therefore the appropriate mixed signature trace required to construct invariants of this algebra. The field $g$ is a $G = \USp(2,2) \X \USp(4)$ group-valued field and the gauge fields $A_\pm \in \mfh = \mfsu(2)^{\oplus\,4}$. The fermionic fields $\YR$ and $\YL$ have canonical kinetic terms, a mass term and a Yukawa-type coupling with $g$. They take values in certain Grassmann-odd subspaces of the superalgebra $\mffh = \mfpsu(2,2|4)$. In particular, $\YR \in \mffh_1^\pa$ and $\YL \in \mffh_3^\pa$ (for a definition of these spaces see appendix \ref{app:psu224}).

\subsection[Reduced theory corresponding to the \texorpdfstring{$\m \ra 0$}{m -> 0} limit]{Reduced theory corresponding to the $\mathbf{\boldsymbol{\m} \ra 0}$ limit}

The action \eqref{eq:rtaction} has a well-defined and non-trivial $\m \ra 0$ limit so long as we follow the same idea as in the bosonic construction and parametrize
\eq{
A_\pm = B_\pm + \m \td C_\pm \ , \qquad \qquad \quad B_\pm \in \mfk \ , \quad \td C_\pm \in \mfl \ .}
Here $\mfl$ is the orthogonal complement of $\mfk$ in $\mfh$. We then have the following useful limits
\eq{\label{psit}
\lim_{\m\ra 0} \, \m \, \e_\upm(T) = T_\upm \ ,
\qquad\qquad & \lim_{\m \ra 0} \, \m^{\fr12}\e_\up(\YL) = \tYL \in \mffh^\pa_{3\,\up} \ ,
\\
\lim_{\m \ra 0} \, \m \, \e_\upm(\td C_\pm) = C_\pm \in \mfl_\upm \ ,
\qquad\qquad & \lim_{\m \ra 0} \, \m^{\fr12}\e_\um(\YR) = \tYR \in \mffh^\pa_{1\,\um} \ ,}
where all of the objects on the right-hand side are finite, non-zero and contain the same number of degrees of freedom as those on the left-hand side. The explicit forms of the generators and spaces on the right-hand side are given in detail in appendix \ref{app:psu2243}.

Recalling once again the subtlety in the bosonic construction (see section \ref{sec:sar}) that in the $\m \ra 0$ limit of the Pohlmeyer reduction of $\ads_n \X S^1$ we needed to gauge-fix $C_\pm = 0$ off-shell\,\footnote{This was needed to get the correct equations of motion, degrees of freedom and to match the action found by taking the $\m \ra 0$ limit first and then partially gauge-fixing.} the final action the $\m \ra 0$ limit of the Pohlmeyer-reduced $\ads_5 \X S^5$ superstring is given by
\eq{\label{eq:rtactionfin}
\Act = \fr{\coup}{4\pi} \STr\Big[\fr{1}{2} & \int d^2x \; \inv g \dpl g \inv g \dm g - \fr{1}{3} \int d^3x \; \e^{mnl}\inv g \del_m g \inv g \del_n g \inv g \del_l g
\\ + & \int d^2x \; \big(B_+ \dm g \inv g - B_- \inv g \dpl g - \inv g B_+ g B_- + B_+B_-\big)
\\ + & \int d^2x \; \big(\YL T \dpl \YL + \YL T \com{B_+}{\YL} +\YR T \dm \YR + \YR T \com{B_-}{\YR}\big)
\\ + & \int d^2x \; \big(\; \inv g T_\up g T_\um + \; \inv g \tYL g \tYR\big) \Big] \ ,}
where $\tYL,\tYR$ are defined in \eqref{psit}.

In the original $\mu\not=0$ reduction \cite{Grigoriev:2007bu} the potential and Yukawa-type fermionic term in the reduced theory action can be identified with the original superstring action for the specific gauge-fixing of the coset currents used in the reduction. The same is true here as well: if we substitute the field redefinitions \eqref{eq:p1cur}--\eqref{eq:definitions} and \eqref{ylyrdef} into the superstring action and take the $\m \ra 0$ limit we find just the final line of \eqref{eq:rtactionfin} with an appropriate identification of $\coup$ and the string tension. This is also true in the bosonic construction discussed in section \ref{sec:gtp}.

One important feature of this action is that the ``sphere'' part of the potential is vanishing in the $\m \ra 0$ limit as $T_\upm$ are only non-zero in the $\mfsu(2,2)$ subalgebra of $\mfpsu(2,2|4)$. This should be expected as the algebra $\mfsu(4)$ has a definite signature. Therefore, demanding that
\eq{
\Tr(\Pc^S_\pm \Pc^S_\pm) = 0\ ,}
implies that $\Pc^S_\pm$ vanish, and it is these fields that appear in the ``sphere'' part of the potential.

\

As suggested by the notation, the Grassmann-odd subspaces in which $\tYLR$ take values satisfy the following commutation relations
\eq{
\com{\mffh_{1\,\um}^{\ph{\pa}}}{\mfl_\um^{\ph{\pa}}} = 0 \ , \qquad \com{\mffh_{3\,\up}^{\ph{\pa}}}{\mfl_\up^{\ph{\pa}}} = 0 \ .}
Therefore, extending the parametrization of section \ref{sec:acpoft}
\eq{\label{eq:parametrization}
& g = k^{-1} e^{\yp}e^{-2\p R} e^{\ym}g_{_S}\ , \qquad k \in K_{_A} = \SU(2)\ , \qquad g_{_S}\in \USp(4) \ ,
\\
& B_\pm = B_{_A\,\pm} + B_{_S\,\pm} \ , \qquad B_{_A\,\pm} \in \mfk_{_A} = \mfsu(2) \ , \qquad B_{_S\,\pm} \in \mfk_{_S} = \mfsu(2)^{\oplus 2} \ ,}
we find that $\yp$ and $\ym$ drop out of the fermionic potential term. To clarify what the groups $K_{_A}$ and $K_{_S}$ are, we note that $K = \SU(2)^3$ contains one $\SU(2)$ subgroup of $\USp(2,2)$ and two of $\USp(4)$. $K_{_A}$ is then just the single $\SU(2)$ that is a subgroup of $\USp(2,2)$, while $K_{_S}$ is the $\SU(2)^2$ that is a subgroup of $\USp(4)$. $\mfk_{_A}$ and $\mfk_{_S}$ are the corresponding algebras.

Using the parametrization \eqref{eq:parametrization} along with the field redefinitions\,\footnote{Note that these field redefinitions of $\YL$ and $\tYL$, which, as one may recall, contain the same degrees of freedom, are consistent as $\com{R}{K} = 0$ by the definition of $K$.}
\eq{
B_{_A\,+} \ra \inv k B_{_A\,+} k +\inv k \dpl k \ , \ \ \ \ \ \ \ \ \ \ \ \YL \ra \inv k \YL k \ , \qquad \qquad \tYL \ra \inv k \tYL k \ ,}
we find that \eqref{eq:rtactionfin} can be rewritten as
\eq{\label{final}
\Act = \ \fr{\coup}{4\pi}\bigg( \, 2 \int d^2 x \; & \big[ \dpl\p\dm\p -\fr14 e^{2\p}\big] + \int d^2x\; e^{2\p} \STr\big[ D_-\ym D_+\yp \big]
\\ + \STr\Big[ \,-\fr{1}{2}&\,\int d^2x \ \ \inv k \dpl k \inv k \dm k \ + \fr{1}{3}\,\int d^3x \; \ \e^{mnl} \ \inv k \del_m k \ \inv k \del_n k \ \inv k \del_l k
\\ - & \,\int d^2x \; \ \big(B_{_A\,+}\dm k \inv k - B_{_A\,-}\inv k \dpl k - \inv k B_{_A\,+} k B_{_A\,-} + B_{_A\,+} B_{_A\,-}\big)
\\ + \,\fr{1}{2}&\,\int d^2x \ \ \inv g_{_S} \dpl g_{_S} \inv g_{_S} \dm g_{_S} \ - \fr{1}{3}\,\int d^3x \; \ \e^{mnl} \ \inv g_{_S} \del_m g_{_S} \ \inv g_{_S} \del_n g_{_S} \ \inv g_{_S} \del_l g_{_S}
\\ + & \,\int d^2x \; \ \big(B_{_S\,+}\dm g_{_S} \inv g_{_S} - B_{_S\,-}\inv g_{_S} \dpl g_{_S} - \inv g_{_S} B_{_S\,+} g_{_S} B_{_S\,-} + B_{_S\,+} B_{_S\,-}\big)
\\ + & \int d^2x \; \big(\YL T \Ds_+ \YL + \YR T \Ds_- \YR + e^\p \inv g_{_S} \tYL g_{_S} \tYR\big)\Big] \ ,}
where $\tYL,\,\tYR$ are defined in \eqref{psit} and
\eq{
D_\pm \ypm = \dpm \ypm + \com{B_{_A\,\pm}}{\ypm} \ , \ \ \ \ \ \ \ \ \ \ \ \ \Ds_\pm \Psi = \dpm \Psi + \com{B_{\pm}}{\Psi} \ .}
As for the reduction of bosonic strings in $\ads_n$ this action contains the Liouville action for the field $\p$. The second and third lines represent the gauged WZW action for the coset $\fk{\SU(2)}{\SU(2)}$, which is coupled to the Liouville part through the $\ypm$ term in the first line. As in the bosonic construction $\ypm$ take values in two $\R^{3}$ subspaces of $\mfg=\mfusp(2,2)$ and transform as vectors under the adjoint action of $K_{_A}=\SU(2)$. The fourth and fifth lines represent a gauged WZW action for the coset $\fk{\USp(4)}{[\SU(2)]^2}$, describing the part of the reduced model coming from the five-sphere. The last line is the fermionic part of the action.

The action \eqref{final} has a manifest $K$ gauge symmetry under which the bosonic fields from the ``AdS'' sector and the full $B_\pm$ transform as in \eqref{gsymtr}, while the fermionic fields and $g_{_S}$ transform as
\eq{
\YLR \ra \inv \Kc \YLR \Kc \ , \qquad g_{_S} \ra \inv \Kc g_{_S} \Kc \ .}
The action \eqref{final} is also invariant under the conformal reparametrizations \eqref{lcr} with the fermions and $g_{_S}$ transforming as
\eq{
\YR \ra \Lu^{\sfr12} \, \YR \ , \qquad \YL \ra \Ld^{\sfr12} \, \YL \ , \qquad g_{_S} \ra g_{_S} \ .}

\subsection[\texorpdfstring{$\m \ra 0$}{m -> 0} limit of reduced theory for \texorpdfstring{$\ads_2 \X S^2$}{AdS2 x S2} truncation: \texorpdfstring{$\Nc = 2$}{N = 2} super Liouville theory]{$\mathbf{\boldsymbol{\m} \ra 0}$ limit of reduced theory for $\mathbf{\ads_2 \X S^2}$ truncation: \\ $\mathbf{\boldsymbol{\Nc}=2}$ super Liouville theory}

To provide a simple example, let us now consider the $\m=0$ limit of the reduced theory for the $\ads_2 \X S^2$ truncation of the general $\ads_5 \X S^5$ theory.

The actions describing the $\ads_2 \X S^2$ superstring\footnote{By the \adss2 superstring theory here we mean the formal supercoset subsector obtained by truncation of the full 10-dimensional superstring theory on $\ads_2 \X S^2 \X T^6$, see \cite{Berkovits:1999zq,Sorokin:2011rr}.} and the corresponding Pohlmeyer-reduced theory \cite{Grigoriev:2007bu} are formally identical to those in the $\ads_5 \X S^5$ case except the underlying supercoset is now
\eq{
\fr{\Fh}{G} = \fr{PSU(1,1|2)}{SO(1,1) \X SO(2)} \ .}
The $\m\ra 0$ limit of the Pohlmeyer-reduced $\ads_2 \X S^2$ superstring is thus found in the same way as in the $\ads_5 \X S^5$ construction above. Parametrizing the group-valued and algebra-valued fields in terms of components as in \cite{Grigoriev:2007bu} one finds that the resulting reduced-theory Lagrangian is given by
\eq{\label{eq:p2susy2}
\Lag =  \dpl\p\dm\p - \fr14 e^{2\p} + \dpl \vp \dm \vp + \a\,\dm\a &+ \,\d\,\dm\d + \,\n\,\dpl\n + \,\r\,\dpl\r
\\ & -  e^{\p} \big[\cos\vp \ (\n\d + \r\a) + \sin\vp \ (-\r\d + \n\a) \big] \ .}
Here $\p$ and $\vp$ are real bosonic fields, while $\a,\d,\n,\r$ are real (hermitian) fermions.

\

As was shown in \cite{Grigoriev:2007bu}, the Pohlmeyer reduction of the $\ads_2 \X S^2$ superstring for finite $\m$ is given by the $\Nc=2$ 2-d supersymmetric sine-Gordon model defined by the superpotential
\eq{\label{sp}
W(\P) = \m \cos \P \ , }
i.e. the corresponding Lagrangian is given by
\eq{\label{neq2}
\Lag = \dpl \P \dm \P^* - |W'(\P)|^2 + \psi_{_L}^* \dpl \psi_{_L} + \psi_{_R}^* \dm \psi_{_R} + \big[ W''(\P)\psi_{_L} \psi_{_R} + {W^*}''(\P^*)\psi_{_L}^* \psi_{_R}^* \big] \ ,}
where $\P$ is a complex scalar and $\psi_{_{L,R}}$ are 2-d complex Weyl fermions. Writing this in real components ($\P = \vp + i\p$, $\psi_{_R} = \d + i \a$, $\psi_{_L} = \n - i \r$)
we find
\eq{\label{susysine}
\Lag = \dpl\p\dm\p + \dpl\vp\dm\vp + \fr{\m^2}{2}(\cos 2\vp &- \cosh 2\p) + \,\a\,\dm\a + \,\d\,\dm\d + \,\n\,\dpl\n + \,\r\,\dpl\r
\\ &- 2\m \big[\cosh\p\cos\vp \ (\n\d + \r\a) + \sinh\p\sin\vp \ (-\r\d + \n\a) \big] \ .}
It is then easy to see that after shifting $\p \ra \p - \log \m$ as in \eqref{phi-shift} the $\m \ra 0$ limit of \eqref{susysine} gives precisely \eqref{eq:p2susy2} which is thus the same as the Lagrangian of $\Nc=2$ super Liouville theory \cite{Ivanov:1983wp} 
(cf. \cite{Ridout:2011wx}). Indeed, performing this shift of $\p$ in \eqref{sp} and then taking the $\m \ra 0$ limit we find the $\Nc=2$ supersymmetric model with the following superpotential
\eq{
W(\P) = \fr12e^{-i \P}\ .}
Expanding the Lagrangian \eqref{neq2} with this superpotential we then get \eqref{eq:p2susy2}. This demonstrates that this AdS reduced theory \eqref{eq:p2susy2} has $\Nc = 2$ 2-d supersymmetry and also provides a new interpretation to $\Nc=2$ supersymmetric Liouville theory which was discussed, for example, in \cite{Hori:2001ax,Creutzig:2011qm,Nakayama:2004vk,Ridout:2011wx}.

The realisation of 2-d supersymmetry in $\mu\not=0$ Pohlmeyer-reduced $\ads_3 \X S^3$ and $\ads_5 \X S^5$ superstring models is known to be more subtle than in the $\ads_2 \X S^2$ case: it is effectively non-local in the action \cite{Goykhman:2011mq,Hollowood:2011fq} and deformed in the S-matrix \cite{Hoare:2011fj}. We would expect the same to be true in the $\m \ra 0$ limit. Indeed, the algebra $\mfh = \mffh_0^\pe$ that plays an important r\^ole in the identification of 2-d supersymmetry for finite $\m$ is modified by the automorphisms $\e_\upm$ that underlie the $\m \ra 0$ limit. For $\ads_2 \X S^2$ case $\mfh$ is empty and therefore it is perhaps not surprising that the $\Nc=2$ supersymmetry of the non-zero $\m$ case is present also in the $\m\ra 0$ limit. However, for the $\ads_3 \X S^3$ and $\ads_5 \X S^5$ cases $\mfh$ is non-trivial and thus the relationship between the two cases may be more subtle.

\renewcommand{\theequation}{5.\arabic{equation}}
\setcounter{equation}{0}
\section{Quadratic fluctuation spectra near simple classical solutions\label{sec:quafluc}}

In this section we investigate some quantum properties of the action \eqref{final} for the Pohlmeyer-reduced $\ads_5 \X S^5$ superstring in the $\m \ra 0$ limit. For non-zero $\m$ the one-loop partition function of the Pohlmeyer-reduced $\ads_5 \X S^5$ superstring, found by fluctuating the action around a particular classical solution, is equal to the one-loop partition function of the Green-Schwarz superstring around the corresponding string theory solution \cite{Hoare:2009rq,Iwashita:2010tg}. Here we will test the equivalent statement for the reduction found in the $\m \ra 0$ limit.

 The $\m \ra 0$ limit by construction is suitable only for describing classical string solutions in $\ads_5$ (just as for non-zero $\m$ one is restricted to classical solutions which are non-vanishing both in $\ads_5$ and in $S^5$). Here we will discuss the reduced theory solutions corresponding to the two simple string solutions: (i) the massless geodesic in $\ads_2 \subset \ads_5$, and (ii) the large spin (or long string) limit of the folded spinning string \cite{Gubser:2002tv,Frolov:2002av,Frolov:2006qe}, which is equivalent also to the ``null cusp'' Euclidean world-sheet solution \cite{Kruczenski:2002fb,Kruczenski:2007cy,Roiban:2007dq}. The first one is the trivial vacuum background of this AdS reduced theory, just like the BMN geodesic is the vacuum of the original $\ads_5\times S^5$ reduced theory \cite{Grigoriev:2007bu}, while the second is related to a non-trivial vacuum background of the corresponding generalized sinh-Gordon model \cite{Jevicki:2007aa}.

\subsection[Fluctuations around vacuum corresponding to massless geodesic in \texorpdfstring{$\ads_5$}{AdS5}]{Fluctuations around vacuum corresponding to massless geodesic in $\mathbf{\ads_5}$\label{sec:qfmgeo}}

Let us start with a massless geodesic in $\ads_5$ and its counterpart which is the trivial vacuum background of the reduced theory. It is sufficient to restrict the geodesic to be in the $\ads_2$ subspace of $\ads_5$
\eq{
Y_0 = 1 \ , \qquad Y_{-1} = \t \ , \qquad Y_1 = \t \ , \qquad Y_{i+1} = 0 \ \qquad i = 1,2,3 \ ,}
where we are using the usual embedding coordinates. To construct the reduced theory solution we may use the coordinate-based parametrization described in appendix \ref{app:cr}. This gives $\p$, $u_i$, $v_i$ and $B_{\pm\,ij}$, from which we can reconstruct $\p$, $\ypm$ and $B_\pm$ using \eqref{uvypm} and the parametrization given in \eqref{foot}. Given that $\dpl Y \cdot \dm Y = 0$, using equation \eqref{definitions} we find
\eq{\label{rtsp1}
\p = - \infty \ , \ \ \ \ \ \ \ \ u_i = v_i = B_{\pm\,ij} = 0 \ ,}
It is easy to check that this is indeed the simplest vacuum solution of the set of equations \eqref{17}, \eqref{row1} and \eqref{row2}.

Using \eqref{uvypm}, along with the parametrization in equation \eqref{foot} we find that the extension of this solution to the Pohlmeyer-reduced superstring defined by \eqref{final} is given by
\eq{\label{sollosmg}
\p & = -\infty \ , \quad \ \ \ypm = 0 \ ,
\\ k = g_{_S} = \id \ , \qquad B_{_A\,\pm} & = B_{_S\,\pm} = 0 \ , \qquad \YL = \tYL = \YR = \tYR = 0 \ .}
Due to its singular nature, we need to expand the action \eqref{final} around this solution with some care. In particular, when considering the terms involving the fields $\ypm$ we should first put the kinetic term $ e^{2\p} \STr\big[D_- \ym D_+ \yp\big]$ into a canonical form by the field redefinition $\ypm \ra e^{-\p}\ypm$. The resulting set of fluctuation frequencies is then given by (here $\mmm$ is an integer representing the 2-d spatial momentum)
\eq{\label{y}
& 4 \ \X \ \pm \mmm \qquad \text{from the first three lines of \eqref{final} -- the reduced AdS sector} \ ,
\\ & 4 \ \X \ \pm \mmm \qquad \text{from the lines four and five of \eqref{final} -- the reduced sphere sector} \ ,
\\ & 8 \ \X \ \pm \mmm \qquad \text{from the final line of \eqref{final} -- the fermions} \ .}
We have dropped three of the massless fluctuations that naively appear in the reduced AdS sector: one should consider the fields $u$ and $v$ with first-order equations of motion as fundamental and hence integrate over $\ypm$ with an appropriate measure cancelling the contribution from these extra massless modes (see section \ref{lslfss} for a detailed discussion).

The spectrum of $8 + 8$ massless modes matches that found from the superstring world-sheet action expanded around the massless geodesic in an $\ads_2$ subspace of $\ads_5 \X S^5$ \cite{Giombi:2009gd}.

\subsection[Fluctuations around background corresponding to long folded spinning string]{Fluctuations around background corresponding to \\ long folded spinning string\label{sec:qfgkp}}

In general, the folded spinning string \cite{Gubser:2002tv} is described by the following background in $\ads_3 \subset \ads_5$:
\eq{\label{eq:ell}
Y_0 + i Y_{-1} = e^{i\k\t} \cosh \r(\s) \ , \qquad Y_1 + i Y_2 = e^{i\w\t} \sinh \r(\s) \ , \ \ \ \ \ \ \ \ Y_{\ISC+1} = 0 \ , \ \ \ \ \ \ISC = 2,3 \ ,}
where
\eq{
\r'{}^2 = \k^2 \cosh^2 \r - \w^2 \sinh^2 \r \ ,}
so that $\r$ varies from $0$ to its maximal value $\r_*$
\eq{
\coth^2 \r_* = \fr{\w^2}{\k^2} \ .}
Defining $\ve^{-2} = \w^2\k^{-2} - 1$, $\r(\s)$ can be written implicitly in terms of the Jacobi $\text{sn}$ function\,\footnote{Note that the periodicity in $\s$ requires the following relation on the parameters $\k$ and $\ve$
\aln{
\k = \fr{2 \ve}{\pi} \mathbf{K}(-\ve^2) \ .}}
\eq{
\sinh \r = \ve \ \text{sn} \, (\k\inv \ve\s,-\ve^2) \ .}
Using the relations in appendix \ref{app:cr} it is easy to find the corresponding reduced theory solution given in terms of $\p$, $u_i$, $v_i$, $B_{\pm\,ij}$ (see also \cite{Jevicki:2007aa})
\eq{\label{fssgen}
\p =\log (2\r') \ , \qquad u_1 = v_1 = 4\k\w \ ,\qquad & u_\ISC = v_\ISC = B_{\pm\,ij} = 0 \ .}
Reversing the parametrization in \eqref{foot} we find that the corresponding algebra/group valued fields of the group-theoretic construction are
\eq{\label{fssgenn}
& \p = \log ( 2\r') \ , \qquad \ypm = - \Big[\frac{(\log\r')'}{\k \w } + \fr i\w \mathbf{E}(i \r,-\e^{-2})\Big]\S_{\upm\,1} \ , \qquad B_{_A\,\pm} = 0 \ , \qquad k = \id \ .}
Here the subscript $A$ on the gauge field indicates that it is from the reduced $\ads_5$ sector of the full reduced superstring theory. The extension to a classical solution of the Pohlmeyer-reduced superstring, defined by the action \eqref{final}, is then given by \eqref{fssgenn} together with
\eq{\label{extsol}
g_{_S} = \id \ , \qquad B_{_S\,\pm} = 0 \ , \qquad \YL = \tYL = \YR = \tYR = 0 \ .}

An important simple limit of the folded spinning string solution \cite{Frolov:2002av,Frolov:2006qe} is found by taking the spin to be very large in which case the string becomes very long and reaches the boundary of AdS, i.e. $\r_* \ra \infty$, $\ve \ra \infty$, $ \w \ra \k \ra \infty$. Then
\eq{
\r(\s) = \k \s \ ,}
and the reduced theory solution \eqref{fssgenn}, \eqref{extsol} becomes\,\footnote{Here we have used the property \aln{\lim_{\k \ra \infty} - \fr i\k \mathbf{E}(i\k \s,0) = \s \ .}}
\eq{\label{sollos}
\p & = \log (2\k) \ , \quad \ \ \ypm = \s \S_{\upm\,1} \ ,
\\ k = g_{_S} = \id \ , \qquad B_{_A\,\pm} & = B_{_S\,\pm} = 0 \ , \qquad \YL = \tYL = \YR = \tYR = 0 \ .}
Note that written in terms of $u,v \sim \del_\pm \ypm$ this ``string-like'' ($\ypm \sim \s$) background becomes homogeneous and may be interpreted as a non-trivial ``vacuum'' of the associated sinh-Gordon model \cite{Jevicki:2007aa} (it corresponds to $e^{4\p} =UV=$ constant in \eqref{sec0e2}).\footnote{Interestingly, as was found in \cite{Jevicki:2007aa}, the soliton solution of the sinh-Gordon model describes the reduced theory counterpart of the leading correction to the long string limit of the above folded string solution.}

\subsubsection[Bosonic fluctuations from the reduced \texorpdfstring{$\ads_n$}{AdSn} sector]{Bosonic fluctuations from the reduced $\mathbf{\ads_n}$ sector\label{lslfss}}

Let us start with the bosonic fluctuations from the reduction of the string in AdS space. To highlight various features we will consider the reduction of the string in $\ads_n$ described by the action \eqref{gwzwc}. To generalize the classical solution \eqref{eq:ell} we allow $\ISC$ to run from $2$ to $n-2$. In the superstring case we have $n=5$.

When considering the reduction of strings in $\ads_n$ we expect to have $n-2$ physical fluctuations, so the corresponding path integral is to be defined accordingly. We will consider the two approaches based on the actions \eqref{gwzwc} and \eqref{gwzwc2} respectively.

The first approach is based on the direct expansion of the action \eqref{gwzwc} around the solution \eqref{sollos} to quadratic order in the fields (we set $k = \id \cdot\exp\, \z\, $)\,\footnote{Note that we have used the algebra relation $\com{\S_{\um\,1}}{\S_{\up\,1}} \propto R$ and the following parametrization in terms of components
\aln{
& \ypm = \xi_{\upm\,1}\S_{\pm\,1} + \xi_{\upm\,\ISC}\S_{\pm\,\ISC} \ ,
\quad \ B_\pm = - B_{\pm\,1\ISC} K_{1\ISC} - B_{\pm\,\ISC\JSC} K_{\ISC\JSC} \ ,
\quad \ \z = \z_{1\ISC} K_{1\ISC} + \z_{\ISC\JSC} K_{\ISC\JSC} \ ,
\quad \ \JSC >\ISC = 2,\ldots,n \ .}}
\aln{\Act = \ \fr{\coup}{2\pi}\Big( & \int d^2 x \; \big[\dpl \p \dm \p - \k^2 \dm \xi_{\um\,1} \dpl \xi_{\up\,1} + 2\k^2 \p (\dpl \xi_{\up\,1} - \dm \xi_{\um\,1})\big]
\\ + & \int d^2x \; \big[- \k^2 \dpl \xi_{\up\,\ISC} \dm \xi_{\um\,\ISC} - \k^2 B_{+\,1\ISC} \xi_{\up\,\ISC} + \k^2 B_{-\,1\ISC} \xi_{\um\,\ISC} - \k^2 \sigma B_{+\,1\ISC} \dm \xi_{\um\,\ISC} - \k^2 \sigma B_{-\,1\ISC} \dpl \xi_{\up\,\ISC}
\\ & \hspace{125pt} + \frac{1}{4} \dpl \zeta_{1\ISC} \dm \zeta_{1\ISC} - \frac{1}{2} B_{+\,1\ISC} \dm \zeta_{1\ISC} + \frac{1}{2} B_{-\,1\ISC} \dpl \zeta_{1\ISC}- \k^2 \sigma ^2 B_{-\,1\ISC} B_{+\,1\ISC} \big]
\\ + \fr{1}{4} & \int d^2x\;\big[\dpl \z_{\ISC\JSC} \dm \z_{\ISC\JSC} - 2 B_{+\,\ISC\JSC} \dm \z_{\ISC\JSC} + 2 B_{-\,\ISC\JSC} \dpl \z_{\ISC\JSC}\big]\Big) \ .\ta{gwzwcflucp1}}
An apparent dependence of this action on the 2-d space coordinate $\s$ that could complicate the computation of the fluctuation spectrum can be removed by field redefinitions. Replacing $B_{\pm\,ij}$ by derivatives of two sets of scalars
\eq{\label{re}
B_{+\,ij} = \dpl b_{ij} \ , \qquad B_{-\,ij} = \dm \bar b_{ij} \ , \qquad j>i=1,\ldots,n-2 \ ,}
we find (after integrating by parts) that the fields $\xi_{\upm\,i}$ only appear in the action as $\dpm \xi_{\upm\,i}$. Therefore, we can introduce the new fields
\eq{\label{ree}
\Uc_i = \dpl \xi_{\up\,i} \ , \qquad \Vc_i = \dm \xi_{\um\,i} \ , \qquad i=1,\ldots,n-2 \ ,}
and immediately integrate them out. A further integration by parts then allows one to completely remove the $\s$-dependence. Such field redefinitions give rise to Jacobian factors in the path integral, an issue we will return to later in this section.

The resulting action depends on $b_{ij}$ and $\bar b_{ij}$ only through the linear combination $c_{ij} = b_{ij} - \bar b_{ij}$. This is a direct consequence of the $\SO(n-2)$ gauge symmetry and the integral over the decoupled gauge degree of freedom ($b+\bar b$) is cancelled as usual by dividing by the volume of the gauge group. Finally, using the field redefinition $\z_{ij} \ra \z_{ij} + c_{ij}$ we end up with the following quadratic action
\eq{\label{thisaction}
\Act = \ \fr{\coup}{2\pi} \int d^2 x \; \Big(\dpl \p \dm \p & - 4 \k^2 \phi^2 -\fr14 \dpl c_{1\ISC} \dm c_{1\ISC} + \fr{\k^2}2 c_{1\ISC} c_{1\ISC}
\\ & +\fr14 \dpl \z_{1\ISC} \dm \z_{1\ISC} + \fr14 \dpl \z_{\ISC\JSC} \dm \z_{\ISC\JSC} - \fr14 \dpl c_{\ISC\JSC} \dm c_{\ISC\JSC} \Big) \ .}
Here we again see the that $n-3$ of the physical fluctuation modes are ghost-like (the sign of the kinetic term for $c_{1\ISC}$ is negative). However, all the modes would have physical signs had we started with an equivalent \cite{Kruczenski:2007cy} ``null cusp'' solution with Euclidean world-sheet and expanded the corresponding reduced action near its counterpart background. For this reason we shall ignore this sign peculiarity of the time-like version of the AdS reduction by doing a formal imaginary rotation $c_{1\ISC} \to i \,c_{1\ISC}$. It is then straightforward to compute the corresponding spectrum of fluctuation frequencies:
\eq{\label{thissof}
1 \ \X & \ \pm \sqrt{\mmm^2 + 4 \k^2} \ , \qquad (n-3) \ \X \ \pm \sqrt{\mmm^2 +2 \k^2} \ , \ \ \ \ \ \ \ \ (n-3)^2 \ \X \ \pm \mmm \ ,}
where as in \eqref{y} $\mmm$ is the momentum in the compact $\sigma$ direction. We thus end up with a spectrum of $n-2$ physical massive modes, which is what we would expect to find in the corresponding string theory in $\ads_n$.

To understand the meaning of the additional $(n-3)^2$ massless modes in \eqref{thissof} we recall that we used various non-trivial field redefinitions to end up with the action \eqref{thisaction}. In particular, the $\fr12(n-2)(n-3)$ fields $B_{\pm\,ij}$ were redefined as in \eqref{re} while the $n-2$ fields $\xi_{\upm\,i}$ underwent the ``opposite'' transformation \eqref{ree}. The corresponding Jacobian factor in the partition function is then
\eq{\label{contrib2}
\big[\det\dpl\dm\big]^{\fr12(n-2)(n-5)} \ .}
Multiplying this by the contribution of the $(n-3)^2$ massless modes gives
\eq{
\big[\det\dpl\dm\big]^{-\fr12(n-1)} \ ,}
which is the same as the contribution of $n-1$ extra massless modes.

One of these modes has an obvious interpretation: it should correspond to the residual conformal transformations \eqref{lcr}, \eqref{lcruv}. These were left unfixed in the above construction. Whether we should include this massless mode as a physical one or not depends on which is the full string-theory target space we are starting with. We could either choose to describe fluctuations around a string living in $\ads_n$ or a string living in an $\ads_n$ subspace of $\ads_n \X S^1$. The latter is the target space we started with before taking the $\m \ra 0$ limit and therefore we should account for the corresponding massless mode (associated to the $S^1$). This agrees with the usual count of degrees of freedom: $\ads_n$ should correspond to $n-2$ physical degrees of freedom, while string in $\ads_n \X S^1$ should have $n-1$ physical degrees of freedom.

We are then left with $n-2$ massless modes, which we still need to remove to match the string spectrum. This requires a proper understanding of which are the fundamental fields of the reduced theory. In particular, in the coordinate-based reduction described in appendix \ref{app:cr} the usual fields of the reduced theory are $u$ and $v$ rather than $\y_{\upm}$. Each of these fields has precisely $n-2$ components and they are related to $\y_\upm$ by the field redefinition \eqref{uvypm}, involving $\dpm$. This field redefinition amounts to adding an extra $n-2$ massless degrees of freedom, which are precisely those that we have found above. This implies that the corresponding part of the reduced theory path integral measure should contain an extra ``ghost'' determinant factor, i.e.
\eq{\label{measure}
\int [d\y_\up][d\y_\um] \; \big[\det\dpl\dm\big]^{\fr12(n-2)} \ ,}
which precisely cancels the contribution of unphysical $n-2$ massless modes.

Let us mention also that there is an alternative approach which leads directly to the expected physical massive mode spectrum when considering string theory in $\ads_n$. Instead of directly expanding the action \eqref{gwzwc}, we may expand the gauge-fixed action \eqref{gwzwc2} found after solving for $B_\pm$. Here the classical solution is given by
\eq{
\p = \log (2 \k) \ , \ \ \ \qquad U = V = 4\k^2 \ , \ \ \ \qquad \hat k = \id \ .}
Using the residual conformal diffeomorphisms to fix $U = V = 4\k^2$, the quadratic expansion of the action \eqref{gwzwc2} is given by ($\ISC=2,\ldots,n-2$)
\eq{\label{gwzwc3} \Act = \ \fr{\coup}{2\pi} & \int d^2 x \; \big[\dpl \p \dm \p - 4\k^2 \p^2 - \dpl X_\ISC \, \dm X_\ISC + 2\k^2 X_\ISC X_\ISC\big] \ ,}
where the fluctuation of $\hat k$ is parametrized as
\eq{
\hat k = \id \cdot \exp(2X_\ISC K_{1\,\ISC}) \ .}
Here we need again to apply an imaginary rotation $X_\ISC\to i X_\ISC$ to get the quadratic action that directly follows from fluctuating the space-like AdS reduction. Then \eqref{gwzwc3} describes the same spectrum of massive modes
\eq{\label{thissofb}
1 \ \X & \ \pm \sqrt{\mmm^2 + 4 \k^2} \ , \qquad (n-3) \ \X \ \pm \sqrt{\mmm^2 +2 \k^2} \ ,}
in agreement with the corresponding string theory in $\ads_n$.

\subsubsection{Complete fluctuation spectrum around long folded spinning superstring\label{512}}

Let us now return to the reduction of the full $\ads_5 \X S^5$ superstring, described by the action \eqref{final}. As discussed above, the spectrum of 8 bosonic frequencies is given by
\eq{\label{bfluc}
1 \ \X \pm \sqrt{\mmm^2 + 4\k^2} \ , \qquad 2 \ \X \pm \sqrt{\mmm^2+2\k^2} \ , \qquad 5 \ \X \pm \mmm \ .}
Here we have included the counterparts of the massless $S^5$ modes.\footnote{Once again, recall that in section \ref{lslfss}, when discussing the reduction of string theory in $\ads_n$ as a limit of string theory in $\ads_n \X S^1$, we found an additional massless mode which should correspond to the massless fluctuation along the $S^1$ direction if one expands around a classical solution in an $\ads_n$ subspace of $\ads_n \X S^1$. Consequently, this mode, related to local conformal reparametrizations, should be ignored when describing strings in $\ads_n$ but included when describing strings in $\ads_n$ subspace of $\ads_n \X S^1$. Here, as we are considering the bosonic space $\ads_5 \X S^5$, we should then include this massless mode. Another, more heuristic, argument why this should be the case is that this mode corresponds to allowing the stress-energy tensor of the $\ads_5$ and $S^5$ sectors to fluctuate $T^{A}_{\pm\pm} = 0 - \d \Ts \ , \quad T^S_{\pm\pm} = 0 + \d \Ts \ .$ In the reduction procedure we solved the Virasoro constraints by demanding both $T^A_{\pm\pm}$ and $T^S_{\pm\pm}$ should vanish separately. However, the Virasoro constraints of the $\ads_5 \X S^5$ superstring theory only tell us that the total stress-tensor should vanish.}

In general, the action describing the $\m \ra 0$ limit of the Pohlmeyer-reduced $\ads_5 \X S^5$ superstring is, by construction, only suitable for describing fluctuations around classical solutions that live in the $\ads_5$ subspace. From superstring theory, for these solutions we always expect to find five massless modes corresponding to the transverse directions belonging to the five-sphere. The above discussion implies that we will also find these five massless modes when fluctuating the reduced theory action \eqref{final} around the corresponding reduced theory solution.

\

The part of the action \eqref{final} determining the fermionic fluctuation spectrum is given by
\eq{\label{fermfluct}
\fr{\coup}{4\pi} \STr \int d^2x \; (\YL T \dpl \YL + \YR T \dm \YR + 2 \k \tYL \tYR) \ .}
Using the identity
\eq{
\STr\big({\rm T}^\pa_{3\,\up}(\vec\a_{_L}) {\rm T}^\pa_{1\,\um}(\vec\a_{_R})\big) = \fr12 \STr\big({\rm T}^\pa_3 (\vec\a_{_L}){\rm T}^\pa_1(\vec\a_{_R})\big) \ ,}
where the matrices are defined in appendix \ref{app:psu2243}, and expanding \eqref{fermfluct} in terms of the component fields we find
\eq{\label{fermfluct2}
\fr{\coup}{4\pi} \, \int d^2x \; \big[\fr12 \a_{_Lp}\dpl\a_{_Lp} + \fr12 \a_{_Rp}\dm\a_{_Rp} + \k \, \a_{_Lp}\a_{_Rp} \big] \ , \qquad \qquad p = 1,\ldots,8 \ .}
This action describes eight fermionic fluctuations with the frequencies
\eq{\label{ffluc}
8 \ \X \pm \sqrt{\mmm^2 + \k^2} \ .}
We have therefore reproduced the spectrum of fluctuations, \eqref{bfluc} and \eqref{ffluc}, following directly from the superstring action \cite{Frolov:2002av,Frolov:2006qe}.

\renewcommand{\theequation}{6.\arabic{equation}}
\setcounter{equation}{0}
\section[Pohlmeyer reduction of string in AdS in the case of Euclidean world-sheet \texorpdfstring{\\}{} signature]{Pohlmeyer reduction of string in AdS in the case of \\ Euclidean world-sheet signature\label{app:glf}}

A puzzling feature of the actions for the physical degrees of freedom of Pohlmeyer reduced string theory in AdS with Minkowski world-sheet signature is the presence of both positive and negative signs in the kinetic terms. In particular, this is seen in the coordinate-based reduction outlined in appendix \ref{app:cr}, which leads to \eqref{4thlag} for $\ads_4$ and \eqref{5thlag} for $\ads_5$, and also in the group-theoretic reduction procedure, with the gauge-fixing described in section \ref{fsota}, where we found \eqref{gwzwc2} for $\ads_n$ with the explicit expression for $\ads_5$ given by \eqref{gwzwc2ads5}. Related opposite signs are seen explicitly in the kinetic terms for massive fluctuation modes in, e.g., \eqref{thisaction} and \eqref{gwzwc3}.

However, in all these cases, it is possible to apply an imaginary rotation a subset of fields to find a real ghost-free action. For example, if we set $\b \ra i\b$, $U \ra i U$ and $V \ra i V$ in \eqref{4thlag}, \eqref{5thlag} and \eqref{gwzwc2ads5} we find
\eq{\label{4thlagmod}
\tilde\Lag_4 = \dpl\p\dm\p + \dpl\b\dm\b - \fr14(e^{2\p} - U V \cosh 2\b \, e^{-2\p}) \ ,}
\eq{\label{5thlagmod}\qquad
\tilde\Lag_5 = \dpl\p\dm\p + \dpl\b\dm\b + \tanh^2\b\dpl\c\dm \c - \fr14(e^{2\p} - U V \cosh 2\b\, e^{-2\p}) \ ,}
\eq{\label{gwzwc2ads5mod}
\tilde\Lag'_{5} = \dpl\p\dm\p + \dpl\b\dm\b +\fr14\dpl\c\dm\c - \fr14(\sech2\b & \, \dpl \c) \, \fr{\dm}{\dpl}(\sech2\b\,\dpl \c) \\ &
- \fr14(e^{2\p} - U V \cosh2\b \, e^{-2\p}) \ .}
Furthermore, for the action in equation \eqref{gwzwc2}, the relevant imaginary rotation is given by \eqref{wr}. This has the effect of modifying the group of the WZW action in \eqref{gwzwc2} from $SO(n-2)$ to $SO(n-3,1)$ and flipping the sign of the kinetic and mass terms for all the ghost-like degrees of freedom.

\subsection{Euclidean rotation of reduced equations in coordinate parametrization}

Let us first recall the equations of motion \eqref{17}, \eqref{row1}, \eqref{row2} for the Pohlmeyer reduction of strings in $\ads_n$ with time-like world-sheet derived in the coordinate-based approach in appendix \ref{app:cr}
\al{\label{17mod} & \dpl\dm \p + \fr14(e^{2\p} - u_i v^i e^{-2\p}) = 0 \ ,
\\ \label{row1mod} & \dm u_i = B_{-\,i}{}^ju_j \ , \qquad \dpl v_i = B_{+\,i}{}^jv_j \ ,
\\ \label{row2mod} F_{-+\,ij} \equiv \dm B_{+\,ij} - & \dpl B_{-\,ij} - B_{-\,i}{}^k B_{+\,kj} + B_{+\,i}{}^kB_{-\,kj} = \fr12 e^{-2\p}(u_i v_j - u_j v_i) \ .}
In \eqref{17}, \eqref{row1}, \eqref{row2} the indices $i,j,\ldots$ are contracted with $\d_{ij}$, hence there was no difference between raised and lowered indices.

It turns out that all the imaginary rotations mentioned above can be universally described by a certain imaginary rotation of the fields $u_i$, $v_i$ and $B_{ij}$. In particular, let us redefine
\eq{
u_1 \ra i u_1 \ , \qquad v_1 \ra i v_1 \ , \qquad B_{1 \JSC} \ra i B_{1 \JSC} \ .}
The equations of motion are then given by \eqref{17mod}, \eqref{row1mod} and \eqref{row2mod}, with the $\SO(n-2)$ indices $i,j,\ldots$ now contracted with the metric
\eq{\label{ha}
\h_{ij} = \diag(-1,1,\ldots,1) \ .}
To find the reduced equations in the space-like world-sheet case we first need to make the obvious replacements
($\tau$ now is Euclidean time)
\eq{\label{modmodmod}
\dpl \ra \del = i\del_\tau + \del_\s \ , \qquad \dm \ra -\bar \del = i\del_\tau - \del_\s \ , \qquad B_{+\,ij} \ra B_{ij} \ , \qquad B_{-\,ij} \ra - \bar B_{ij} \ .}
Also, $u_i$, $v_i$, $B_{ij}$, $\bar B_{ij}$ are to be considered now as complex-valued functions satisfying the reality conditions\,\footnote{While these reality conditions are consistent with the definitions \eqref{definitions}, they are not consistent with the gauge-fixing $u_\ISC = 0$ ($\ISC = 2,\ldots, n-2$) used in appendix \ref{app:sc} to construct actions for just the physical degrees of freedom -- see equations \eqref{sec0e4} and \eqref{seeq}. Therefore, to construct such actions in this Pohlmeyer reduction for space-like world-sheets alternative gauge-fixings need to be used. However, in general it should be possible to find such gauge-fixings achieving the same effect as the imaginary rotations of the fields $\b$, $U$, $V$ and $\Xs$ described above (see, for example \cite{Dorn:2009gq,Burrington:2009bh}).}
\eq{
u^* = - v \ , \ \ \ \ \ \ \ \ \qquad B_{ij}^* = \bar B_{ij} \ .}
Then \eqref{17mod}, \eqref{row1mod} and \eqref{row2mod} become a set of equations for the Pohlmeyer reduction of strings with space-like world-sheets in $\ads_n$ (which itself is always assumed to have Minkowski signature in this paper).

The difference of the Euclidean world-sheet case compared to the Minkowski one is related to modification of the Virasoro constraints \eqref{22}, in which we need to replace $\dpl$ with $\del$ and $\dm$ with $ -\bar \del$ as in \eqref{modmodmod}. Then the signature of the three-space with the real basis $\{Y, \fr12(\del Y + \bar \del Y), \fr{1}{2i}(\del Y - \bar \del Y)\}$ is $(-,+,+)$ and the complementary basis is given by $n-2$ vectors $N_i$ which should satisfy
\eq{
N_i \cdot N_j = \h_{ij} \ ,}
with $\h_{ij}$ in \eqref{ha}.

Two simple examples of Euclidean AdS surfaces are the null cusp \cite{Kruczenski:2002fb} mentioned in section 5 (see also \cite{Jevicki:2007aa,Alday:2009yn}) and the surface corresponding to circular Wilson loop \cite{Berenstein:1998ij}. The first is the $\ads_3$ solution ending on the null cusp line at the boundary and corresponds in the reduced theory to the non-trivial vacuum of the Euclidean version of the generalized sinh-Gordon equation \eqref{sec0e2} with $\phi=\const$ and $UV = e^{4 \phi}$. The second solution (non-trivial only in a Euclidean $\ads_2$ subspace of the full Minkowski $\ads_n$) describes a 2-d Euclidean-signature world-sheet ending on the unit circle at the boundary. Explicitly, in the Poincar\'e coordinates it is given by $z = \tanh\t \, , \ x_1 = \sech \tau \, \cos \sigma \, , \ x_2 = \sech \tau \, \sin \sigma \, ,$ while in the embedding coordinates
\eq{
Y_0 = \coth \t \ , \qquad Y_{-1} = 0 \ , \qquad Y_1 + i Y_2 = \csch \t \, e^{i\s} \ , \qquad Y_i = 0 \ , \qquad i = 1,\ldots,n-2 \ ,}
where $0 < \t < \infty $ is the Euclidean world-sheet time. One finds then that the reduced theory solution has all the fields vanishing apart from $\p$, which solves the Euclidean Liouville equation,
\eq{
\p = \log(2\csch \t) \ , \ \ \ \ \qquad u_i = v_i = B_{ij} = {\bar B}_{ij} = 0 \ .}

\subsection{Action for AdS reduced theory in the case of Euclidean world-sheet\label{app:glfs}}

The action \eqref{gwzwactmzalt} and \eqref{gwzwc} can be suitably modified to give an action for the reduction of strings in AdS with space-like world-sheet surface. We start by introducing the metric in $\R^{n-1,2}$ as
\eq{\label{theta}
\vs = \diag(-1,1,1,-1,1,\ldots,1) \ ,}
instead of $\diag(-1,-1,1,\ldots,1)$ implicitly assumed in appendix \ref{app:alg}. This metric is useful to define the generators of $\mfso(n-1,2)$ ($\mfJ = -\vs\, \mfJ\, \vs$) used to describe the reduction in the Euclidean world-sheet case. We are simply to replace the generators in appendix \ref{app:alg} by those (which are non-zero in the same row and column) constructed using \eqref{theta}. The reality conditions for the fields of the reduced theory then also need to be suitably modified.

Using this prescription the constant matrices $T_\upm$ are given by
\eq{\label{ttttt}
T_\upm = \fr12 \MAT{ 0     & \mp i &    -1 & \mon
\\                   \mp i & 0     &     0 & \mon
\\                   -1    & 0     &     0 & \mon
\\                   \mon  & \mon  & \mon  & \monn } \ ,}
and satisfy the relation $T_\up^\dag = - \vs T_\um \vs$. The factors of $i$ in \eqref{ttttt} preserve the property $\Tr(T_\upm^2) = 0$, which is linked to solving the Virasoro constraints of the string theory. Furthermore, it is clear from \eqref{theta} that the group $K$, whose algebra is spanned by those generators which are only non-zero in the bottom right $(n-2)^2$ block, is modified to $\SO(n-3,1)$. The group $G$ remains as $\SO(n-1,1)$.

Considering the action \eqref{gwzwactmzalt}, taking $T_\upm$ to be given by \eqref{ttttt}, changing the coset for the gauged WZW model to $\fk{\SO(n-1,1)}{\SO(n-3,1)}$ and Wick-rotating to Euclidean space \eqref{modmodmod} (i.e. $\dpl \ra \del \, , \ \dm \ra - \bar \del \, , \ B_+ \ra B \, , \ B_- \ra - \bar B$) we almost arrive at the reduced theory action for the Euclidean case. The final step is to specify the reality conditions for the fields $g$, $B$, $\bar B$: we let them take values in the complexified group/algebra and then restrict them as follows
\eq{\label{realityaa}
g^\dag = \vs \, g \, \vs \ , \ \ \ \ \ \ \qquad B^\dag = - \vs \, \bar B \, \vs \ .}
The reality condition for $g$ implies that it does not take values in a real subgroup of $\SO(n,\C)$, which is the complexification of $\SO(n-1,1)$. These conditions (which follow from the derivation of the reduction below) can be found also by demanding that the Euclidean counterpart of the action \eqref{gwzwactmzalt} is real.

For the space-like world-sheet the Maurer-Cartan one-form of the string AdS coset sigma model should satisfy the following reality condition
\eq{\label{reality1111}
\Pc^\dag = - \vs \, \bar{\Pc} \, \vs \ , \qquad \qquad \Ac^\dag = -\vs \, \bar{\Ac} \, \vs \ .}
As in the time-like case we solve the Virasoro constraints $\Tr(\Pc^2) = \Tr(\bar{\Pc}^2) = 0$ in terms of $T_\upm$\,\footnote{As before, we could alternatively solve the Virasoro constraints for finite $\m$ and introduce two algebra automorphisms to find a non-trivial $\m \ra 0$ limit
\aln{
\bar{\Pc} = \m g_1^{-1} \e_{\up}(\bar T) \gol \ , \qquad & \Pc = \m g_2^{-1} \e_{\um}(T) \gtl \ ,
\\  T =       \MAT{ 0 & i & 0 & \mon
     \\             i & 0 & 0 & \mon
     \\             0 & 0 & 0 & \mon
     \\    \mon & \mon & \mon & \monn }\ , \qquad & \bar T = - \vs T^\dag \vs \ ,
\qquad \e_{\upm}(\mfJ) = \m^{\mp i R} \mfJ \m^{\pm i R} \ .}
A matrix form  of  the generator $R$ is specified below in \eqref{rr}.}
\eq{
\bar{\Pc} = g_1^{-1} T_\up \gol \ , \qquad \qquad \Pc = g_2^{-1} T_\um \gtl \ .}
The reality condition \eqref{reality1111} implies that $g_2^\dag = \vs \, g_1^{-1} \, \vs$. Recalling that $g = \gol g_2^{-1}$ we get the condition on $g$ in \eqref{realityaa}.

The resulting action for the reduction of AdS strings with Euclidean world-sheet is therefore given by\,\footnote{Note that we have modified the action by an overall minus sign. An extra minus sign has also been introduced in potential term -- this accounts for the minus sign in the redefinition $\Pc_- \ra -\bar \Pc$ when Wick rotating to the Euclidean space.}
\eq{\label{gwzwspace} \Act = \ \fr{\coup}{8\pi} \Tr \Big[ \fr{1}{2} & \,\int d^2x \ \ \inv g \del g \ \inv g \bar \del g \ - \fr{i}{3}\,\int d^3x \; \ \e^{mnl} \ \inv g \del_m g \ \inv g \del_n g \ \inv g \del_l g
\\ + & \,\int d^2x \; \ \big( B \bar \del g \inv g - \bar B \inv g \del g - \inv g B g \bar B + B \bar B \big) + \,\int d^2x \ \; \inv g T_\up g T_\um \Big] \ ,}
where $T_\upm$ are defined in \eqref{ttttt}, and $g \in \SO(n,\C)$ and $B,\,\bar B \in \mfso(n-1,\C)$ satisfy the reality conditions \eqref{realityaa}.

\

The action \eqref{gwzwc} was derived from \eqref{gwzwactmzalt} using the parametrization of $g$ given in \eqref{gparam}. This parametrization is not consistent with the reality condition \eqref{realityaa}. However, using the gauge transformation\,\footnote{Here we have schematically used the notation $k^\fr12$ to denote an element that squares to $k$. As we are working with connected groups the square root can be defined by first rotating $k$ to live in the Cartan subgroup and then introducing the appropriate branch cuts for each factor of $U(1)$, $\R$ or $\C$.}
\eq{
g \ra k^{\fr12} g k^{-\fr12} \ , \qquad B_\pm \ra k^{\fr12} B_\pm k^{-\fr12} - \dpm k^{\fr12} k^{-\fr12} \ ,}
under which \eqref{gwzwc} is invariant, the parametrization \eqref{gparam} is put into a form that can be made consistent with the reality condition \eqref{realityaa}, i.e.
\eq{\label{555}
g = k^{-\fr12} e^{\yp} e^{-2i \p R} e^{\ym} k^{-\fr12} \ , \qquad \yp^\dag = \vs \, \ym \, \vs \ ,\qquad k^\dag = \vs \, k \, \vs \ .}
Here $R$ is the element of the real algebra $\mfso(n-1,2)$ ($R^\dag = - \vs\,R\,\vs$)
\eq{\label{rr} R = \MAT{
   0 & 0  & 0 & \mon
\\ 0 & 0  & 1 & \mon
\\ 0 & -1 & 0 & \mon
\\ \mon & \mon & \mon & \monn} \ .}
Assuming that $\p$ is real, the factor of $i$ in front of $\p R$ in \eqref{555} is required to satisfy the reality condition \eqref{realityaa}. The fields $\ypm$ take values in the complexification of the abelian algebras $\mfl_\upm$, spanned by the generators
\eq{\S_{\upm\,1} = \fr12 \MAT{
0 & 0 & 0 & 0 & \mont
\\
0 & 0 & 0 & 1 & \mont
\\
0 & 0 & 0 & \mp i & \mont
\\
0 & 1 & \mp i & 0 & \mont
\\
\mont & \mont & \mont & \mont & \monnt}
\ , \ \ \ \ \
\S_{\upm\,\ISC} = \fr12 \MAT{
0 & 0 & 0 & 0 & \mont
\\
0 & 0 & 0 & 0 & \iont
\\
0 & 0 & 0 & 0 & \mp i \, \iont
\\
0 & 0 & 0 & 0 & \mont
\\
\mont & - \iont & \pm i \, \iont & \mont & \monnt}}
where $E_{\ISC}$ has zeroes in all entries apart from 1 in the $\ISC$-th entry. These matrices satisfy the following relations ($i=(1,\ISC)$)
\eq{
\S_{\up\,i}^\dagger = -\vs \, \S_{\um\,i} \, \vs \ , \qquad \Tr(\S_{\up\,i}\S_{\um\,j}) = -h_{ij} \ , \qquad \com{T_\upm}{\S_{\upm\,i}} = 0 \ .}
Furthermore, as $\ypm$ take values in the complexification of $\mfl_\upm$, the fields parametrizing them should be complex-valued. These should then satisfy a reality condition coming from the reality condition for $\ypm$ given in \eqref{555}. Finally, the field $k$ takes values in $\SO(n-2,\C)$ (i.e. complexification of $\SO(n-3,1)$) such that it satisfies the reality condition in \eqref{555}.

Substituting the parametrization \eqref{555} into \eqref{gwzwspace} along with the field redefinitions\,\footnote{These redefinitions are consistent with the reality conditions for $B$, $\bar B$ \eqref{realityaa} and $k$ \eqref{555}.}
\eq{B \ra k^{-\fr12} B k^{\fr12} + k^{-\fr12} \del k^{\fr12} \ , \qquad \qquad \bar B \ra k^{\fr12} \bar B k^{-\fr12} - \bar \del k^{\fr12} k^{-\fr12} \ ,}
we find the Euclidean world-sheet version of the reduced theory action \eqref{gwzwc}:
\eq{\label{gwzwcspace}
\Act = \ \fr{\coup}{8\pi}\bigg( & 4 \int d^2 x \; \big[ \del \p \bar \del \p + \fr14 e^{2\p}\big] + \int d^2x\; e^{2\p} \Tr\big[\bar D\ym D\yp\big]
\\ & - \Tr \Big[\,\fr{1}{2}\,\int d^2x \ \ \inv k \del k \inv k \bar\del k \ - \fr{i}{3}\,\int d^3x \; \ \e^{mnl} \ \inv k \del_m k\ \inv k \del_n k \ \inv k \del_l k
\\ & \qquad \qquad \, + \,\int d^2x \; \ \big(B \bar \del k \inv k - \bar B \inv k \del k - \inv k B k \bar B + B \bar B\big)\Big]\bigg)\ .}
where $D \yp = \del \yp + \com{B}{\yp}$ and $\bar D \ym = \bar \del \ym + \com{\bar B}{\ym}$. To summarize, the fields $\ypm$ take values in the complexification of $\mfl_\upm$, $k \in \SO(n-2,\C)$ and $B,\bar B \in \mfso(n-2,\C)$, subject to the following reality conditions
\eq{
\p^* = \p \ , \qquad \yp^\dag = \vs\,\ym\,\vs \ ,\qquad k^\dag = \vs\,k\,\vs \ , \qquad B^\dag = -\vs\,\bar B\,\vs \ .}
This action can be again rewritten in a simplified form following the discussion in section \ref{fsota}.

\renewcommand{\theequation}{7.\arabic{equation}}
\setcounter{equation}{0}
\section{Summary and some open problems\label{sum}}

In this paper we have constructed an action \eqref{gwzwc} that leads to the Pohlmeyer-reduced equations for a string moving in $\ads_n$. This action takes the form of a Liouville part plus a gauged WZW model for the coset $K/K$, where $K=\SO(n-2)$, coupled through a term containing two extra fields transforming as vectors under $K$. This action was found by taking a limit of the previously known action for Pohlmeyer reduction of strings in $\ads_n \X S^1$ (which is given by a $\SO(n-1)/\SO(n-2)$ gauged WZW model with a potential). The limit involved is similar to the group contraction leading from orthogonal to Euclidean groups and thus model has some similarity with gauged WZW models based on non-semisimple groups (see, e.g., \cite{Nappi:1993ie,FigueroaO'Farrill:1994he,Klimcik:1994wp}). We furthermore showed that after a change of variables, one can integrate out the vector fields to get an action of the form of a non-abelian Toda model coupled in a specific way to the Liouville model.

We generalized this limiting procedure to the Pohlmeyer reduction of the $\ads_5 \X S^5$ superstring and thus found the extension \eqref{final} of the action \eqref{gwzwc}, which includes the contribution of the $S^5$ part and the fermions. A particular truncation of this action that describes the Pohlmeyer-reduced $\ads_2 \X S^2$ superstring was shown to be precisely the $\Nc = 2$ supersymmetric Liouville model. A realisation of 2-d supersymmetry (possibly in a non-local or deformed form) in the $\m \ra 0$ limit of the Pohlmeyer-reduced $\ads_3 \X S^3$ and $\ads_5 \X S^5$ superstring models remains an interesting open question.

Starting from the action \eqref{final} we studied fluctuations around two simple classical solutions, corresponding to a massless geodesic in $\ads_2 \subset \ads_5$ and to the long folded spinning string, and found that the spectra of fluctuation frequencies match those found directly from string theory. With a proper choice of fundamental fields (or integration measure) in the reduced theory this implies that the one-loop part of its partition function should match the one-loop partition function found in string theory. Like in the case of $\mu\not=0$ reduction \cite{Hoare:2009rq,Iwashita:2010tg} this should be true for generic string backgrounds, with time-like (Minkowski) or space-like (Euclidean) world-surface.

One of the peculiarities of the AdS reduction in the case of Minkowski-signature world-sheet is the presence of the opposite-sign kinetic terms in the Pohlmeyer-reduced action \eqref{gwzwc}. The ghost-like modes appear in the gauge-fixed action containing $n-2$ physical degrees of freedom (the same as the number of transverse degrees of freedom for a string in $\ads_n$) found in section \ref{fsota} (see also appendix \ref{app:sc}). One could wonder if this is a consequence of a particular gauge fixing procedure used, but, as can be seen from the analysis of fluctuation spectrum in section \ref{sec:quafluc}, this appears not to be the case. In general, $n-3$ out of $n-2$ modes appear with the same sign, which is opposite to that of the Liouville mode.

The reason for this strange feature remains to be understood, but it can be taken care of by a formal analytic continuation (imaginary rotation) of ``ghost'' modes which keeps the action real. This imaginary rotation (along with the rotation of the world-sheet time) leads precisely to the action of the Pohlmeyer reduction for strings in AdS with Euclidean signature world-sheets, which was found in section \ref{app:glfs}. In this case all fields have same physical sign, thus the action is positive and the Euclidean path integral is well defined.\footnote{In section \ref{app:glfs} we considered only the bosonic case but the full superstring action should be straightforward to find either using an analytic continuation of the action \eqref{eq:rtactionfin} or by direct construction. In the latter case some care needs to be taken over the reality conditions for the fermionic fields. In particular, considering the matrix representation of $\mfpsu(2,2|4)$ described in appendix \ref{app:psu224} we have \ $\W(\mffh_r{}^\dag) = e^{\fr{i\pi}{2}r} \, \mffh_r{}^\dag $, \ where the $\Z_4$ automorphism $\W$ and hermitian conjugation are defined in equations \eqref{eq:st} and \eqref{eq:reality} respectively. This relation is consistent with the string equations for a Minkowski world-sheet \eqref{eq:p1fulleom1}, \eqref{eq:p1fulleom2}, \eqref{eq:p1decompmc1}--\eqref{eq:p1decompmc4}, but not for a Euclidean world-sheet. For example, in the latter case, if one considers the Wick rotation of the two equations \eqref{eq:p1fulleom2} -- $\com{\Pc}{\bar \Qc_{1}} = 0$ and $\com{\bar \Pc}{\Qc_{3}} = 0$ -- one should find that these equations are conjugate to each other. This implies that either the definition of the $\Z_4$ automorphism or hermitian conjugation should be modified such that \ $\W(\mffh_r{}^\dag) = e^{-\fr{i\pi}{2}r} \, \mffh_r{}^\dag $. \ In the end, for the Euclidean-case reduced model the fermionic fields $\YL$ and $\YR$ should be complexified and related to each other by conjugation.}

This suggests that it is for the case of strings with Euclidean world-sheets that the AdS Pohlmeyer reduction may play a more fundamental r\^ole. Surfaces with Euclidean signature appear in the context of the string path integral with Dirichlet boundary conditions at the boundary of AdS, i.e. in the string (strong-coupling) representation of Wilson loop expectation values in the context of AdS/CFT. In fact, it is in this context that the classical equations of the AdS Pohlmeyer reduction were recently actively used (see, e.g., \cite{Jevicki:2007aa,Alday:2009ga,Alday:2009yn,Alday:2009dv,Dorn:2009kq,Dorn:2009gq,Jevicki:2009bv,Burrington:2009bh,Dorn:2009hs,Dorn:2010xt,Ishizeki:2011bf,Papathanasiou:2012hx}).

For computing the superstring path integral over Euclidean surfaces ending at a closed contour at the AdS boundary in the semiclassical (large string tension) approximation, the leading contribution $Z_0 = \exp(-\fr{\sqrt\lambda}{2\pi} \Ac)$ is given by the exponent of the classical string action, proportional to the area of the corresponding minimal surface $\Ac = \fr12 \int d^2 x \ \del Y \cdot \bar \del Y$ (in conformal gauge). This area has a simple representation in the AdS reduced theory, i.e. $\fr{1}{4} \int d^2 x \ e^{2\p}$ (see \eqref{definitions}). The one-loop correction $Z_1$ is expressed in terms of determinants of quadratic fluctuation operators as in, e.g., \cite{Drukker:2000ep}. In view of the above remarks, $Z_1$ should be equal to the one-loop partition function in the reduced theory found by expanding near the background which is a counterpart of the corresponding minimal surface in string theory.

Since the structure of the Pohlmeyer-reduced theory is, in general, simpler than that of the original string theory, this suggests that one of the applications of the reduced-theory action found in this paper may be the computation of 1-loop corrections to minimal surfaces ending on null polygons (which represent null Wilson loops related to gluon scattering amplitudes at strong coupling \cite{Alday:2007hr,Alday:2009ga,Alday:2009yn,Alday:2009dv}). An example is the class of minimal surfaces inside $\ads_3$ bounded by regular null $2k$-polygons. They are described by solutions of the generalized sinh-Gordon equation -- the Euclidean counterpart of \eqref{sec0e2} with $UV \to |p(z)|^2$, where $p(z)$ (with $z= \sigma + i \tau$) is a holomorphic function -- in this case a homogeneous polynomial of degree $k-2$ \cite{Alday:2009ga,Alday:2009yn}. The simplest case of a four-cusp surface \cite{Kruczenski:2002fb,Alday:2007hr}, for which $p$ and $\p$ are constant, is equivalent, in the absence of regularization, to the null cusp (or long folded string \cite{Kruczenski:2007cy}) mentioned in sections \ref{sec:quafluc} and \ref{app:glf}. Detailed study of these and other Wilson-loop related surfaces in the context of the Pohlmeyer-reduced theory remain problems for the future.

\section*{Acknowledgements}

We would like to thank Y.~Iwashita, M.~Kruczenski and R.~Roiban for useful discussions. This work was supported by the ERC Advanced Grant No. 290456.


\appendices

\renewcommand{\theequation}{A.\arabic{equation}}
\setcounter{equation}{0}
\section[Coordinate formulation of Pohlmeyer reduction for bosonic strings \texorpdfstring{\\}{} in AdS space]{Coordinate formulation of Pohlmeyer reduction \\ for bosonic strings in AdS space\label{app:cr}}

In this appendix we review the coordinate formulation of the Pohlmeyer reduction of strings in AdS space, see, for example, \cite{Barbashov:1980kz,Barbashov:1982qz,DeVega:1992xc,Jevicki:2007aa,Dorn:2009kq,Dorn:2009gq,Alday:2009dv}.

\subsection{Basic set of reduced equations of motion\label{bsrem}}

The equations of motion for strings moving in $\ads_n$ can be written in terms of an $\R^{n-1,2}$ vector-valued embedding coordinate field $Y$ (we shall use the signature choice $(-,-,+,\ldots,+)$)
\eq{\label{2}
Y \cdot Y = -1 \ , \qquad \qquad  Y \in \R^{n-1,2} \ .}
In the conformal gauge where
\eq{\label{22}
\dpm Y \cdot \dpm Y = 0 \ ,}
the equations of motion of the AdS sigma model are
\eq{\label{equationsofmotion}
\dpl\dm Y - (\dpl Y \cdot \dm Y) Y = 0 \ .}
Considering a time-like world-sheet, the signature of the vector space spanned by $\{Y,\dpl Y, \dm Y\}$ is $(-,-,+)$ and therefore we can introduce the basis of vectors in the orthogonal space
\eq{
N_i \ , \qquad i = 1, \ldots, n-2 \ ,}
with the following inner products
\eq{\label{2222}
N_i \cdot Y = N_i \cdot \dpl Y = N_i \cdot \dm Y = 0 \ , \qquad \qquad N_i \cdot N_j = \d_{ij} \ ,}
to construct a basis for $\R^{n-1,2}$.

We may then define the fields $\p, \ u_i, \ v_i, \ B_{ij}$ as\,\footnote{Denoting $\del_\t$ by $\dot{}$ and $\del_\s$ by $'$ the Virasoro constraints \eqref{22} imply that $\dot Y \cdot \dot Y + Y' \cdot Y' = 0$. Therefore,
\aln{
\dpl Y \cdot \dm Y = \dot Y \cdot \dot Y - Y' \cdot Y' = - 2 Y' \cdot Y' = 2 \cosh^2\r\  t'{}^2 - 2 \r'{}^2 - 2 \sinh^2\r \ n_{\!_A}'n_{\!_A}' \ ,}
where we have introduced global AdS coordinates:
$Y_0 + i Y_{-1} = \cosh \r \, e^{it} \ , \ \ Y_i = \sinh \r \, n_{\!_A} \ , n_{\!_A} n_{\!_A} = 1 \ , \ \ A = 1 , \ldots , n-1$. It is then clear that a class of physical string solutions, for which we can use the local conformal reparametrizations, $x^\pm \ra f_{_\pm}(x^\pm)$, to choose $t$ to be independent of $\s$, will satisfy $\dpl Y \cdot \dm Y \leq 0$.}
\eq{\label{definitions}
e^{2\p} = - 2 \dpl Y \cdot \dm Y \ , \ \ \ & \qquad u_i = 2 N_i \cdot \dpl^2 Y \ , \ \ \ \qquad v_i = - 2 N_i \cdot \dm^2 Y \ , \quad
\\ B_{\pm ij} & = \del_\pm N_i \cdot N_j \ = - \del_\pm N_j \cdot N_i \ ,}
which will end up as the fields of the reduced theory.

Since $\{Y,\ \dpl Y,\ \dm Y,\ N_i\}$ form a basis for $\R^{n-1,2}$, any other vectors can be written as linear combinations, in particular,
\eq{\label{3333}
\dpl^2 Y = 2\dpl \p \, \dpl Y + \fr12 u_i\, N_i \ , \quad & \text{ } \quad \dm^2 Y = 2 \dm \p \, \dm Y - \fr12 v_i N_i \ ,
\\ \dpl N_i = u_i e^{-2\p} \dm Y + B_{+\,ij} N_j \ , \quad & \text{ } \quad \dm N_i = - v_i e^{-2\p}\dpl Y + B_{-\,ij}N_j \ .}
Taking the inner product of the $\dpl^2 Y$ and $\dm^2 Y$ and using the equations of motion \eqref{equationsofmotion} and constraint equations \eqref{2} and \eqref{22}, we find that the field $\p$ satisfies
\eq{
\label{17} \dpl\dm \p + \fr14(e^{2\p} - u_i v_i e^{-2\p}) = 0 \ .}
Furthermore, it is possible to show that $u_i, \ v_i \ \text{and} \ B_{ij}$ satisfy the following equations
\al{\label{row1} \dm u_i = B_{-\,ij} u_j \ , & \qquad \qquad \dpl v_i = B_{+\,ij} v_j \ ,
\\  \label{row2} F_{-+\,ij} \equiv \dm B_{+\,ij} - \dpl B_{-\,ij} - B_{-\,ik} & B_{+\,kj} + B_{+\,ik} B_{-\,kj} = \fr12 e^{-2\p}(u_i v_j - u_j v_i)\ .}
The system of equations \eqref{17}, \eqref{row1}, \eqref{row2} has an obvious $\SO(n-2)$ gauge invariance ($\Kc(x) \in \SO(n-2)$, cf. \eqref{gsymtr})
\eq{\label{kc}
u_i'= \Kc_{ij} u_j \ , & \ \ \ \ \ \ \ \ \ \ \ \ v_i'= \Kc_{ij} v_j \ , \ \ \ \ \ \ \ \ \ \ \ \Kc_{ik}\Kc_{jk}= \delta_{ij} \ , \\
& B'_{\pm\,ij} = \Kc_{ik} B_{\pm\,kl} \Kc_{jl} + \del_\pm \Kc_{ik} \Kc_{jk} \ .}
As we discuss below, one can simplify the system \eqref{17}--\eqref{row2} by using special $\SO(n-2)$ gauge choices.

While it is not obvious a priori if the full set of equations \eqref{17}--\eqref{row2} can be in general derived from a local Lagrangian, we have shown in section \ref{sec:eom} that they follow, in fact, from the action \eqref{gwzwc}, where $\xi_\pm$ are related to $u_i,\,v_i$ by \eqref{uvypm} and \eqref{foot}.

\subsection{Gauge choices\label{app:scc}}

The $\SO(n-2)$ gauge invariance allows us to solve the equations for $u_i,v_i$ in \eqref{row1} explicitly. For example, we may choose the following gauge
\eq{\la{be}
B_{-\,ij}=0 \ , \ \ \ \ \ \ \ \ \ \ \ B_{+\,ij} = - \kK_{ki} \del_+ \kK_{kj} \ , \ \ \ \ \ \ \ \ \ \ \kK_{ij} = \kK_{ij}(x^+, x^-) \in \SO(n-2) \ ,}
so that the general solutions of \eqref{row1} for $u_i,\, v_i$ may be written as
\eq{\la{ut}
u_i = \tyu_i(x^+) \ , \ \ \ \ \ \ \ \ \ \ \ \ v_i = \kK_{ji}(x^+,x^-)\,\tyv_j(x^-) \ .}
Then \eqref{row2} becomes
\eq{\la{yu}
\dm(\kK_{ki} \del_+ \kK_{kj}) = - \fr12 e^{-2\p}(\tyu_i\kK_{kj} - \tyu_j\kK_{ki})\tyv_k \ .}
This equation follows from the $\SO(n-2)$ WZW action for $\kK$ with a potential $e^{-2\p} \, \tyv_i \kK_{ij} \tyu_j$. Including the Liouville term, the resulting action that leads to both \eqref{17} and \eqref{yu} is
\eq{\label{gwcoc}
\Act = \ \fr{\coup}{8\pi}\Big( 4 \int d^2 x \; \big[\dpl \p \dm \p - \fr14 e^{2\p}\big]- \int d^2x\; e^{-2\p} \, \tyv_i \kK_{ij} \tyu_j \ - \ I (\kK)\Big) \ ,}
where the WZW term is defined in \eqref{wwzz}. This action is equivalent to \eqref{gwzwcoa} with $\tilde k \to \kK$. Note that the WZW term enters with the opposite (``ghost'') sign compared to the Liouville term.\footnote{For example, expanding $\kK= \delta_{ij} + \zeta_{ij} + \ldots$, where $\zeta_{ij} = - \zeta_{ji}$, for the Lagrangian in \eqref{gwcoc} we get
\aln{
{\cal L} = \dpl \p \dm \p - \fr14 e^{2\p} - \fr14 e^{-2\p} (\tyv_i\tyu_i + \tyv_i\zeta_{ij}\tyu_j + \ldots) - \fr18(\dpl\zeta_{ij}\dm\zeta_{ij} + \ldots) \ ,}
so that the variation over $\p$ leads to \eqref{17} while the variation over $\zeta_{ij}$ gives the leading term in \eqref{yu}.}

Furthermore, since the WZW term has a residual chiral $\SO(n-2) \times \SO(n-2)$ symmetry under $\kK \to \Kc_-(x^-)\kK(x^+, x^-)\Kc_+(x^+)$, we may further simplify this action by rotating $\tyu_i$ and $\tyv_i$ to fixed directions; denoting their norms as $U=U(x^+),\ V=V(x^-)$ we end up with
\eq{\label{gwoc}
\Act' = \ \fr{\coup}{8\pi}\Big( 4 \int d^2 x \; \big[\dpl \p \dm \p - \fr14 e^{2\p}\big] - \int d^2x\; e^{-2\p} \, U V \, \uu^{(0)}_i\kK_{ij}\vv^{(0)}_j - I(\kK)\Big) \ ,}
where $\uu^{(0)}_i$ and $\vv^{(0)}_i$ are fixed constant unit vectors. This action is essentially equivalent to \eqref{gwzwc2}, and takes the form of a non-abelian Toda model coupled to the Liouville model.

\

In general, one may always parametrize $u_i$ and $v_i$ as
\eq{\label{eqs23}
u_i \equiv U \uu_i \ , \ \ \ \qquad v_i \equiv V \vv_i \ , \ \ \ \ \ \ \ \ \uu_i\uu_i=1 \ , \ \ \ \ \ \ \vv_i\vv_i=1 \ ,}
and then, in view of the antisymmetry of $B_{ij}$, the first-order equations for $u_i$ and $v_i$ \eqref{row1} imply that
\eq{\label{eqs223}
\dm U = 0 \ , \qquad \text{ } \qquad \dpl V = 0 \ ,}
i.e.
\eq{\label{xxp}
U = U(x^+) \ , \ \ \ \ \ \ \ \ \ \ V = V(x^-) \ .}
For topologically trivial situations one can use the residual conformal diffeomorphism freedom to fix $U$ and $V$ to be constants, so that $u_i$ and $v_i$ are then parametrized by two unit vectors $\uu_i,\,\vv_i$.

An alternative (partial) $\SO(n-2)$ gauge to \eqref{be} that we will use in the examples below is to fix
\eq{\la{utu}
\uu_i = (1, 0, 0, \ldots, 0) \ .}
The unit vector $\vv_i$ can be parametrized, e.g., by coordinates of an $(n-3)$-dimensional sphere as in the explicit examples discussed below. The gauge-fixing \eqref{utu} leaves a residual $\SO(n-3)$ gauge symmetry and also breaks the global part of the gauge symmetry to $\SO(n-3)$. It is therefore convenient to split $\vv_i$ as
\eq{
\vv_i = (\vv_1,\vv_\ISC) \ , \ \ \ \ \ \ \ \ \ \ \vv_1 = \sqrt{1-\vv_\ISC\vv_\ISC} \ , \ \ \ \ \ \ISC = 2,\ldots,n-2 \ .}
The first-order equations for $u_i$ and $v_i$ in \eqref{row1} then reduce to
\eq{\label{uvfo}
B_{-\,i1} = 0 \ , \ \ \ \qquad \dpl \vv_i - B_{+\,ij} \vv_j= 0 \ , \ \ \ \qquad \dm U = \dpl V = 0 \ ,}
while the equation for $F_{-+}(B)$ in \eqref{row2} projected onto the $\mfso(n-3)$ subalgebra becomes
\eq{\label{flatfalt}
F_{-+\,\ISC\JSC} = 0 \ .}
Hence, using the residual $\SO(n-3)$ gauge symmetry it is possible to set the components $B_{\pm\,\ISC\JSC}$ to vanish on shell. Then the gauge symmetry is completely fixed and the equations \eqref{uvfo} lead to
\eq{
B_{-\,ij} = 0 \ ,\ \ \ \qquad \dpl \vv_1 - B_{+\,1\ISC} \vv_\ISC = 0 \ , \ \ \ \qquad \dpl \vv_\ISC - B_{+\,\ISC 1} \vv_1 = 0 \ ,}
implying (since $\vv_i\dpl\vv_i = 0$) that
\eq{
B_{+\,1\ISC} = - \fr{\dpl\vv_\ISC}{\sqrt{1-\vv_\ISC\vv_\ISC}} \ .}
Finally the $(1,\ISC)$ component of \eqref{row2} leads to
\eq{\label{eqeq1}
\dm(\fr{\dpl \vv_\ISC}{\sqrt{1-\vv_\ISC\vv_\ISC}}) + \fr12 e^{-2\p} U V \, \vv_\ISC = 0 \ .}
In addition to these $n-3$ second-order equations for $\vv_\ISC$ we also have the equation for $\p$ \eqref{17}
\eq{\label{eqeq2}
\dpl\dm \p + \fr14(e^{2\p} - U V \sqrt{1-\vv_\ISC\vv_\ISC} \, e^{-2\p}) = 0 \ ,}
where $U,V$ are given by \eqref{xxp}.

While the set of equations of motion \eqref{eqeq1}, \eqref{eqeq2} describes just physical $n-2$ degrees of freedom, it cannot be found from a local Lagrangian. As for the $\ads_5$ case discussed below, this is a consequence of the gauge choice \eqref{flatfalt}; we expect there should be a non-local action, related to \eqref{gwzwc2}, whose variational equations are equivalent to \eqref{eqeq1} and \eqref{eqeq2}.

As we review below, for a few low-dimensional examples one can find a simpler non-abelian Toda-like Lagrangian describing the subset of $n-2$ fields not including $U,V$. As discussed above, to derive equations also for the latter requires starting with the full action \eqref{gwzwc} containing also extra degrees of freedom.

\subsection{Special cases\label{app:sc}}

Let us discuss explicitly the examples of $\ads_n$ with $n=2,3,4,5$.

\

$\bf \tads_2$:
\nopagebreak

For strings moving in $\ads_2$ there are no functions $u_i$, $v_i$ and $B_{ij}$ and we end up with just the Liouville equation for $\phi$
\eq{\label{sec0e1}
\dpl \dm \p + \fr14 e^{2\p} = 0 \ ,}
which can be derived from
\eq{
\Lag_2 = \dpl \p \dm \p - \fr14 e^{2\p} \ .}
Since strings in two-dimensional space carry no dynamical degrees of freedom this case is ``topological'': one can fix $\p$ using residual conformal diffeomorphisms under which it transforms so that the Liouville model is conformally invariant (cf. \eqref{definitions}).

\

$\bf \tads_3$:
\nopagebreak

For strings moving in $\ads_3$ the index $i$ takes just one value and $u_1 = U, \ v_1 = V$, i.e. the equations for $u_i$ and $v_i$ in \eqref{row1} are solved already by \eqref{eqs223} and \eqref{xxp}. Then \eqref{17} becomes a (generalized) sinh-Gordon equation
\eq{\label{sec0e2}
\dpl\dm\p + \fr14(&e^{2\p} - U V \, e^{-2\p}) = 0 \ ,}
which follows from the following Lagrangian
\eq{
\Lag_3 = \dpl \p \dm \p - \fr14\big(e^{2\p} + U V \, e^{-2\p}\big) \ .}

\

$\bf \tads_4$:
\nopagebreak

Here the index $i$ runs over $1,\,2$. We shall use the $\SO(2)$ gauge \eqref{utu} and parametrize $\vv_i$ in \eqref{eqs23} as
\eq{\label{sec0e4}
\vv_1 = \cos 2\b \ , \quad \quad \vv_2 = - \sin 2\b \ ,}
The equations \eqref{eqs223} are solved again by \eqref{xxp}. Solving \eqref{row1} for $B_{\pm\,12}$ we find
\eq{
B_{+\,12} = 2 \dpl \b \ , \ \ \ \ \qquad B_{-\,12} = 0 \ .}
Substituting into \eqref{17} and \eqref{row2} we get the following set of equations for $\p$ and $\b$
\eq{
\dpl\dm \p + \fr14(e^{2\p} - U V \cos2\b\, e^{-2\p}) = 0 \ , \qquad \qquad \dpl\dm\b + \fr14 e^{-2\p} U V \sin2\b = 0 \ ,}
which can be derived from the following Toda-like Lagrangian
\eq{\label{4thlag}
\Lag_4 = \dpl\p\dm\p - \dpl\b\dm\b - \fr14(e^{2\p} + U V \cos 2\b \, e^{-2\p}) \ .}
The $\ads_4$ case is the first one where we see the appearance of a ghost-like field ($\beta$) in the reduced action. However, the action remains real if we formally apply the rotation $\beta \to i \beta$ corresponding to switching to the space-like world-sheet version of the reduction.

\

$\bf \tads_5:$
\nopagebreak

For strings moving in $\ads_5$ the index $i$ runs over $1,2,3$. Here we again fix the $\SO(3)$ gauge symmetry as in \eqref{utu} and also parametrize $\vv_i$ in \eqref{eqs23} as
\al{\label{seeq}
\vv_1 = \cos 2\b \ , \ \ \ \ \ \ \ \ \ \ \ \ \vv_2 = - \sin 2\b \, \cos\c \ , \ \ \ \ \ \ \ \ \ \ \ \ \vv_3 = - \sin 2\b \, \sin\c \ ,}
with $U,V$ again satisfying \eqref{xxp}. Solving the equations \eqref{row1} for $B_{\pm\,ij}$ we find
\al{\no & B_{+\,12} = 2 \dpl \b \cos \c - 2\sin \c \,(\dpl \c \tan \b + \l_+ \tan 2\b) \ , && B_{-\,12} = 0 \ ,
\\  \no & B_{+\,13} = 2 \dpl \b \sin \c + 2\cos \c \,(\dpl \c \tan \b + \l_+ \tan 2\b) \ , && B_{-\,13} = 0 \ ,
\\  \label{seeabove} & B_{+\,23} = 2 \l_+ -\tan^2\b \dpl \c \ , && B_{-\,23} = 2 \l_- + \tan^2\b\dm\c \ ,}
where $\b,\c$ and $\l_\pm$ are four new fields. There is a residual $\SO(2)$ gauge symmetry under which $\c \to \c + \epsilon $ and $\l_\pm \to \l_\pm + \fr12(\pm \tan^2\b - 1)\dpm \epsilon$. Equation \eqref{row2} gives three independent relations for $\b,\c,\l_\pm$ and using the residual gauge symmetry and an appropriate linear combination of these equations it is possible to fix $\l_\pm = 0$ on-shell.

Equation \eqref{row2} together with \eqref{17} then lead to the following equations for $\p$, $\b$ and $\c$
\eq{\label{t}
\dpl\dm \p + \fr14(e^{2\p} - & U V \cos 2\b \, e^{-2\p}) = 0 \ , \ \ \ \ \qquad \dm(\tan^2\b \, \dpl \c) + \dpl(\tan^2\b \, \dm \c) = 0 \ ,
\\ & \dpl\dm\b - \sec^2\b \, \tan\b \, \dpl\c\dm\c + \fr14 e^{-2\p} U V \sin2\b = 0 \ ,}
which can be derived from
\eq{\label{5thlag}
\Lag_5 = \dpl\p\dm\p - \dpl\b\dm\b - \tan^2\b \, \dpl\c\dm\c - \fr14(e^{2\p} + U V \cos 2\b \, e^{-2\p}) \ .}
As in the $\ads_4$ case we can get rid of the ghost-like signs by the analytic continuation $\b \to i \b$. Note that without
 the Liouville term $e^{2\p}$ this is equivalent to the Lagrangian of the $B_2 = \mfso(5)$ non-abelian Toda model discussed,
 e.g., in \cite{Gervais:1992bs,Bilal:1993rg,Bakas:1996np}.

Let us note that an alternative to the (on-shell) gauge-fixing \eqref{seeabove} is
\al{\no & B_{+\,12} = 2 \dpl \b \cos \c - \dpl \c \tan2\b \sin\c \ , && B_{-\,12} = 0 \ .
\\  \no & B_{+\,13} = 2 \dpl \b \sin \c + \dpl \c \tan2\b \cos\c \ , && B_{-\,13} = 0 \ .
\\  \label{seeabove1} & B_{+\,23} = 0 \ , && B_{-\,23} = 0 \ .}
Here instead of \eqref{t} we get from \eqref{row2} and \eqref{17}
\eq{\label{eqeqeqeq}
\dpl\dm \p + \fr14(e^{2\p} - & U V \cos2\b\, e^{-2\p}) = 0 \ , \ \ \qquad \dpl(\cos2\b \, \dm \c) - \dm(\sec2\b \, \dpl \c) = 0 \ ,
\\ & \dpl\dm\b - \fr12\tan2\b \, \dpl\c\dm\c + \fr14 e^{-2\p} U V \sin2\b = 0 \ .}
Unlike the system \eqref{t} corresponding to the choice in \eqref{seeabove} these equations cannot be derived from a local Lagrangian. Indeed, in the discussion in section \ref{fsota}, which describes a Lagrangian version of a similar gauge-fixing procedure, we found a non-local action \eqref{gwzwc2ads5} whose variational equations are equivalent to \eqref{eqeqeqeq}. One advantage of this gauge-fixing is that it admits a natural generalization to arbitrary $n$ described in the previous subsection.

\subsection{Residual conformal reparametrizations and degrees of freedom count\label{app:dof}}

The Pohlmeyer reduction of strings in $\ads_n$ described above, was based on using the conformal gauge and solving the resulting Virasoro conditions \eqref{22}. This still leaves part of the original diffeomorphism invariance -- conformal reparametrizations -- unfixed. Indeed, the equations \eqref{17}, \eqref{row1} and \eqref{row2} are invariant under the transformations
\eq{\label{lst}
& \qquad (\dpl, B_+) \ra \Lu (\dpl, B_+) \ , \qquad (\dm, B_-) \ra \Ld (\dm, B_-) \ ,
\\ & \qquad \, \quad e^{2\p} \ra \Lu \Ld e^{2\p} \ , \qquad u_i \ra {\Lu^2} u_i \ , \qquad v_i \ra {\Ld^2} v_i \ ,
\\ & \xxx^\pm \ra \, f_{_{\pm}} (\xxx^\pm) \ , \qquad \Lu = f'_{_{+}} = \Lu(\xxx^+) \ ,\qquad \Ld = f'_{_{-}} = \Ld(\xxx^-) \ .}
As solving the Virasoro condition eliminates only one spurious massless degree we should therefore expect the reduced system of equations to describe the $n-1$ degrees of freedom, rather than the $n-2$ transverse degrees of freedom for a string moving in an $n$-dimensional space.\footnote{In the procedures outlined in section \ref{app:sc} certain equations were understood as constraint equations that should be solved, whereas others were taken to be fundamental. Classically, such choices lead to equivalent results, but they may matter at the quantum level. For example, an alternative approach to that used implicitly in section \ref{app:sc} is to consider the second-order equation for $\p$ and the first-order equations for $u_i$ and $v_i$ as fundamental. Equation \eqref{row2} should then be solved for the field strength, and, along with the gauge symmetry, this would give all the components of $B_{\pm ij}$ in terms of $u_i,\ v_i$ and $\p$. Interpreting $u_i$ and $v_i$ as ``half'' degrees of freedom we then find $n-1$ degrees of freedom.}

We saw a reflection of this in the examples described in appendix \eqref{app:sc}: one may interpret $U$ and $V$, satisfying the first-order equations in \eqref{eqs223}, as describing together one massless degree of freedom.\footnote{In general, $u_i$ and $v_i$ in \eqref{row1} may be interpreted as a generalization of chiral scalars each representing ``half'' of a bosonic scalar degree of freedom.} As in string theory, to get a description only in terms of the physical transverse modes, we are to fix the local conformal reparametrizations \eqref{lcr} and thus eliminate one massless degree of freedom. In view of \eqref{xxp} a natural choice is to fix the freedom \eqref{lst} by fixing $U$ and $V$.

An exceptional case is the Pohlmeyer reduction of strings in $\ads_2$ where we find the Liouville theory \eqref{Liou1eom}, \eqref{Liou1lag}. Equation \eqref{Liou1eom} is solved by
\eq{
\p = \fr12 \log\Big[\fr{16 \a_+' \a_-'}{1+\a_+^2\a_-^2}\Big] \ \ , \ \ \ \ \ \ \a_\pm = \a_\pm(x^\pm) \ ,}
and using the local conformal reparametrizations \eqref{lcr} it is possible to choose $\a_\pm = x^\pm$, thus completely fixing $\p$.

For higher-dimensional reduced AdS models $\p$ no longer satisfies the simple Liouville equation, but rather a generalized sinh-Gordon equation \eqref{eq1}, \eqref{eqeq2}. As a result, fixing $\p$ completely using the freedom in \eqref{lst} is not possible. Hence, while this choice is allowed for the simpler $\ads_2$ reduced theory it is not for the higher dimensional theories. In these cases we can instead use \eqref{lst} to fix the norms of $u_i$ and $v_i$ as discussed above (and also mentioned at the end of section \ref{sec:eom}).

\subsection[Generalization of the coordinate formulation to non-zero \texorpdfstring{$\m$}{m}: string in \texorpdfstring{$\ads_n \X S^1$}{AdSn x S1}]{Generalization of the coordinate formulation to non-zero $\boldsymbol{\m}$: string in $\mathbf{\tads_n \X S^1}$}

We conclude this appendix by briefly discussing the direct generalization of the coordinate formulation of the Pohlmeyer reduction outlined in appendix \ref{bsrem} to non-zero $\m$. In this case we consider a string moving in $\ads_n \X S^1$, for which the equations of motion can again be written in terms of an $\R^{n-1,2}$ vector-valued $\ads_n $ embedding coordinate $Y$ \eqref{2} satisfying \eqref{equationsofmotion} in the conformal gauge as before. Setting the angle on the $S^1$ equal to $\m$ times the world-sheet time to fix the residual conformal reparametrization symmetry, the Virasoro condition is now modified from \eqref{22} to
\eq{\label{note22}
\dpm Y \cdot \dpm Y = -\m^2 \ .}
For a string with a time-like world-sheet embedded in $\R^{n-1,2}$ we find that $\big|\dpl Y \cdot \dm Y\big| \geq \m^2$ so that the signature of the vector space spanned by $\{Y,\dpl Y, \dm Y\}$ is $(-,-,+)$.\footnote{If $\big| \dpl Y \cdot \dm Y\big| \leq \m^2$ the signature of the vector space spanned by $\{Y,\dpl Y,\dm Y\}$ is $(-,-,-)$, and the string cannot be embedded in $\R^{n-1,2}$.} Therefore, we can again introduce the basis of vectors in the orthogonal space $N_i$ ($i = 1, \ldots, n-2 $) with the same inner products as in \eqref{2222} thus getting a basis for $\R^{n-1,2}$. We may then define the fields $\p, \ u_i, \ v_i, \ B_{ij}$ as  (cf. \eqref{definitions})\,\footnote{As for $\m = 0$, we assume that we are considering classical strings where the $\ads$ time in global coordinates is independent of $\s$. This implies that $\dpl Y \cdot \dm Y \leq -\m^2$.}
\eq{\label{notedefinitions}
e^{2\p }\,  +\,  \m^4\, e^{-2\p}\equiv  2 f(\p)    = - 2 \dpl Y & \cdot \dm Y \ ,
\ \ \ \qquad u_i = 2 N_i \cdot \dpl^2 Y \ , \ \ \ \qquad v_i = - 2 N_i \cdot \dm^2 Y \ , \quad
\\ & B_{\pm ij} = \del_\pm N_i \cdot N_j \ = - \del_\pm N_j \cdot N_i \ .}
Since $\{Y,\ \dpl Y,\ \dm Y,\ N_i\}$ form a basis for $\R^{n-1,2}$, other vectors can be written as linear combinations of them (cf. \eqref{3333})
\eq{
\dpm^2 Y = - \frac{f' \, f}{\m^4-f^2} \dpm \p \, \dpm Y + \frac{\m^2 f'}{\m^4-f^2} \dpm \p \, \dmp Y  - \m^2 Y \pm \fr12 \left(\ARR{u_i \\ v_i}\right) \, N_i \ ,
\\
\dpm N_i = \pm \fr12 \left(\ARR{u_i \\ v_i}\right) \frac{\m^2}{\m^4-f^2} \dpm Y \mp \fr12 \left(\ARR{u_i \\ v_i}\right) \frac{f}{\m^4-f^2} \dmp Y + B_{\pm\,ij} N_j \ .}
Taking the inner product of the $\dpl^2 Y$ and $\dm^2 Y$ and using the equations of motion and constraint equations \eqref{2} and \eqref{note22}, we find that the field $\p$ satisfies
\eq{
\label{note17} \dpl\dm \p + \frac{f'}{4} - \frac{u_i v_i}{4f'} = 0 \ .}
One finds also  that  $u_i, \ v_i \ \text{and} \ B_{ij}$ satisfy  (cf. \eqref{row1}, \eqref{row2})
\al{\label{noterow1} \dm u_i + \frac{4 \m^2}{f'} \dpl \p\, v_i  = B_{-\,ij} u_j \ , & \qquad \qquad \dpl  v_i + \frac{4\m^2}{f'} \dm \p\,  u_i = B_{+\,ij} v_j \ ,
\\  \label{noterow2} F_{-+\,ij} \equiv \dm B_{+\,ij} - \dpl B_{-\,ij} - B_{-\,ik} & B_{+\,kj} + B_{+\,ik} B_{-\,kj} = \frac14 \frac{f}{f^2-\m^4}(u_i v_j - u_j v_i)\ .}
The system of equations \eqref{note17}, \eqref{noterow1}, \eqref{noterow2} has an obvious $\SO(n-2)$ gauge invariance given by \eqref{kc}, and reduces to \eqref{17}, \eqref{row1} and \eqref{row2} in the $\m \ra 0$ limit as expected.
One can generalize the various gauge-fixings discussed in appendices \ref{app:scc} and \ref{app:sc} to this case of non-zero $\m$. However, unlike the $\m = 0$ case, the resulting set of equations in general cannot be derived from an action: to find an action one needs to use the construction based on the $\fk FG = \fk{\SO(n-1,2)}{\SO(n-1,1)}$ coset description of the $\ads_n$ sigma model in first-order form leading to the $\fk GH = \fk{\SO(n-1,1)}{\SO(n-1)}$ gauged WZW theory with a potential \cite{Bakas:1995bm,Grigoriev:2007bu}.

\renewcommand{\theequation}{B.\arabic{equation}}
\setcounter{equation}{0}
\section{Definitions of relevant algebras and matrix representations\label{app:algebras}}

\subsection[The algebra \texorpdfstring{$\mfso(n-1,2)$}{so(n-1,2)}]{The algebra $\mathbf{\mfso\boldsymbol{(}n\boldsymbol{-}1\boldsymbol{,}2\boldsymbol{)}}$\label{app:alg}}

For the Pohlmeyer reduction of strings in $\ads_n$ the algebra of interest is $\mfso(n-1,2)$. We choose the following basis for its matrix representation
{\allowdisplaybreaks \aln{ &
T = \MAT{
0 & 1 & 0 & \mon
\\
-1 & 0 & 0 & \mon
\\
0 & 0 & 0 & \mon
\\
\mon & \mon & \mon & \monn
}\ ,
&&
S = \MAT{
0 & 0 & 1 & \mon
\\
0 & 0 & 0 & \mon
\\
1 & 0 & 0 & \mon
\\
\mon & \mon & \mon & \monn
}
\\ &
R = \MAT{
0 & 0 & 0 & \mon
\\
0 & 0 & 1 & \mon
\\
0 & 1 & 0 & \mon
\\
\mon & \mon & \mon & \monn
} &&
\\ &
N_i = \MAT{
0 & 0 & 0 & \ion
\\
0 & 0 & 0 & \mon
\\
0 & 0 & 0 & \mon
\\
\ion & \mon & \mon & \monn
}\ ,
&&
M_i = \MAT{
0 & 0 & 0 & \mon
\\
0 & 0 & 0 & \ion
\\
0 & 0 & 0 & \mon
\\
\mon & \ion & \mon & \monn
}
\\ &
H_i = \MAT{
0 & 0 & 0 & \mon
\\
0 & 0 & 0 & \mon
\\
0 & 0 & 0 & \ion
\\
\mon & \mon & - \ion & \monn
} &&
\\ &
K_{ij} = \MAT{
0 & 0 & 0 & \mon
\\
0 & 0 & 0 & \mon
\\
0 & 0 & 0 & \mon
\\
\mon & \mon & \mon & \ionn
} \ , & \qquad & \text{where} \qquad j > i = 1,\ldots,n-2 \ .}}
\hspace{-3.5pt}Here $\ion$ has zeroes in all entries apart from $1$ in the $i$-th entry, while $\ionn$ has zeroes in all entries apart from 1 in the $i$-th row and $j$-th column and $-1$ in the $j$-th row and $i$-th column.

\subsection[The superalgebra \texorpdfstring{$\mfpsu(2,2|4)$}{psu(2,2|4)}]{The superalgebra \tpsuttf\label{app:psu224}}

Here we present the matrix representation of $\mfpsu(2,2|4)$ which we used in the main text. In particular, we shall make explicit the identification of the $\mfg=\mfusp(2,2) \oplus \mfusp(4)$ subalgebra whose corresponding group is the subgroup $G\subset \Fh $ in the $\fk{\Fh}{G}$ supercoset, and also the group $G$ in the $\fk GH$ gauged WZW model.

Let us define the following $8 \X 8$ matrices
\eq{\label{psu224metric}
& \mathbf{\Q}=\MAT{\Q & \ze \\ \ze & \id } \ , \quad \mathbf{K}=\MAT{ K & \ze \\ \ze & K } \ , \qquad \Q^2 = \id \ , \quad K^2 = -\id \ ,
\\ & \quad \Q = \MAT{1&0&0&0\\0&1&0&0\\0&0&-1&0\\0&0&0&-1} \ , \qquad K = \MAT{0&-1&0&0\\1&0&0&0\\0&0&0&-1\\0&0&1&0}\ .}
A generic element of the algebra $\mfpsu(2,2|4)$ can then be written as follows
\eq{\label{eq:psu}
\mffh = \MAT{ \mfA & \mfX \\ \mfY & \mfB } \ ,\qquad \qquad \Tr \mfA = \Tr \mfB = 0 \ ,}
satisfying the reality condition\,\footnote{Our convention for the conjugation of Grassmann-odd numbers is $(ab)^* = a^* b^*$. The definition of hermitian conjugation in \eqref{eq:reality} is therefore consistent with $(\mffh_1 \, \mffh_2)^\dag = \mffh_2{}^\dag\mffh_1{}^\dag$ as required. Note also that $(\mffh{}^\dag)^\dag = \mffh$.}
\eq{\label{eq:reality}
\mffh = - \inv{\mathbf{\Q}} \mffh^\dag \mathbf{\Q} \ , \qquad \qquad \mffh^\dag = \MAT{\mfA^\dag & - i \mfY^\dag \\ - i \mfX^\dag & \mfB^\dag} \ .}
Here $\mfA$ and $\mfB$ are $4 \X 4$ matrices whose components are commuting while $\mfX$ and $\mfY$ are $4 \X 4$ matrices whose components are anticommuting. The reality condition \eqref{eq:reality} then gives the following conditions on $\mfA$, $\mfB$, $\mfX$ and $\mfY$
\eq{
 \S \mfA^\dag \S = - \mfA \ , \qquad
    \mfB^\dag    = - \mfB \ , \qquad
i\S \mfY^\dag    =   \mfX \ , \qquad
i   \mfX^\dag \S =   \mfY \ .}
Thus $\mfA \in \mfsu(2,2)$ and $\mfB \in \mfsu(4)$.

The algebra $\mfpsu(2,2|4)$ can be extended by two central elements. For the matrix representation described above these are given by the identity matrix multiplied by the imaginary unit $i$, which we denote $\mfI$, and the fermionic parity matrix $\mfW$,
\eq{\label{defofiw}
\mfI = \ i \ \diag(1,1,1,1,1,1,1,1) \ , \qquad \qquad \mfW = \ i \ \diag(1,1,1,1,-1,-1,-1,-1) \ .}
The supertrace of $\mfI$ vanishes, hence its union with $\mfpsu(2,2|4)$ gives the superalgebra $\mfsu(2,2|4)$. If $\mfW$ is also included then one finds the superalgebra $\mfu(2,2|4)$.

The algebra $\mfpsu(2,2|4)$ admits a $\Z_4$ grading and the matrix representation that we are using can be decomposed under this grading as follows\,\footnote{The definition of supertransposition in \eqref{eq:st} is consistent with $(\mffh_1 \, \mffh_2)^{st} = \mffh_2^{\,st} \mffh_1^{\, st}$ as required. Note also that $(\mffh{}^{\, st})^{st} = -\mfW \, \mffh \, \mfW$ where $\mfW$ is given in \eqref{defofiw}.}
\eq{\label{eq:az4decomp}
\mffh = \mffh_0 \oplus \mffh_1 \oplus \mffh_2 \oplus \mffh_3 \ ,}
\eq{\label{eq:st}
\W(\mffh) = -\inv{\mathbf{K}} \ \mffh_r^{\, st} \ \mathbf{K} = e^{\fr{i\pi}{2}r} \ \mffh_r \ , \qquad \mffh^{\, st} = \MAT{\mfA^t & -\mfY^t \\ \mfX^t & \mfB^t} \ .}
General elements of $\mff_{0,2}$ take the form
\eq{\mffh_{0,2} = \MAT{\mfA_{0,2} & \ze \\ \ze & \mfB_{0,2}} \ , \qquad \ARR{\mfA_0 = K \mfA_0^t K \ , & \quad \mfA_2 = - K \mfA_2^t K \ , \\ \mfB_0 = K \mfB_0^t K \ , & \quad \mfB_2 = -K \mfB_2^t K \ ,}}
and general elements of $\mffh_{1,3}$ are
\eq{\mffh_{1,3} = \MAT{\ze & \mfX_{1,3} \\ \mfY_{1,3} & \ze} \ , \qquad i \mfX_1 = - K \mfY_1^t K \ , \quad i \mfX_3 = K \mfY_3^t K \ .}
The subspaces of this decomposition satisfy the following commutation relations ($m,n=0,1,2,3$)
\eq{\label{eq:compsu}
\com{\mffh_m}{\mffh_n} \subset \mffh_{\, m + n \! \mod 4} \ .}
Identifying $\mffh_0 = \mfg$ and $\mffh_2=\mfp$, then $\mfg$ forms a subalgebra. It is this algebra $\mfg$ whose corresponding group is $G$ in the $\fk \Fh G$ supercoset and in the $\fk GH$ gauged WZW model.

It is possible to perform a further $\Z_2$ decomposition, allowing one to define the group $H$ in the $\fk GH$ gauged WZW model. To do this we identify the following fixed element $T\in \mffh_2$
\eq{
T = \fr i2 \ \diag (1, 1, -1, -1, 1, 1, -1, -1 ) \ ,\ \ \ \ \ \ \qquad \STr(T^2) = 0 \ ,}
which is non-degenerate in both the $\mfsu(2,2)$ and $\mfsu(4)$ bosonic subalgebras. The $\Z_2$ decomposition is then given by
\eq{
\mffh^\pa_r = - \com{T}{\com{T}{\mffh_r}} \ , \qquad \quad \mffh^\pe_r = - \acom{T}{\acom{T}{\mffh_r}} \ .}
It should be noted that this is an orthogonal decomposition, that is
\eq{\label{eq:apa1}
\mffh = \mffh^\pa \oplus \mffh^\pe \ , \qquad \qquad \STr(\mffh^\pa \ \mffh^\pe) = 0 \ .}
Then
\eq{\label{eq:apa2}
\com{\mffh^\pe}{\mffh^\pe} \subset \mffh^\pe \ , \qquad
\com{\mffh^\pe}{\mffh^\pa} \subset \mffh^\pa \ , \qquad
\com{\mffh^\pa}{\mffh^\pa} \subset \mffh^\pe \ . }
We identify $\mfh = \mffh_0^\pe$, $\mfm = \mffh_0^\pa$, $\mfa = \mff_2^\pe$, $\mfn = \mffh_2^\pa$. $\mfh$ is thus a subalgebra and the corresponding subgroup is then identified as the group $H$ in the $\fk GH$ gauged WZW model.

It is possible to show that $\mfh$ has the following form
\eq{\label{eq:h}
\mfh = [\mfsu(2)]^{\oplus 4} = \MAT{
  \mfsu(2)_1  & \ze        & \ze        & \ze
\\\ze         & \mfsu(2)_2 & \ze        & \ze
\\\ze         & \ze        & \mfsu(2)_3 & \ze
\\\ze         & \ze        & \ze        & \mfsu(2)_4 } \ .}

\subsubsection[Subspaces and subalgebras for \texorpdfstring{$\mu\not=0$}{m != 0} reduction]{Subspaces and subalgebras for $\mathbf{\boldsymbol{\mu}\boldsymbol{\neq}0}$ reduction\label{app:psu2242}}

The physical fields of the Pohlmeyer-reduced theory (for $\m \not= 0$), namely, $\h\big|_{\mfm}$, where $g =\exp\h$ (the bosons) and $\YR$ and $\YL$ (the fermions), take values in $\mffh_0^\pa$, $\mffh_1^\pa$ and $\mffh_3^\pa$ respectively \cite{Hoare:2009fs}. Here we explicitly write down the bases of these subspaces of the superalgebra $\mfpsu(2,2|4)$ used in section \ref{sec:sst}.

An arbitrary element of bosonic subspace $\mffh_0^\pa$ can be written as
\footnotesize
\al{\no
\begin{aligned}
{\rm T}_0^\pa(\vec x) = \MAT{
  0&0&x_1+i x_2&-x_3-i x_4&0&0&0&0
\\0&0&-x_3+i x_4&-x_1+i x_2&0&0&0&0
\\x_1-i x_2&-x_3-i x_4&0&0&0&0&0&0
\\-x_3+i x_4&-x_1-i x_2&0&0&0&0&0&0
\\0&0&0&0&0&0&x_5+i x_6&x_7+i x_8
\\0&0&0&0&0&0&-x_7+i x_8&x_5-i x_6
\\0&0&0&0&-x_5+i x_6&x_7+i x_8&0&0
\\0&0&0&0&-x_7+i x_8&-x_5-i x_6&0&0}
\end{aligned} \ ,}
\normalsize
where the components of $\vec x$ are commuting parameters. An arbitrary element of fermionic subspace $\mffh_1^\pa$ is
\footnotesize
\al{\no
\begin{aligned}
{\rm T}_1^\pa(\vec \a) = \MAT{
  0&0&0&0&0&0&\a_1+i\a_2&\a_3+i\a_4
\\0&0&0&0&0&0&-\a_3+i\a_4&\a_1-i\a_2
\\0&0&0&0&\a_5-i\a_6&-\a_7-i\a_8&0&0
\\0&0&0&0&-\a_7+i\a_8&-\a_5-i\a_6&0&0
\\0&0&-i\a_5+\a_6&i\a_7-\a_8&0&0&0&0
\\0&0&i\a_7+\a_8&i\a_5+\a_6&0&0&0&0
\\i\a_1+\a_2&-i\a_3+\a_4&0&0&0&0&0&0
\\i\a_3+\a_4&i\a_1-\a_2&0&0&0&0&0&0}
\end{aligned} \ ,}
\normalsize
where the components of $\vec\a$ are anticommuting parameters. Finally, an arbitrary element of the fermionic subspace $\mffh_3^\pa$ can be written in terms of $f_1^\pa(\vec\a)$ as
\eq{\label{eq:f3pa}
{\rm T}_3^\pa(\vec\a) = \, 2 T \, {\rm T}_1^\pa(\vec \a) \ .}

\subsubsection[Subspaces and subalgebras for \texorpdfstring{$\m \ra 0$}{m -> 0} limit of the reduction ]{Subspaces and subalgebras for the $\mathbf{\boldsymbol{\m} \ra 0}$ limit of the reduction\label{app:psu2243}}

The choice of matrix $R$ we use for explicit computations in section \ref{sec:sst} is
\al{\label{Rx}
\begin{aligned}
R = \fr12 \MAT{
  0&0 &1&0 & \ze
\\0&0 &0&-1& \ze
\\1&0 &0&0 & \ze
\\0&-1&0&0 & \ze
\\\ze & \ze & \ze & \ze & \ze}
\end{aligned} \ .}
We then have
\eq{
T_\upm \equiv \lim_{\m \ra 0} \ \m \, \m^{\mp R} T \m^{\pm R} =
\fr14 \left(\begin{array}{cccccccc}
 i & 0 & \mp i & 0 & \ze \\
 0 & i & 0 & \pm i & \ze \\
 \pm i & 0 & - i & 0 & \ze \\
 0 & \mp i & 0 & - i & \ze \\
 \ze & \ze & \ze & \ze & \ze
\end{array}\right) \ .}

If an arbitrary element of $\mfh$ is written as
\footnotesize
\al{\no
{\rm T}_0^\pe(\vec\x) =
\left(\begin{array}{cccccccc}
 i \x_1 & \x_2+i \x_3 & 0 & 0 & 0 & 0 & 0 & 0 \\
 -\x_2+i \x_3 & -i \x_1 & 0 & 0 & 0 & 0 & 0 & 0 \\
 0 & 0 & i \x_4 & \x_5+i \x_6 & 0 & 0 & 0 & 0 \\
 0 & 0 & -\x_5+i \x_6 & -i \x_4 & 0 & 0 & 0 & 0 \\
 0 & 0 & 0 & 0 & i \x_7 & \x_8+i \x_9 & 0 & 0 \\
 0 & 0 & 0 & 0 & -\x_8+i \x_9 & -i \x_7 & 0 & 0 \\
 0 & 0 & 0 & 0 & 0 & 0 & i \x_{10} & \x_{11}+i \x_{12} \\
 0 & 0 & 0 & 0 & 0 & 0 & -\x_{11}+i \x_{12} & -i \x_{10}
\end{array}\right) \ ,}
\normalsize
then the corresponding elements of $\mfl_\upm$ are
\scriptsize
\al{\no
{\rm T}_{0\,\upm}^\pe(\vec\x) & \equiv \lim_{\m \ra 0} \, \m\, \m^{\mp R}\, {\rm T}_0^\pe(\vec\x)\, \m^{\pm R}
\\ & \!\!\!\!\!\!\!\!\!\!\!\!\!\!\!\!\!\!\! = \fr14 \left(\begin{array}{cccccccc}
 i (\x_1-\x_4) & \x_2+\x_5+i (\x_3+\x_6) & \mp i (\x_1-\x_4) &\pm( \x_2+\x_5) \pm i (\x_3+\x_6) & \ze \\
 -\x_2-\x_5+i (\x_3+\x_6) &- i (\x_1-\x_4) & \pm(\x_2 + \x_5)\mp i (\x_3+\x_6) & \mp i (\x_1-\x_4) & \ze \\
\pm i (\x_1-\x_4) & \pm( \x_2+\x_5 ) \pm i (\x_3+\x_6) & - i (\x_1-\x_4) & \x_2+\x_5+i (\x_3+\x_6) & \ze \\
\pm( \x_2 + \x_5 ) \mp i (\x_3 +\x_6) & \pm i (\x_1-\x_4) & -\x_2-\x_5+i (\x_3+\x_6) & i (\x_1-\x_4) & \ze \\
\ze & \ze & \ze & \ze & \ze \\
\end{array}\right) \ ,\no}
\normalsize
such that the matrices $\S_{\upm\,i}$ in the reduced superstring theory are given by
\eq{
\S_{\upm\,1} = {\rm T}_{0\,\upm}^\pe(1,0,0,0,0,0) \ ,
\\ \S_{\upm\,2} = {\rm T}_{0\,\upm}^\pe(0,1,0,0,0,0) \ ,
\\ \S_{\upm\,3} = {\rm T}_{0\,\upm}^\pe(0,0,1,0,0,0) \ ,}
while the element spanning the subalgebra of $\mfh$ that commutes with $R$, denoted $\mfk$ in the main text, is given by
\footnotesize
\al{\no
{\rm T}_{0\,\ze}^\pe(\vec\x) = \left(\begin{array}{cccccccc}
 i \x_1 & \x_2+i \x_3 & 0 & 0 & 0 & 0 & 0 & 0 \\
 -\x_2+i \x_3 & -i \x_1 & 0 & 0 & 0 & 0 & 0 & 0 \\
 0 & 0 & i \x_1 & -\x_2-i \x_3 & 0 & 0 & 0 & 0 \\
 0 & 0 & \x_2-i \x_3 & -i \x_1 & 0 & 0 & 0 & 0 \\
 0 & 0 & 0 & 0 & i \x_7 & \x_8+i \x_9 & 0 & 0 \\
 0 & 0 & 0 & 0 & -\x_8+i \x_9 & -i \x_7 & 0 & 0 \\
 0 & 0 & 0 & 0 & 0 & 0 & i \x_{10} & \x_{11}+i \x_{12} \\
 0 & 0 & 0 & 0 & 0 & 0 & -\x_{11}+i \x_{12} & -i \x_{10}
\end{array}\right) \ .}
\normalsize
An arbitrary element of the Grassmann-odd subspace $\mffh_{1\,\um}^\pa$ is given by
\scriptsize
\al{\no
{\rm T}_{1\,\um}^\pa(\vec\a) & \equiv \ \lim_{\m \ra 0} \, \m^{\sfr12}\m^R \, {\rm T}_1^\pa(\vec\a) \, \m^{-R} \\ & = \fr12\left(\begin{array}{cccccccc}
 0 & 0 & 0 & 0 & -\alpha_5+i \alpha_6 & \alpha_7+i \alpha_8 & \alpha_1+i \alpha_2 & \alpha_3+i \alpha_4 \\
 0 & 0 & 0 & 0 & -\alpha_7+i \alpha_8 & -\alpha_5-i \alpha_6 & -\alpha_3+i \alpha_4 & \alpha_1-i \alpha_2 \\
 0 & 0 & 0 & 0 & \alpha_5-i \alpha_6 & -\alpha_7-i \alpha_8 & -\alpha_1-i \alpha_2 & -\alpha_3-i \alpha_4 \\
 0 & 0 & 0 & 0 & -\alpha_7+i \alpha_8 & -\alpha_5-i \alpha_6 & -\alpha_3+i \alpha_4 & \alpha_1-i \alpha_2 \\
 -i \alpha_5+\alpha_6 & -i \alpha_7+\alpha_8 & -i \alpha_5+\alpha_6 & i \alpha_7 - \alpha_8 & 0 & 0 & 0 & 0 \\
 i \alpha_7+\alpha_8 & -i \alpha_5- \alpha_6 & i \alpha_7+\alpha_8 & i \alpha_5+\alpha_6 & 0 & 0 & 0 & 0 \\
 i \alpha_1+\alpha_2 & -i \alpha_3+\alpha_4 & i \alpha_1+\alpha_2 & i \alpha_3- \alpha_4 & 0 & 0 & 0 & 0 \\
 i \alpha_3+\alpha_4 & i\alpha_1 - \alpha_2 & i \alpha_3+\alpha_4 & -i \alpha_1+\alpha_2 & 0 & 0 & 0 & 0
\end{array}\right) \ ,\no}
\normalsize
while an arbitrary element of the Grassmann-odd subspace $\mffh_{3\,\up}^\pa$ is given by
\scriptsize
\al{\no
{\rm T}_{3\,\up}^\pa(\vec\a) & \equiv \lim_{\m \ra 0} \, \m^{\sfr 12} \, \m^{-R} \, {\rm T}_3^\pa(\vec\a) \, \m^R \\ & = \fr12\left(\begin{array}{cccccccc}
 0 & 0 & 0 & 0 & -i \alpha_5- \alpha_6 & i \alpha_7 - \alpha_8 & i \alpha_1 - \alpha_2 & i \alpha_3 - \alpha_4 \\
 0 & 0 & 0 & 0 & -i \alpha_7- \alpha_8 & -i \alpha_5+\alpha_6 & -i \alpha_3 - \alpha_4 & i \alpha_1+\alpha_2 \\
 0 & 0 & 0 & 0 & -i \alpha_5- \alpha_6 & i \alpha_7 - \alpha_8 & i \alpha_1 - \alpha_2 & i \alpha_3 - \alpha_4 \\
 0 & 0 & 0 & 0 & i \alpha_7+\alpha_8 & i \alpha_5 - \alpha_6 & i \alpha_3+\alpha_4 & -i \alpha_1 - \alpha_2 \\
 -\alpha_5-i \alpha_6 & -\alpha_7-i \alpha_8 & \alpha_5+i \alpha_6 & -\alpha_7-i \alpha_8 & 0 & 0 & 0 & 0 \\
 \alpha_7-i \alpha_8 & -\alpha_5+i \alpha_6 & -\alpha_7+i \alpha_8 & -\alpha_5+i \alpha_6 & 0 & 0 & 0 & 0 \\
 \alpha_1-i \alpha_2 & -\alpha_3-i \alpha_4 & -\alpha_1+i \alpha_2 & -\alpha_3-i \alpha_4 & 0 & 0 & 0 & 0 \\
 \alpha_3-i \alpha_4 & \alpha_1+i \alpha_2 & -\alpha_3+i \alpha_4 & \alpha_1+i \alpha_2 & 0 & 0 & 0 & 0
\end{array}\right) \ .\no}
\normalsize


\pdfbookmark[0]{References}{Ref}

\end{document}